\documentclass[a4paper,11pt]{article}
\pdfoutput=1 

\usepackage{jinstpub} 
\usepackage[utf8]{inputenc}
\usepackage{lineno}
\usepackage{textcomp}

\usepackage{xcolor}
\usepackage{graphicx}
\usepackage{subcaption}
\usepackage{upgreek}
\usepackage[LGRgreek]{mathastext}
\usepackage{soul}

\newcommand{\epm}{${\it e}^{\pm}$}

\newcommand{\layireco}{$L_{reco}^{shower-start}$}
\newcommand{\layi}{$L_{true}$}

\newcommand{\ratioavg}{$R_i$}
\newcommand{\intL}{$\lambda_{int}$}

\newcommand{\piL}{$\lambda_{\pi}$}
\newcommand{\radL}{$X_{0}$}
\newcommand{\ifb}{$fb^{-1}$ {}}
\newcommand{\zss}{$z^{shower-start}$}
\newcommand{\zsstrue}{$z^{shower-start}_{true}$}
\newcommand{\Lss}{$L^{shower-start}_{reco}$}

\newcommand{\qgsp}{{\tt QGSP\_FTFP\_BERT\_EMN}}
\newcommand{\ftfp}{{\tt FTFP\_BERT\_EMN}}

\newcommand{\cmsorcid}[1]{\href{https://orcid.org/#1}{\hspace*{0.1em\raisebox{0.5ex}{\includegraphics[width=0.7em]{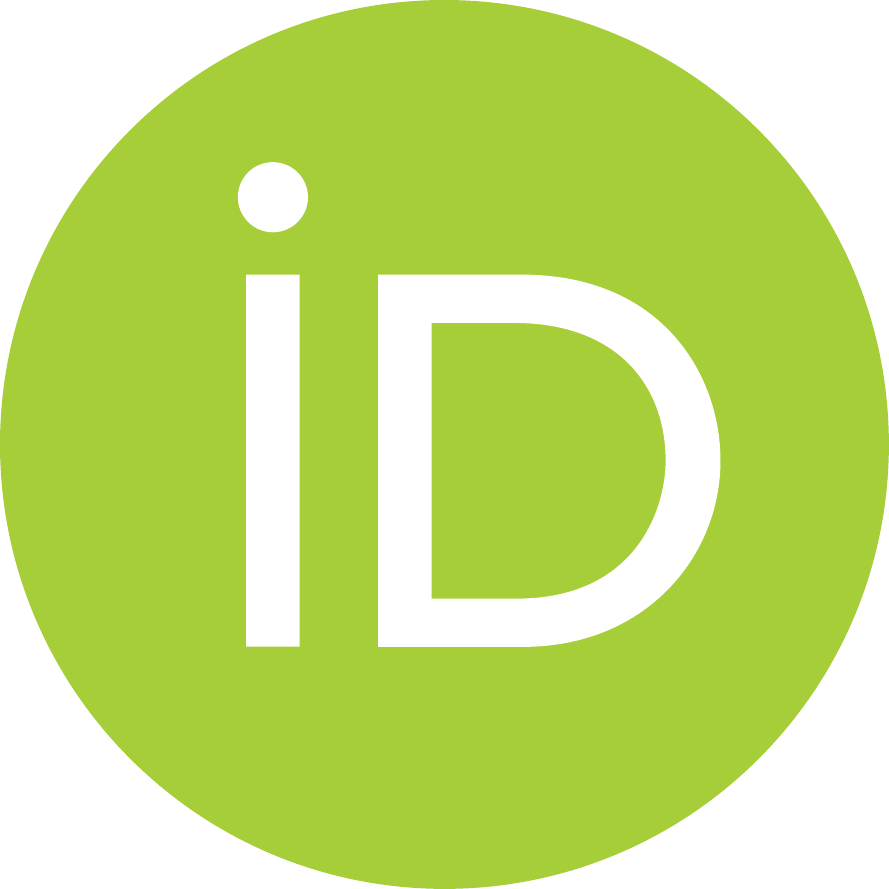}}}}}

\title{\boldmath Performance of the CMS High Granularity Calorimeter prototype to charged pion beams of 20$-$300 GeV/c}

\author[2]{B.~Acar\cmsorcid{0000-0002-4034-7549}}
\author[16]{G.~Adamov}
\author[32]{C.~Adloff}
\author[45]{S.~Afanasiev}
\author[41]{N.~Akchurin\cmsorcid{0000-0002-6127-4350}}
\author[2]{B.~Akg\"{u}n\cmsorcid{0000-0001-8888-3562}}
\author[24]{M.~Alhusseini\cmsorcid{0000-0002-9239-470X}}
\author[6]{J.~Alison\cmsorcid{0000-0003-0843-1641}}
\author[5]{J.~P.~Figueiredo de sa Sousa de Almeida\cmsorcid{0000-0003-4920-8956}}
\author[5]{P.~G.~Dias de Almeida\cmsorcid{0000-0003-4920-8956}}

% ...
\affiliation[2]{Bo\u{g}azi\c{c}i University, \\Bebek 34342, Istanbul, Turkey}
% ...
\affiliation[5]{CERN,\\Espl. des Particules 1, 1211 Geneve 23, Switzerland}
\affiliation[6]{Carnegie Mellon University, \\ 5000 Forbes Ave, Pittsburgh 15213, PA, USA}
% ...
\affiliation[16]{Georgian Technical University, \\ 77 Kostava Str 0175, Tbilisi, Georgia}
% ...
\affiliation[24]{Istanbul Technical University,\\ 34134 Vezneciler-Fatih,  Istanbul, Turkey}
% ...
\affiliation[32]{NICADD, Northern Illinois University, Department of Physics, DeKalb 60115, IL, USA }
% ...
\affiliation[41]{CEA Paris-Saclay, \\ IRFU, Batiment 141,91191, Gif-Sur-Yvette Paris, France}
% ...
\affiliation[45]{Texas Tech University,\\ Lubbock 79409, TX, USA}
% ...

\abstract{The upgrade of the CMS experiment for the high luminosity operation of the LHC comprises the replacement of the current endcap calorimeter by a high granularity sampling calorimeter (HGCAL). The electromagnetic section of the HGCAL is based on silicon sensors interspersed between lead and copper (or copper tungsten) absorbers. The hadronic section uses layers of stainless steel as an absorbing medium and silicon sensors as an active medium in the regions of high radiation exposure, and scintillator tiles directly read out by silicon photomultipliers in the remaining regions. As part of the development of the detector and its readout electronic components, a section of a silicon-based HGCAL prototype detector along with a section of the CALICE AHCAL prototype was exposed to muons, electrons and charged pions in beam test experiments at the H2 beamline at the CERN SPS in October 2018. The AHCAL uses the same technology as foreseen for the HGCAL but with much finer longitudinal segmentation. The performance of the calorimeters in terms of energy response and resolution, longitudinal and transverse shower profiles is studied using negatively charged pions, and is compared to GEANT4 predictions. This is the first report summarizing results of hadronic showers measured by the HGCAL prototype using beam test data. \vspace{10mm}
} 

\graphicspath{{figs/}}

\begin{document}
\maketitle
\flushbottom

\newpage
\section{Introduction}
\label{sec:intro}

The CMS Collaboration is preparing to replace the existing endcap calorimeter (CE) detectors by a high granularity calorimeter (HGCAL) to accommodate the high-luminosity LHC (HL-LHC) operations scheduled to begin later this decade \cite{bib.cms-phase2-techproposal}. The current electromagnetic section, covering the pseudorapidity range 1.479 $<|\eta|<$ 3.0, is a homogeneous calorimeter made of lead tungstate crystals. The hadronic section, covering 1.3 $<|\eta|<$ 3.0, is a sampling calorimeter with plastic scintillators sandwiched between layers of brass absorber \cite{bib.cms-detpaper}. The system is designed to maintain its performance through the ongoing LHC operations, and is expected to accumulate an integrated luminosity of 300 \ifb by the end of Run-3 (2022$-$25) beyond which the physics performance of the system will degrade below an acceptable threshold for offline recovery. During the HL-LHC phase, experiments will accumulate data corresponding to 3000 \ifb of integrated luminosity during a span of ten years, at a leveled instantaneous luminosity of 5 $\times$ 10$^{34}$ cm$^{-2}$s$^{-1}$, resulting in an average number of interactions (pileup) per bunch crossing of 140 to 200 depending on the beam configuration. 

The most salient features of the HGCAL design are its fine transverse and longitudinal segmentation, aimed at facilitating three capabilities: an efficient particle identification, particle-flow reconstruction, and pileup rejection in the high pileup environment at the HL-LHC \cite{bib.cms-hgcalphase2-tdr}. The HGCAL is a sampling calorimeter comprising an electromagnetic section (CE-E) followed by a hadronic section (CE-H), which are longitudinally segmented into 50 layers. It has a total of more than six million channels. The CE-E is composed of 28 layers of silicon modules, sandwiched between layers of Pb and Cu or CuW absorbers. The silicon sensors are hexagonal modules segmented into hexagonal cells ranging in area from about 0.5 cm$^{2}$ to 1.1 cm$^{2}$. The CE-H uses stainless steel as the absorbing medium, longitudinally segmented into 22 layers. The active medium in the hadronic sections are silicon sensors in the regions of high radiation exposure. In the remaining regions of the hadronic calorimeter, approximately square shaped plastic scintillator tiles directly read out by silicon photomultipliers (SiPMs), with sizes approximately ranging from 4 cm$^2$ to 30 cm$^2$, are used. The CE-E corresponds to a depth of $\approx$25 radiation lengths (\radL) for measuring energies deposited by electrons and photons. Together with CE-E, which corresponds to a depth of 1.4 interaction lengths (\intL), the CE-H provides a total depth of $\approx$10 \intL{} for measuring energies deposited by charged and neutral hadrons. 

The CMS HGCAL collaboration has been actively carrying out detailed testing of prototype detector modules and their associated electronic components, using single particles beams at CERN, Fermilab, and DESY. The first prototype silicon modules equipped with the SKIROC2 ASIC were tested in beam test experiments at Fermilab and CERN in 2016 using electron beams \cite{bib.hgcal-2016-tbpaper}. In the 2018 beam test experiments at the H2 beamline \cite{bib.cern-h2-manual} at CERN, a prototype of silicon HGCAL detector equipped with the SKIROC2-CMS ASIC \cite{bib.cern-h1paper} combined with a CALICE Analog Hadronic Calorimeter (AHCAL) \cite{bib.calice-ahcal-construction, bib.calice-ahcal-2018tb-paper} were exposed to beams of positrons and pions in the energy range of 20$-$300 GeV, and muons of 200 GeV. The AHCAL, equipped with the SPIROC2E ASIC \cite{bib.calice-ahcal}, uses the same technology of scintillator tiles directly read out by SiPMs as foreseen for the HGCAL but the size of tiles is $3 \times 3$ cm$^2$ and the longitudinal segmentation is much finer. The response of the combined detector prototype to hadronic showers produced by the negatively charged pions collected in beam test experiments is the main focus of this publication. Various aspects of hadronic shower production simulated with GEANT4 \cite{bib.geant4} are also compared against the data. The prototype performance to positron beams has been reported in \cite{bib.hgcal-2018-positrons}. In this publication, Section \ref{sec:beamtest} describes the detector setup and its simulation, and the datasets used in this analysis. Event reconstruction starting from the raw data and criteria for selecting events for further analysis are summarized in Section \ref{sec:evtreco}. We present an algorithm used to find the depth at which a hadronic shower is initiated in Section \ref{sec:showerstart}, followed by a presentation of the performance of the detector in terms of energy response and resolution in Section \ref{sec:enereco}. Characteristics of the longitudinal and the transverse development of hadronic showers are discussed in Section \ref{sec:long-trans-showers}. This is the first study of the HGCAL prototype to measure performance characteristics using charged pions.

\section{Experimental and simulated detector setups and datasets}
\label{sec:beamtest}

In this section, we describe the experimental setup consisting of the HGCAL detector prototype, its associated data acquisition electronics, and the beamline elements used for characterizing and triggering on the particles arriving at the prototype detector. We provide a brief description of the modeling of the beamline elements and detector prototype in the GEANT4 based simulation. Datasets used to obtain the results presented in this report are briefly summarized at the end of this section.

\subsection{Experimental beam test setup} \label{sec:exptsetup}

Protons accelerated to a momentum of 400 GeV/c by the Super Proton Synchrotron (SPS) are collided with a 500 mm thick Beryllium target. Secondary beams of muons, electrons and pions are extracted from the particles produced in the interaction with the target and are directed to the HGCAL prototype situated almost 600 m downstream via a dedicated beamline of dipole and quadrupole magnets, and collimators. The particles selected in the momenta range of 20$-$300 GeV/c have a momentum spread of 0.2$-$2.0\,\%. The beamline is instrumented with several particle detectors to study the beam purity and to track the beam direction. Hits recorded by a set of four delay wire chambers (DWCs) \cite{bib.cern-h2-dwc}, located at the end of the beamline, are used to measure the trajectory of the beam particles. A coincidence of signal from two scintillator detectors, placed in front of the HGCAL prototype, is used as an external trigger. Two micro-channel plates (MCPs) were used to provide a timing reference for arrival of the incident particles. The relative position of these detectors in the beamline, followed by the HGCAL prototype detectors is shown in the schematic of the experimental setup in Figure \ref{fig:tb-hgcal-exp-schem}. A more detailed description of the integration of signal from these beamline detectors with the HGCAL data acquisition is documented in \cite{bib.cern-h1paper}.

\begin{figure}[htbp]
  \centering
  \includegraphics[width=\linewidth]{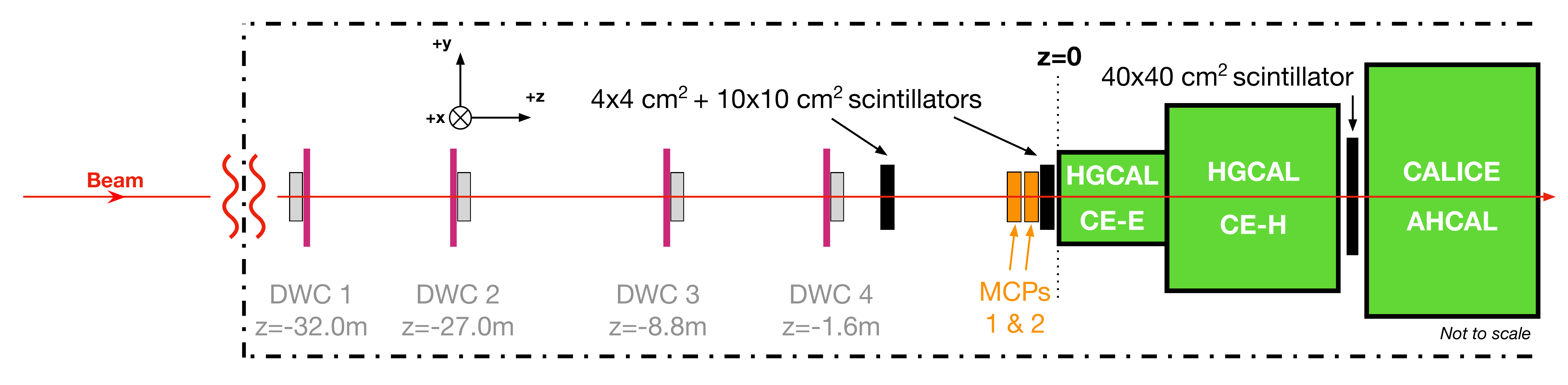}
  \caption{\label{fig:tb-hgcal-exp-schem} Schematic view of the experimental setup of the October 2018 beam test experiment showing DWCs, trigger scintillators and MCPs followed by the CE-E, CE-H, and AHCAL prototypes.}
\end{figure}

The HGCAL prototype used in the 2018 beam test experiments is comprised of an electromagnetic section (CE-E prototype) and a hadronic section including a silicon-based calorimeter (referred as CE-H prototype) followed by a section of the AHCAL prototype \cite{bib.calice-ahcal-2018tb-paper}. The scintillators and the detector setup are mounted on a concrete platform at the H2 beamline area at CERN as shown in Figure \ref{fig:tb-hgcal-exp}. The CE-E prototype makes up the front section of this setup, and is the first detector encountered by the impinging beam particles. It is a sampling electromagnetic calorimeter made of segmented silicon sensors, one per active layer. The absorbers in alternate layers comprise 6 mm thick Cu cooling plates along with 1.2 mm thick CuW baseplates, and 4.9 mm thick Pb plates cladded with 300 $\mu m$ thick stainless steel. With its 28 sampling layers, the total depth of the CE-E corresponds to 26 \radL{} or 1.4 \intL.

The CE-H prototype is a stack of 12 layers of silicon modules, each mounted on a 6 mm thick copper cooling plate, sandwiched between 4 cm thick steel absorber plates. Each of the first 9 layers consists of seven silicon modules arranged in a daisy-like structure, see Figure \ref{fig:tb-fh-ahcal} (left) while the last three layers could only be instrumented with one module. The layers were arranged in two separate boxes for mechanical support which were separated by 4 cm thick steel absorber plates. The total depth of CE-H prototype corresponds to approximately 3.4 \intL. 

A silicon module consists of a 6-inch hexagonal silicon sensor, a copper-tungsten or copper baseplate for mechanical support, and a printed circuit board (PCB) with embedded electronics as described in \cite{bib.cern-h2paper}. Each sensor is subdivided into 128 hexagonal cells of approximately 1.1 $cm^2$ in area resulting in about 3500 channels in the CE-E prototype and about 8500 channels in the CE-H prototype. A total of 94 modules are used in this CE-E and CE-H setup out of which 90 are built from 300 micron-thick sensors. The remaining four sensors are 200 micron thick. These are located at layers 27 and 28 of the CE-E, and at off-centre locations of layers 5 and 6 of the CE-H.

\begin{figure}[htbp]
  \centering
  \includegraphics[width=0.75\linewidth]{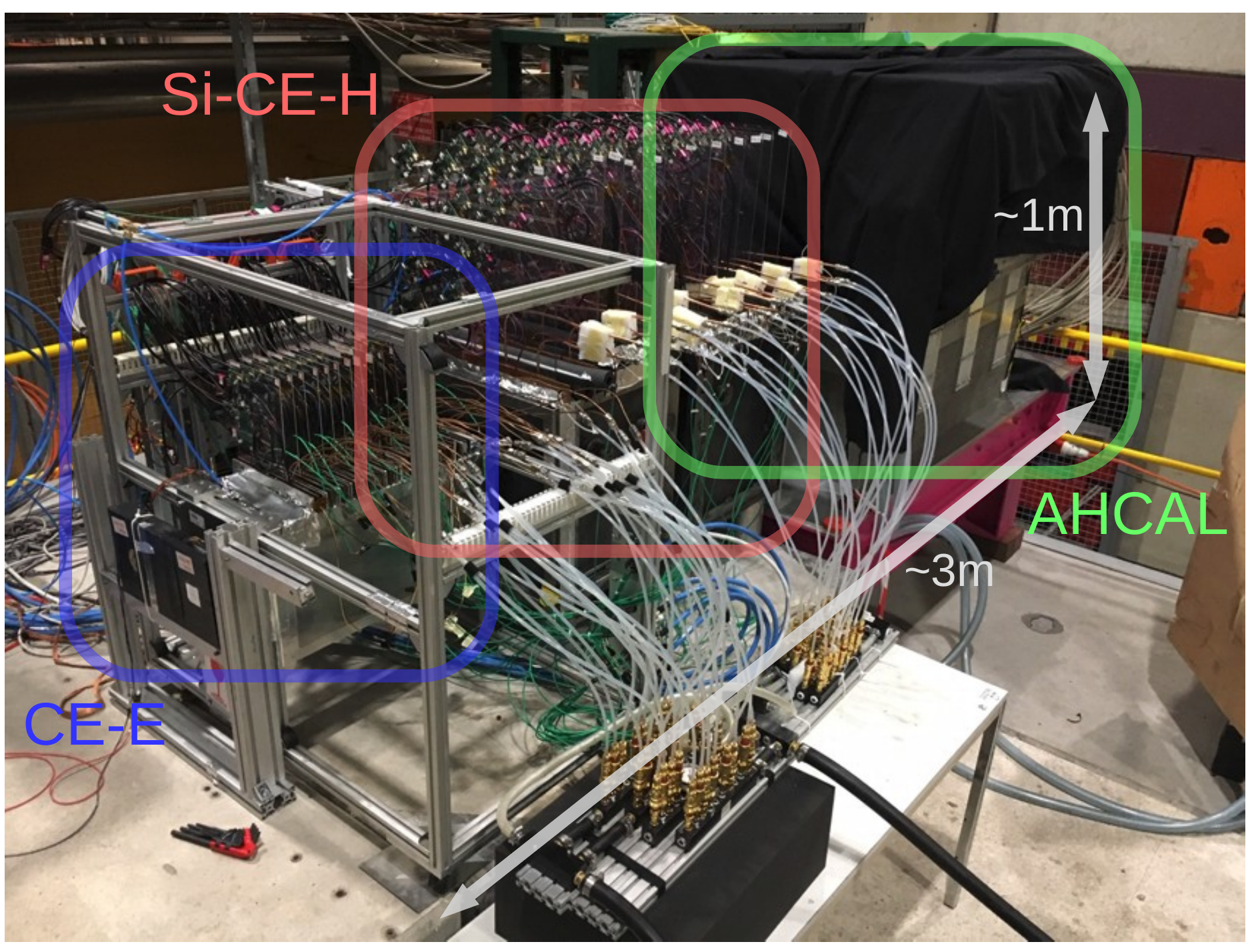}
  \caption{\label{fig:tb-hgcal-exp} A picture of the CE-E, CE-H, and AHCAL prototypes setup mounted on a concrete platform in the H2 experimental area of the CERN SPS.}
\end{figure}

As shown in the Figures \ref{fig:tb-hgcal-exp-schem} and \ref{fig:tb-hgcal-exp}, the hadronic section of the calorimeter is completed by an additional 4.4 \intL{} by a section of the AHCAL prototype \cite{bib.calice-ahcal}. Interleaved between nonmagnetic stainless steel absorber plates of approximately 17 mm thickness are the layers of active elements made of 3\,$\times$\,3\,$\times$\,0.3 $cm^3$ scintillator tiles, individually read out by SiPMs. The scintillator tiles are injection-moulded from polystyrene and have a small dimple at the center which fits the SiPMs mounted on the PCBs. The SiPMs are Hamamatsu type S13360-1325PE, and have a size of 1.3\,$\times$\,1.3 $mm^2$ with 2668 pixels. An illustration of two scintillator tiles mounted on a base unit with SiPMs is shown in Figure \ref{fig:tb-fh-ahcal} (right) where the left tile is unwrapped while the adjacent tile is wrapped in a reflective foil. The prototype used in this beam test experiment consists of 39 sampling layers and approximately 22,000 scintillator tiles, all wrapped in reflective foils, read out independently. With its finer longitudinal segmentation, the AHCAL prototype provides a unique opportunity to study the latter part of the longitudinal development of the hadronic showers in the detector.

\begin{figure}[htbp]
  \centering
  \includegraphics[width=0.35\linewidth]{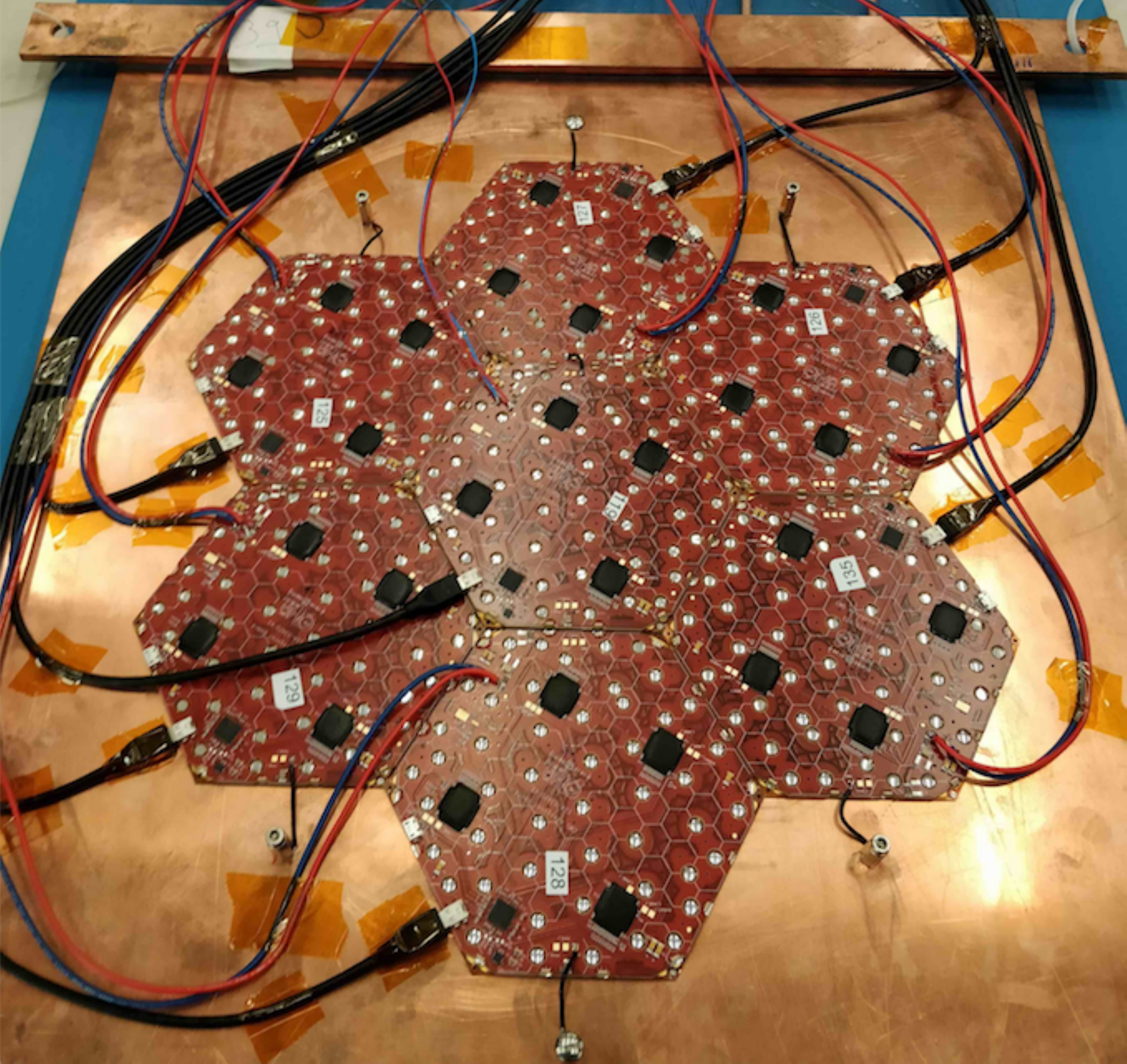}
  \includegraphics[width=0.455\linewidth]{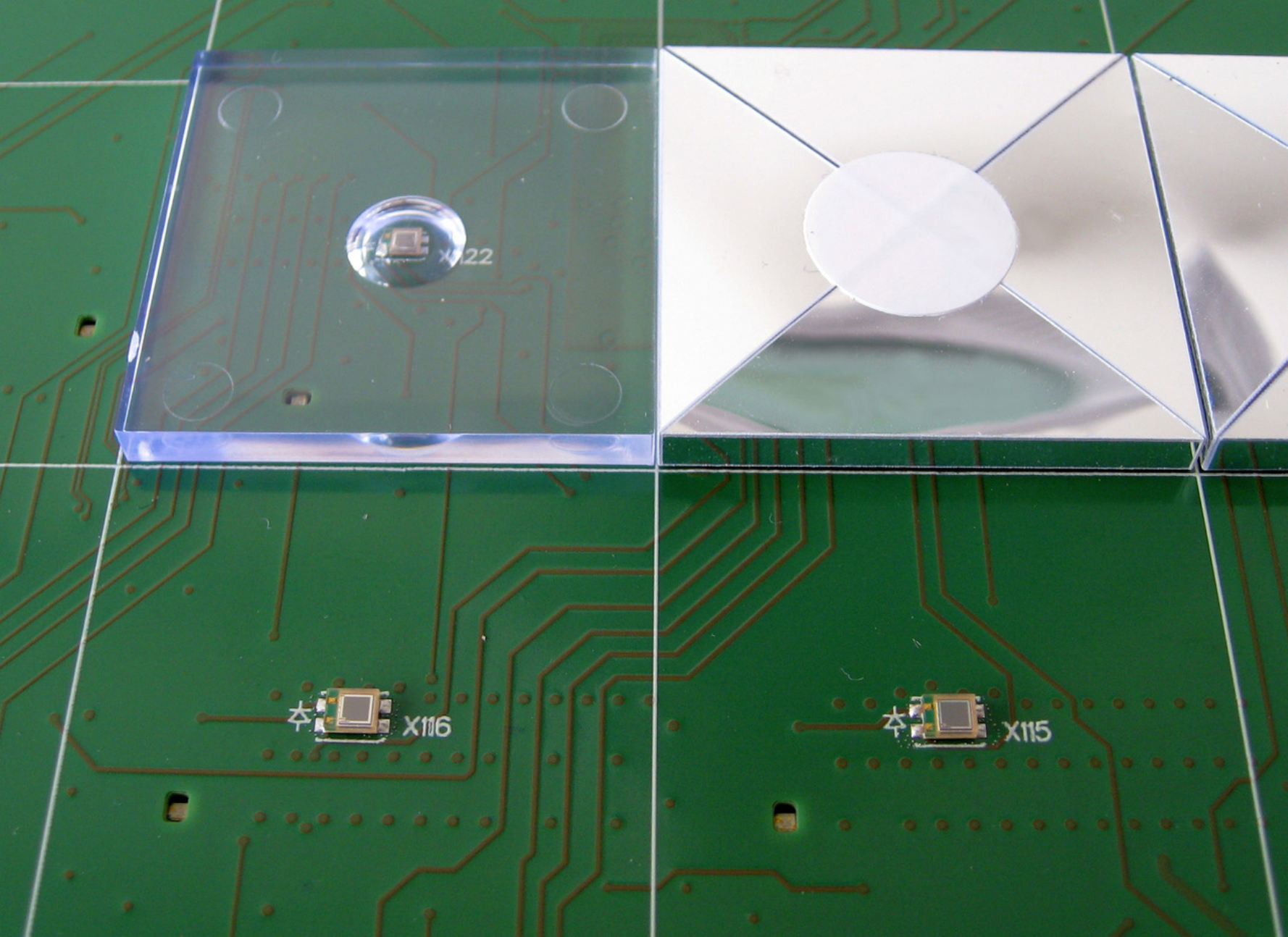}
  \caption{\label{fig:tb-fh-ahcal} Seven silicon modules, arranged in a daisy like structure, which are used to make one layer of the CE-H prototype (left), and scintillator tiles with SiPMs mounted on a base unit (right). The bare tile is shown for illustrative purposes.}
\end{figure}

The signals from the 128 hexagonal silicon cells are collected and processed via four SKIROC2-CMS ASICs \cite{bib.Skiroc2cms} mounted on the PCB as visible in Figure \ref{fig:tb-fh-ahcal} (left), connected to the silicon sensor (not visible in the image) via wire bonds. The signal produced by a charged particle traversing a silicon sensor can range from a few fC, corresponding to a minimum ionizing particle, to 10 pC, corresponding to the signal produced at the core of a shower generated by a highly-energetic electromagnetic particle. The SKIROC2-CMS ASIC achieves this wide dynamic range using two levels of amplification, high-gain and low-gain, and a time-over-threshold (ToT) measurement. It also provides a time-of-arrival (ToA) measurement. The signals corresponding to the two gain settings are sampled at 40 MHz. Thirteen samples of signal digitized with a 12-bit analog-digital-converter (ADC) are saved for the offline analysis. The ToT and the ToA are measured by a TDC with 25 ps time bins \cite{bib.Skiroc2cms}. The low-gain and ToT digitized signal counts are converted offline to an equivalent high-gain ADC count using a dedicated procedure of gain linearization to achieve a consistent signal over the full dynamic range. The front end electronics for silicon-sensor modules and the procedure of gain linearization are described in detail in the reference \cite{bib.cern-h2paper}.

In case of the AHCAL, 144 SiPMs (each collecting light produced in one scintillator tile) are mounted on a base unit of 36$\times$36 $cm^2$. These are electronically read out by four SPIROC2E ASICs \cite{bib.calice-ahcal, bib.calice-ahcal-2018tb-paper}, which operate in a self-triggering mode. This provides a charge measurement in both a high gain mode and a low gain mode, necessary to achieve the large dynamic range required to measure signals of 160 fC up to 320 pC at a SiPM gain of about $10^6$. The SPIROC2E ASIC also provides a time measurement from a TDC. The signal saved in analog memories are internally digitized by ADCs. Out of the three measurements per channel, the high gain, the low gain, and the time, only two can be digitized and read out for further processing. The AHCAL data is synchronized with the HGCAL data using the trigger number and the trigger time-stamp \cite{bib.cern-h1paper}.

\subsection{Beam test setup in simulation} \label{sec:simsetup}

The detector setup along with the beamline elements are simulated using the GEANT4 toolkit \cite{bib.geant4}. Various beamline elements, starting from the production target T2 up to the front of HGCAL prototype, are defined using the G4Beamline simulation framework implemented in GEANT4 version 10.3. The main elements accounted for are bending magnets and pipes, quadrupole magnets, collimators, detectors along the NA61 areas, upstream halo and veto counters, vacuum pipes, and air sections. These elements are defined by their geometry, material, positions and fields. The particles produced at T2 are then tracked through the simulated beamline system using the GEANT4 physics list FTFP\_BERT\_EMZ. All secondary particles that are produced in the interactions of these particles with various materials encountered, as these propagate through the beamline, are passed on to the HGCAL prototype detector simulation along with their corresponding momenta and energies. The simulated position of the beam entering the HGCAL prototype is adjusted to the position measured in data for each beam energy. Energy losses due to synchrotron radiation are negligible for pions at the GeV-scale energies used for the beam test campaign, and hence the main component of momentum spread is that at the T2 target as mentioned in Section \ref{sec:exptsetup}.

The calorimeter sections are simulated using GEANT4 version 10.4.3. A schematic of the simulated HGCAL and AHCAL prototype detectors used in the beam test experiment is shown in Figure \ref{fig:tb-hgcal-sim}. The thickness of the absorbers as seen by particles between the consecutive active layers in units of \intL{} in the simulation is presented in Figure \ref{fig:tb-hgcal-sim-intl}. Along with silicon as active material, the description of the sensor modules include the structural material of the PCB, gold-plated kapton sheet, Cu and CuW baseplates for mechanical support, and cooling structure as well as air gaps as passive materials \cite{bib.cern-h2paper}. In the CE-E prototype, pairs of modules mounted on the copper cooling plates form a mini cassette. These are then interspersed by 4.9\,mm thick lead sheets cladded with 300\,$\mu m$ thick steel. This results in consecutive silicon layers separated by absorbers of thickness ${\it d} \approx$ 0.03 \intL{} and ${\it d} \approx$ 0.06 \intL{} in this section.

\begin{figure}[h]
  \centering
  \includegraphics[width=0.85\linewidth]{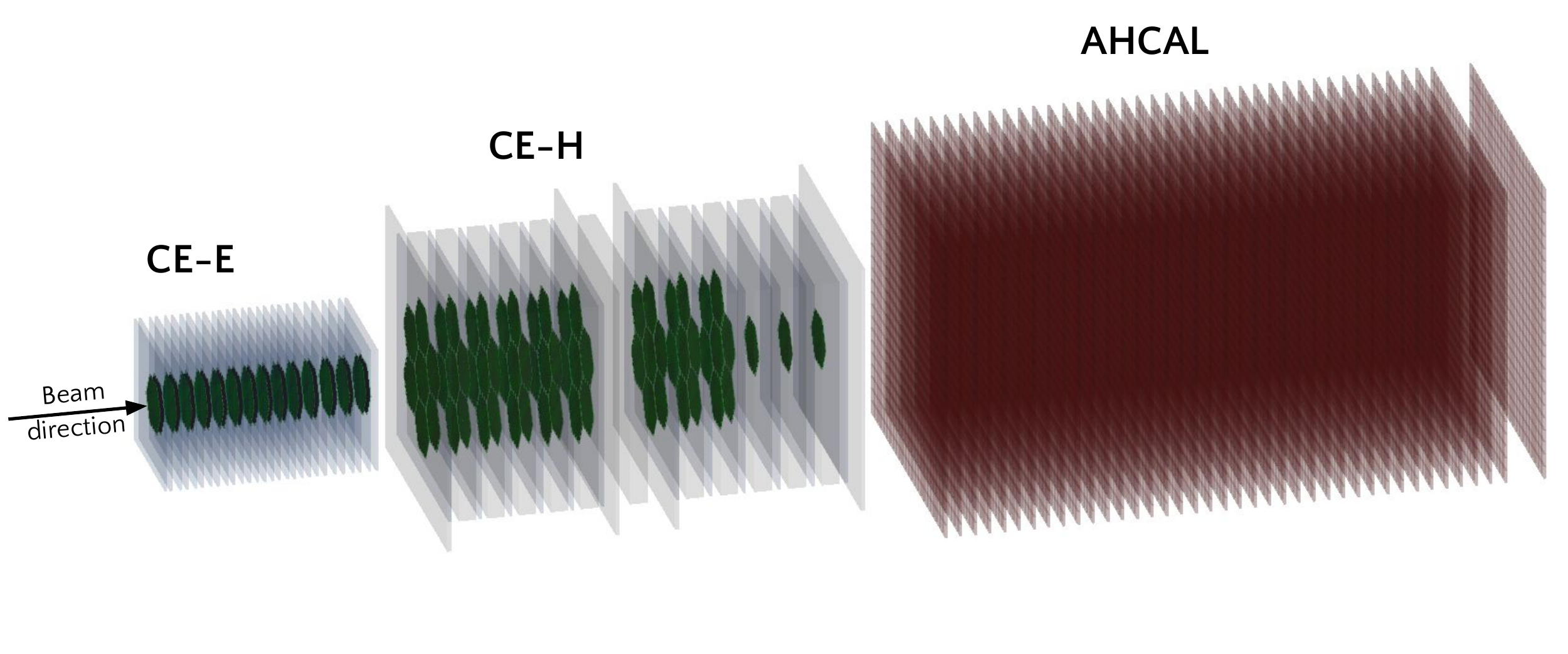}
  \caption{\label{fig:tb-hgcal-sim} Simulated geometry of the CE-E, CE-H and AHCAL detector prototypes depicting the position of active layers and absorbing material.}
\end{figure}

In case of the CE-H prototype, each layer comprises seven modules mounted on a copper plate on one side and steel absorber on the other, resulting in ${\it d} \approx$ 0.3 \intL{} for all the layers. The larger absorber thickness in the middle of the CE-H, at Layer 35, corresponds to extra material due to the walls of iron boxes used to contain the two CE-H prototype sections, as described in Section \ref{sec:exptsetup}. The simulation of the AHCAL prototype proceeds in a similar way with polystyrene scintillator tiles as active elements along with the PCB and mechanical support structure and steel absorber, resulting in ${\it d} \approx$ 0.1 \intL{} for all the layers except the first and the last layers. An additional material of ${\it d} \approx $ 0.1 \intL{} at the Layer 41 in Figure \ref{fig:tb-hgcal-sim-intl} is due to the copper cooling plate and the end wall of the iron box containing the CE-H section. There are three absorber layers upstream of the last active layer of the AHCAL, resulting in an absorber thickness ${\it d} \approx $ 0.3 \intL.

The cumulative depth of the detector in the units of \intL{} is also shown in the Figure \ref{fig:tb-hgcal-sim-intl} as open squares with the corresponding scale on the right y-axis. 

\begin{figure}[h]
  \centering
  \includegraphics[width=0.75\linewidth]{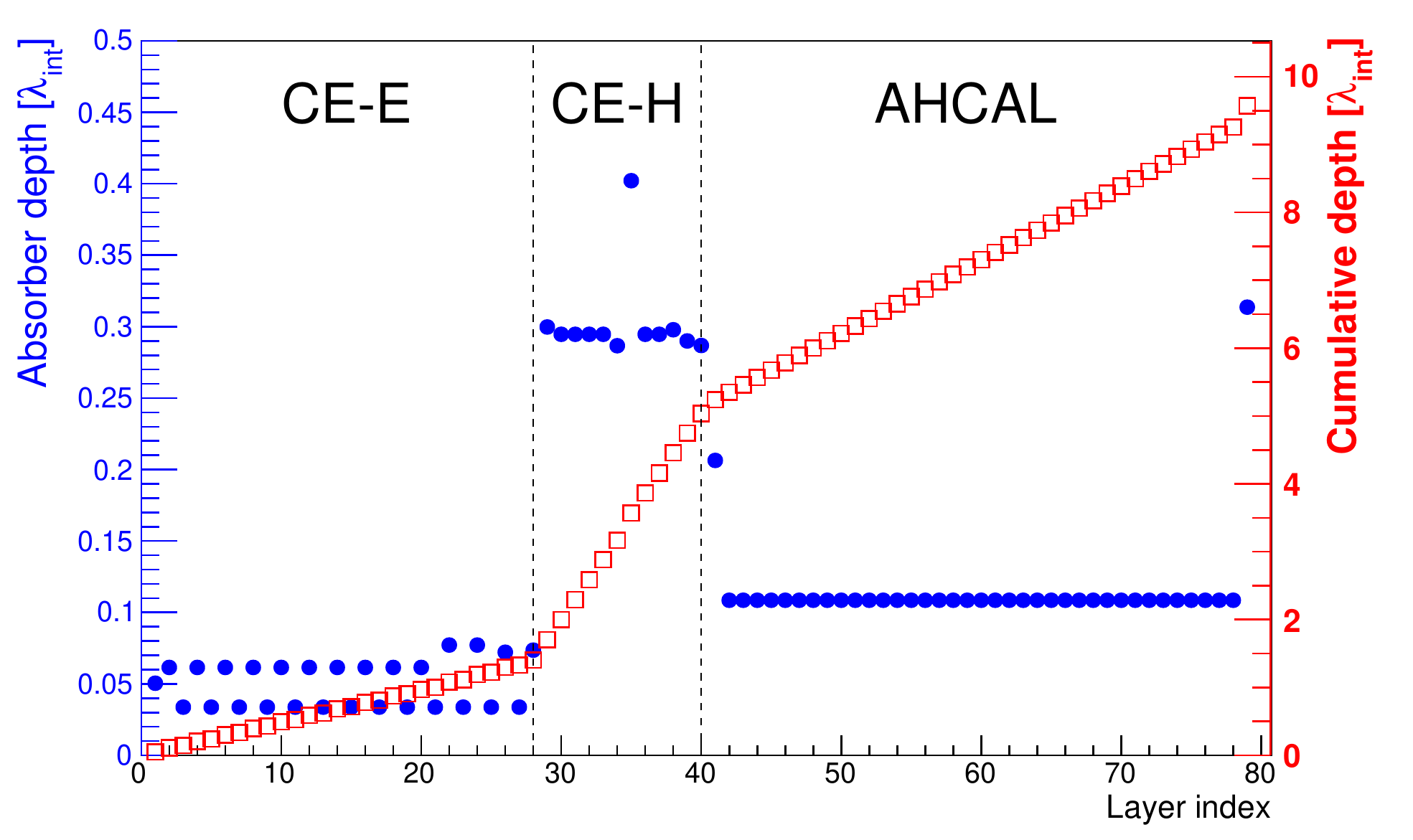}
  \caption{\label{fig:tb-hgcal-sim-intl} The thickness of the passive material between consecutive active layers in the simulated detector geometry expressed in units of \intL{}. The cumulative depth of the calorimeter prototype is also shown.}
\end{figure}

\subsection{Experimental and simulated datasets} \label{sec:datapresel}

The pion samples were collected at eight beam energy values, 20, 50, 80, 100, 120, 200, 250 and 300 GeV. The sample size for each energy varies between 80\,k$-$100\,k events before event selection. The final number of events used in the analysis after event cleaning criteria described in Section \ref{sec:evtsel} correspond to 40\,k$-$80\,k events, allowing a study of various characteristics of hadron showers.

The simulated event samples are produced using the GEANT4 toolkit \cite{bib.geant4} implemented in the CMS software. Two physics lists, namely \ftfp{} and \qgsp{} \cite{bib.g4validation} are used to model the development of hadronic showers in the beam test prototype detectors, and to compare the performance with the data. The term {\tt EMN} specifies the electromagnetic physics model, which includes a detailed modeling of multiple scattering and bremsstrahlung production tuned for the Phase-2 CMS detector components. The hadronic physics lists are a combination of models specified by the terms {\tt QGSP}, {\tt FTFP}, and {\tt BERT}, which dominate in different energy ranges. The \ftfp{} physics list uses the Bertini cascade model for energies less than 12 GeV for pions and less than 6 GeV for all other particles, and the {\tt FTFP} model for energies larger than 3 GeV for all particles. In the case of the \qgsp{} physics list, the {\tt FTFP} model is used between energies of 3 to 25 GeV, and {\tt QGSP} for energies above 12 GeV for all particles. In the overlapping ranges of particle energies, the models are combined using probabilities defined a priori. Approximately 100k events corresponding to the pion beam configuration of data taking are produced for each of the eight energy points.

\section{Event reconstruction in data and simulation}
\label{sec:evtreco}

The signal amplitude, ToT, and ToA for each hit are saved as raw data for the offline analysis as described in Section \ref{sec:exptsetup} along with the physical locations and electronics identifiers of each cell. The various steps involved in converting this raw data to energies measured by the calorimeters, that is, the event reconstruction and calibration of the detectors, are described in this section, both for the beam test data and the simulated data.

\subsection{Signal reconstruction} \label{sec:sigreco}

For each silicon cell, pedestal noise and common mode noise are subtracted from the ADC counts in all the time samples to obtain a waveform, which is fitted as a function of time to obtain the signal amplitudes corresponding to high gain and low gain amplifications~\cite{bib.cern-h2paper}. In case of large energy deposits in a cell, the ToT readout value is used as the respective signal amplitude after subtracting an offset to account for its insensitivity to low charge signals. The signal corresponding to each cell is then converted to equivalent high gain ADC counts ($A^{HG}_{Eqv}$) using a gain linearization procedure. In the next step, the linearized $A^{HG}_{Eqv}$ are converted into the corresponding number of minimum ionizing particles (MIPs) using 200 GeV muons as decribed below. The actual energy deposited by muons of 200 GeV is higher than minimum ionizing particles \cite{bib.pdg}. However, these serve as a robust tool for the detector calibration, and are referred to as MIPs in this context. The $A^{HG}_{Eqv}$ spectrum of muons for a given cell is fitted with a Landau distribution convoluted with a Gaussian distribution, and the maximum value of the fitted function is used as the MIP calibration constant ($C_{MIP}$). Overall, 85\% of the channels from the CE-E and CE-H prototypes are calibrated using the muon data, hence equalizing the channel-to-channel response to MIP energy scale. The noise and calibration have been stable over the course of data taking~\cite{bib.cern-h2paper}. For the AHCAL prototype, a similar procedure is followed for the conversion of the low gain amplitude measurements to the high gain scale, and for the equalization to the MIP energy scale \cite{bib.calice-ahcal-2018tb-paper}.

In simulation, the energy distribution of 200 GeV muons is fitted with the same functional form as used in the data, and the maximum value of the fitted function is used to convert the energy deposited in each cell to the number of MIPs. In the absence of a dedicated modeling of electronics effects and digitization in simulation, a Gaussian smearing is applied to the simulated MIP spectra before the fitting procedure. The channel-to-channel response in the simulation is perfectly equalized for a given thickness of the silicon cells or scintillator tiles. In the CE-H section, the MIP conversion factors are obtained for sensors of thickness 300 $\mu$m and 200 $\mu$m. Example comparisons of energy distributions of 200 GeV muons as measured in data and simulation for the three compartments CE-E, CE-H and AHCAL are shown in Figure \ref{fig:mips-data-mc}. As expected, the distributions peak at one in both data and simulation, and the overall spectrum predicted by simulation matches well with that measured in the data even though it is slightly wider in some cases. For the AHCAL, a threshold of 0.5 MIP is applied in the reconstruction, and it is well above the self-trigger threshold for all channels. These distributions include the signals of all cells in a given layer.

\begin{figure}[h]
  \centering
  \includegraphics[width=0.32\linewidth]{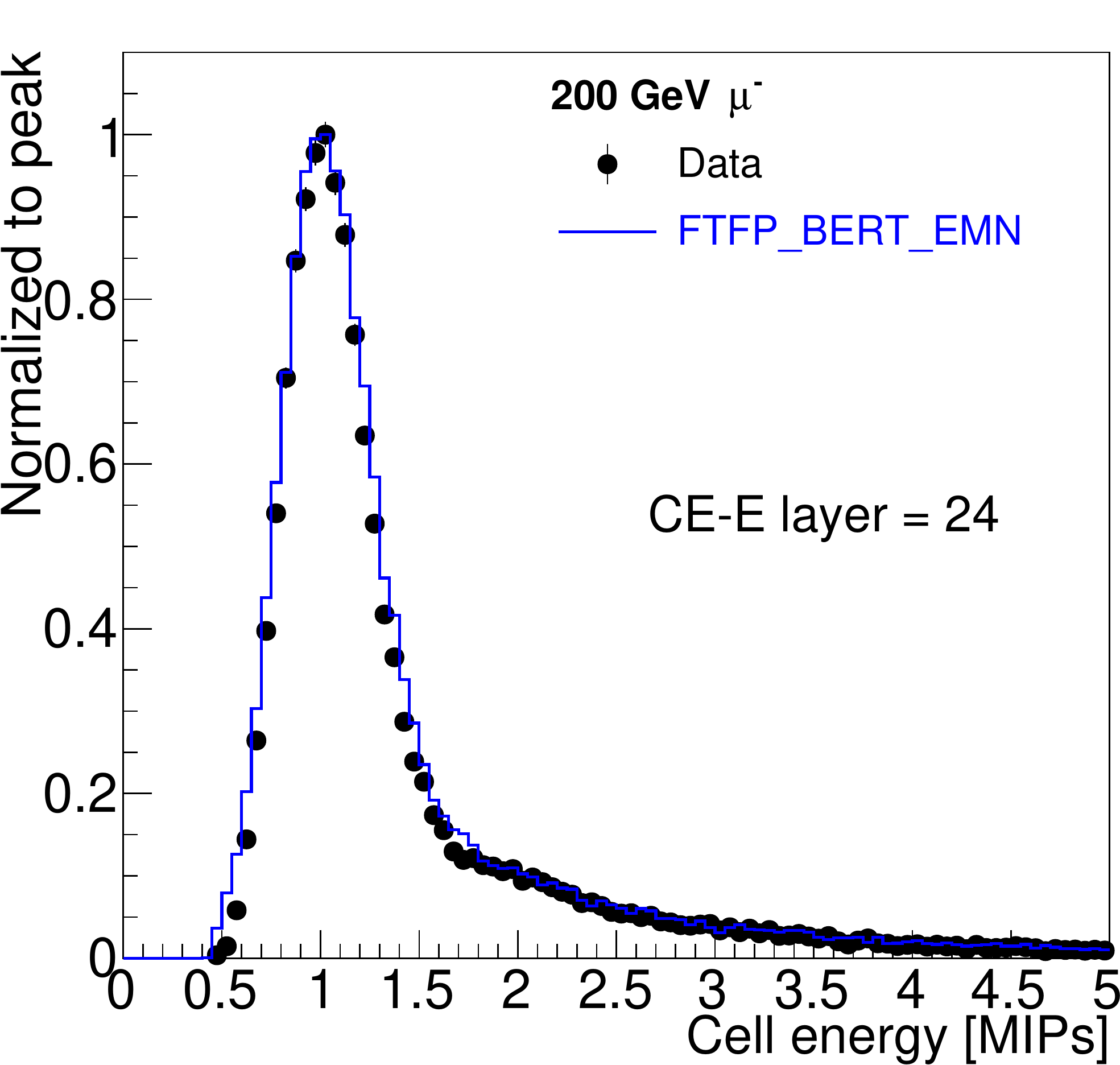}
  \includegraphics[width=0.32\linewidth]{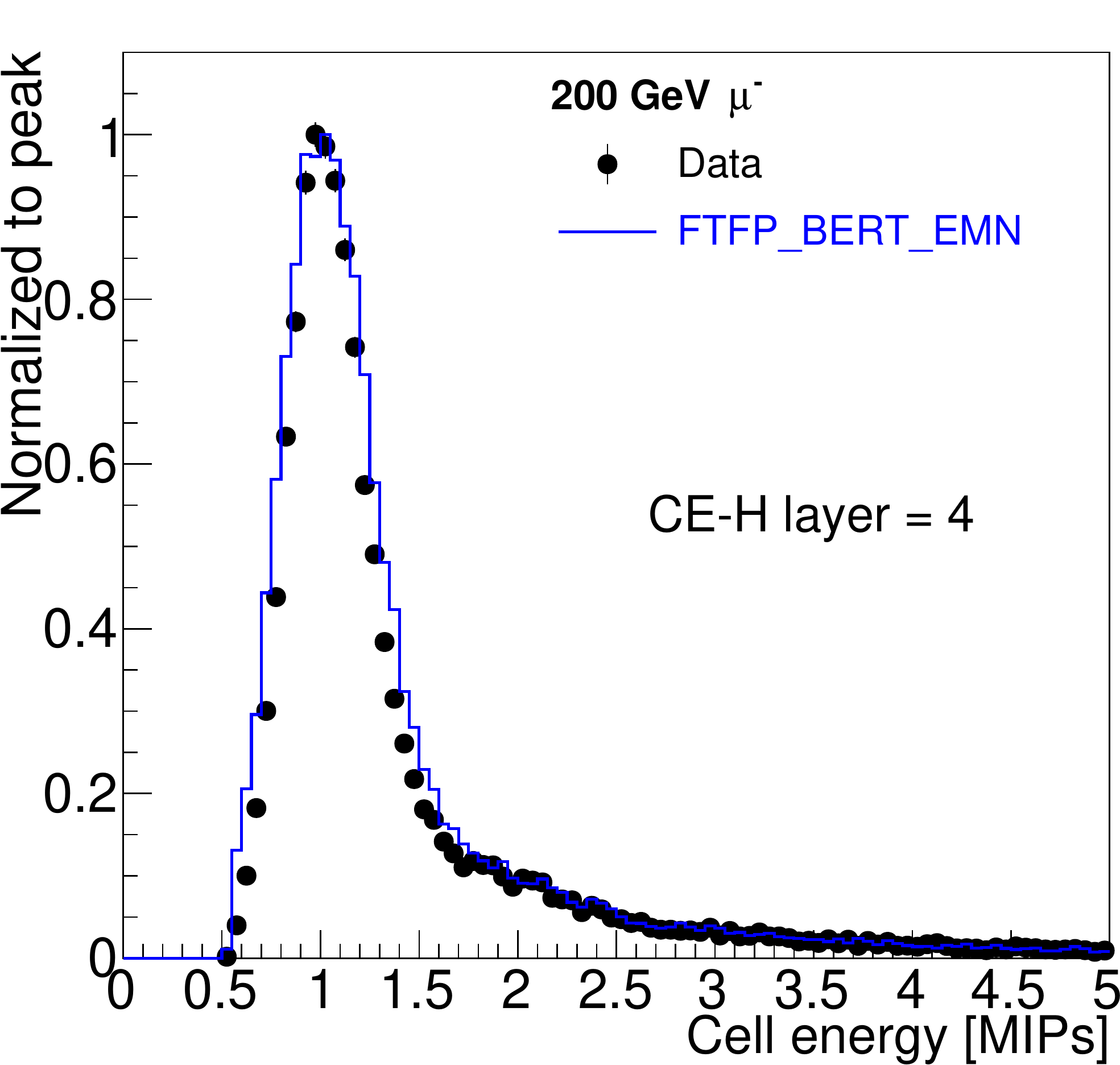}
  \includegraphics[width=0.32\linewidth]{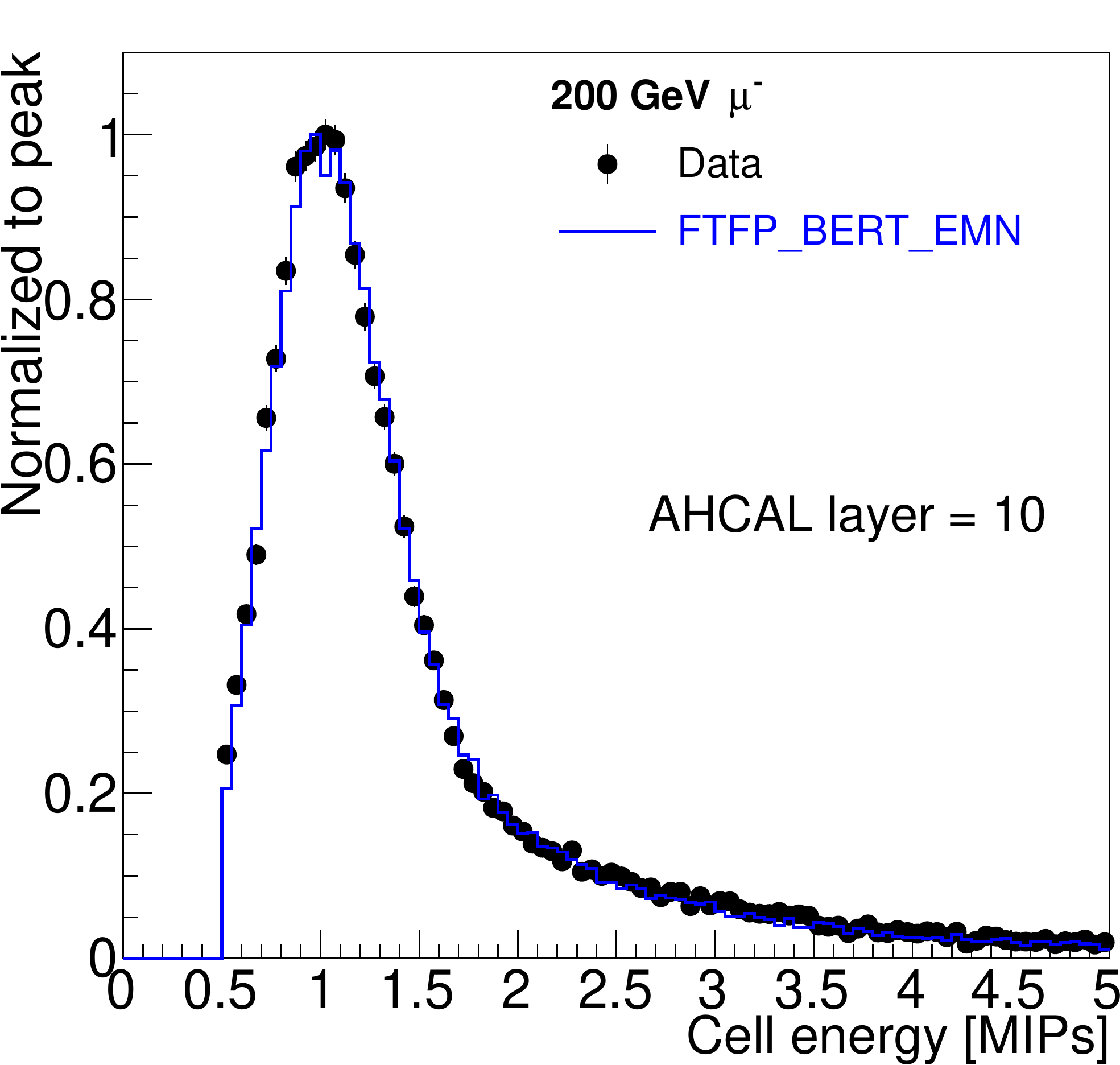}
  \caption{\label{fig:mips-data-mc} Cell energy in units of MIPs deposited by 200 GeV muons in data and simulation for a selected CE-E (left), CE-H (middle), and AHCAL (right) layer.}
\end{figure}

\subsection{Event selection in data and simulation} \label{sec:evtsel}

A set of selection criteria is used to select pion events of high purity to ensure a reliable determination of the physics performance of the HGCAL prototype in terms of energy response and resolution and longitudinal and transverse shower profiles, and to perform a consistent comparison with simulation.

Individual malfunctioning channels are identified through the irregularities in pedestals or signal-to-noise studies or through known hardware failures. These are removed (masked) in the further analysis both in the data and the simulation. Approximately 3.2\% (3.8\%) of channels are masked in CE-E (CE-H), and the number of channels masked in AHCAL is less than one per mille. The channels in which the measured signal is below 3$\sigma$ (4$\sigma$) random noise in CE-E (CE-H) are rejected from further analysis. This is equivalent to $\sim$0.5 MIP or less, and does not result in any loss in signal given the high signal-to-noise ratio for silicon sensors \cite{bib.cern-h2paper}.

The data acquisition from the four DWCs was configured in a way that only one hit per chamber is available for track reconstruction. A track is reconstructed by fitting a straight line through the available hits. An event is selected if the reconstructed track hits at least three wire chambers, and the $\chi^{2}$ per degree of freedom is less than 10. The selected tracks are extrapolated to the detector. The points of intersection at various layers give the direction of the initial beam particle along the calorimeter sections. The projection of the profile of the beam impact point on the initial layers of CE-E in ${\it x}$ and ${\it y}$ directions are used to define a 2\,cm $\times$ 2\,cm window to select pion events for beam energies $\geq$ 200 GeV. The window is $\sim$4\,cm $\times$ 4\,cm for pion beam energies $\leq$ 120 GeV. This selection rejects particles impinging too far away from the beam direction, maintains a good trigger efficiency, and ensures similar beam profiles in data and simulation to allow a faithful comparison of various shower shape observables.

In the absence of dedicated muon veto detectors, potential contamination due to muons originating at the target or in-flight decay of pions are rejected using the differences in patterns of energy deposits in the detector corresponding to muons and pions. The pion events with total reconstructed energy, E $<$ 100\,MIPs in CE-E, E $<$ 60\,MIPs in CE-H, and $\sum E_1$/$\sum E_{25}$ $>$ 0.8 in AHCAL are rejected. Here, $\sum E_1$ and $\sum E_{25}$ refer to the sum of energy of the highest energy cell of each AHCAL layer measured in units of number of MIPs, and the sum of energies measured in 25 nearest cells centered on them, respectively. For the measurement of energy response and resolution (to be presented in Section \ref{sec:enereco-det}), it is important to reject pions that may have interacted with the beamline elements. Since only one particle track could be reconstructed using DWCs, further cleaning of events is done by requiring particles that do not start showering in the first two layers of CE-E using an algorithm described in Section \ref{sec:showerstart}. The effect of these selection criteria, successively applied on total energy measured in data and simulation, are represented in Figure \ref{fig:evt-cleaning}. As expected, the simulation is not much affected by most of the noise or event rejection cuts except for those corresponding to muon veto or preshowering particle rejection. Approximately 75\% (85\%) of data (simulated) pions fulfill these selection criteria across all beam energies except 20 GeV in data for which selection efficiency is $\sim$65\%. The larger event rejection in data is mainly attributed to the track quality and track window selections.

\begin{figure}[h]
  \centering
  \includegraphics[width=0.45\linewidth]{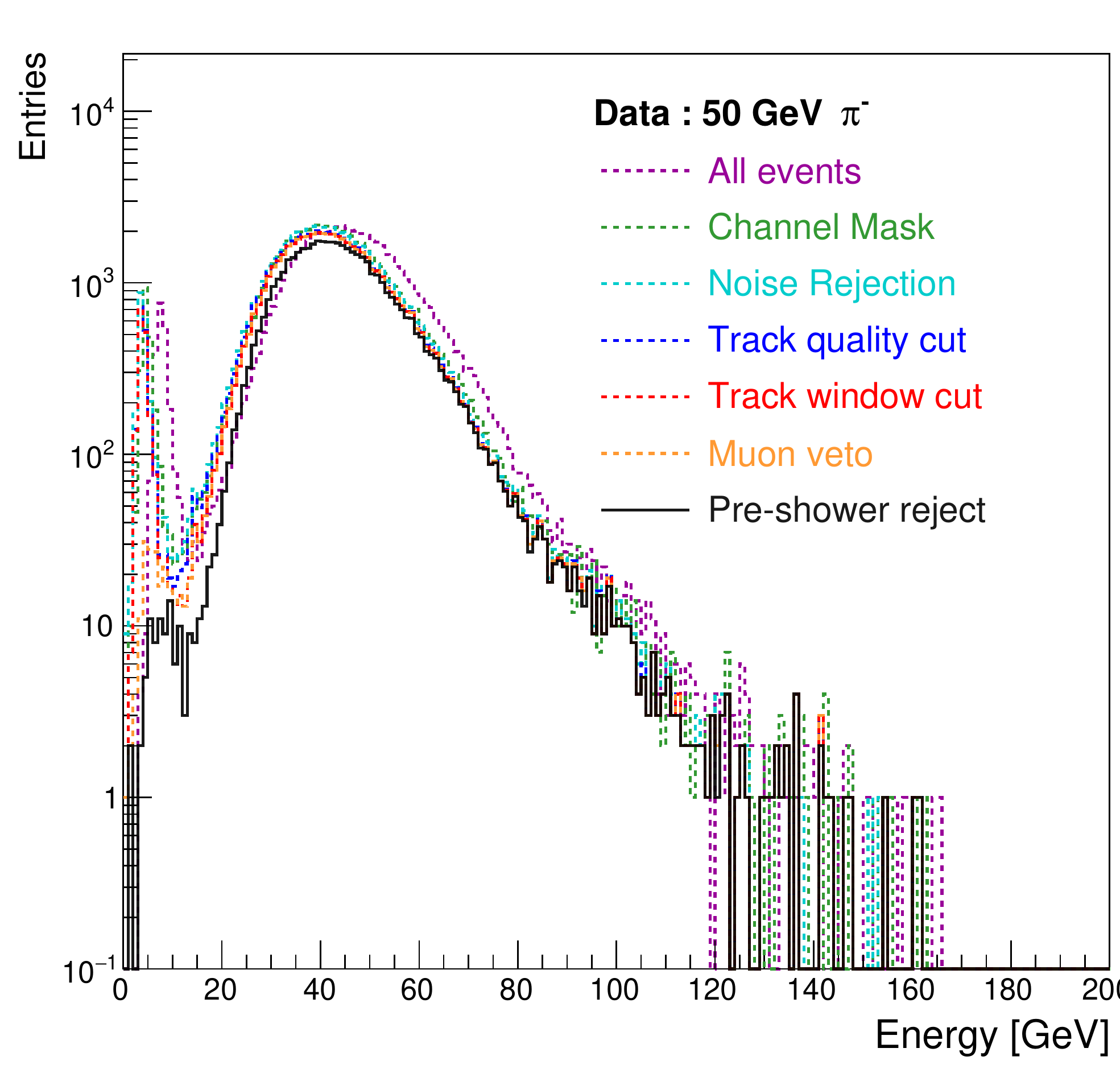}
  \includegraphics[width=0.45\linewidth]{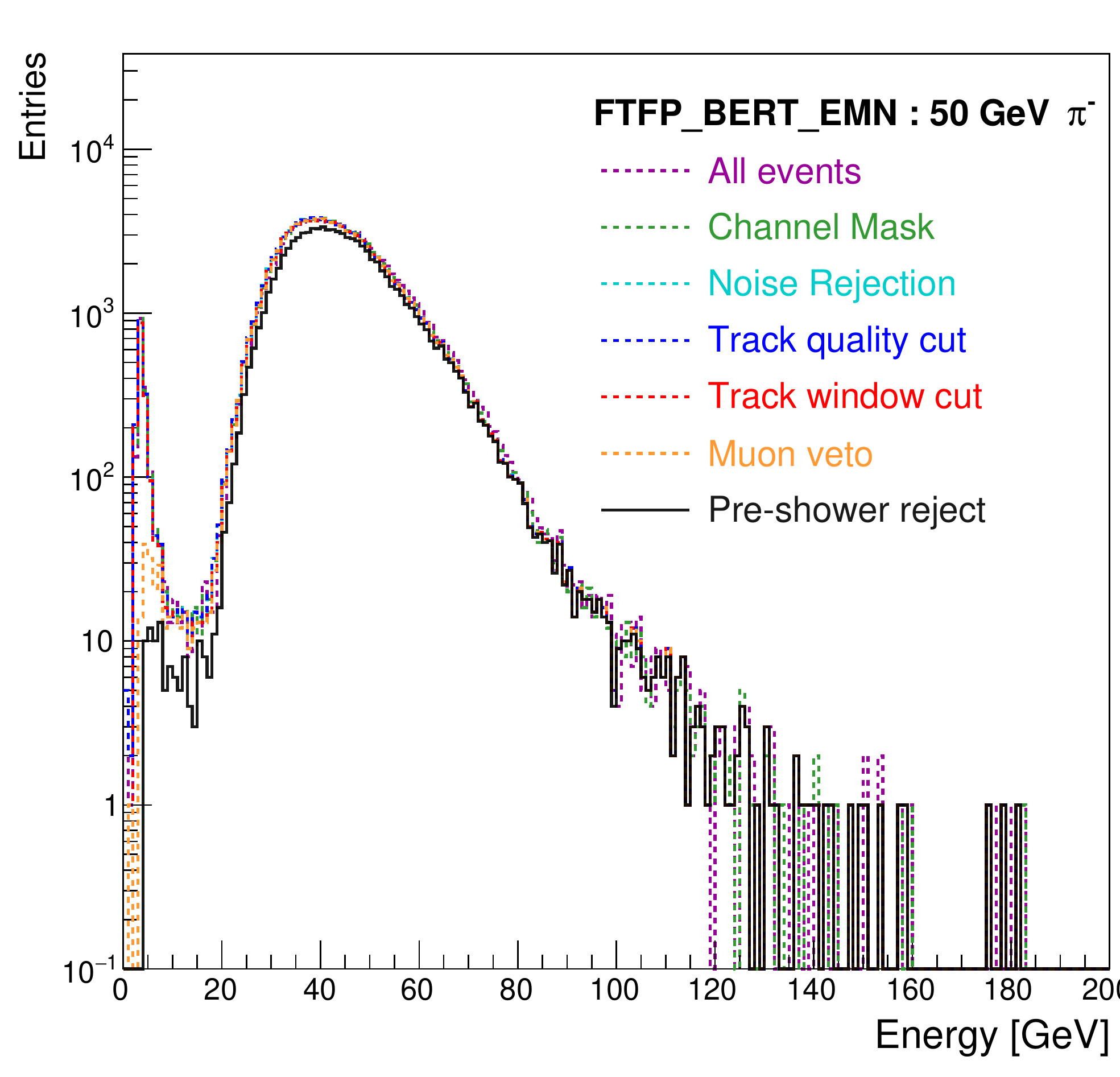}
  \caption{\label{fig:evt-cleaning}{Distribution of total reconstructed energy after applying various data cleaning and event rejection cuts for 50 GeV pions in data (left) and simulation (right). The distribution for a selection as given in the legend includes all previous cuts, i.e. those appearing in the legend above it.}}
\end{figure}

For the events passing all the selection criteria, energies measured in the individual silicon cells and scintillator tiles in units of MIPs are summed up for the CE-E, CE-H, and AHCAL prototype sections, and are compared to simulation. Figure \ref{fig:emips-datasim-EEFHAH} shows measured energy sum distributions for  20 GeV pions (upper row) and 100 GeV pions (lower row) for the CE-E (left column), CE-H (middle column), and AHCAL (right column) prototypes compared with those predicted by the \ftfp{} and \qgsp{} physics lists. Both the physics lists show similar out-of-the-box performance with \qgsp{} showing slightly better agreement with the data at higher beam energies in the CE-E and CE-H. However, the simulation predicts longer tails in the energy deposited in AHCAL for all energies considered in this analysis.  With this cleaned set of measured and simulated datasets, we proceed to develop an algorithm to identify the position of the first hadronic interaction in the detector and measure the energy scales of the three detector sections.

\begin{figure}[h!]
  \centering
  \includegraphics[width=0.32\linewidth]{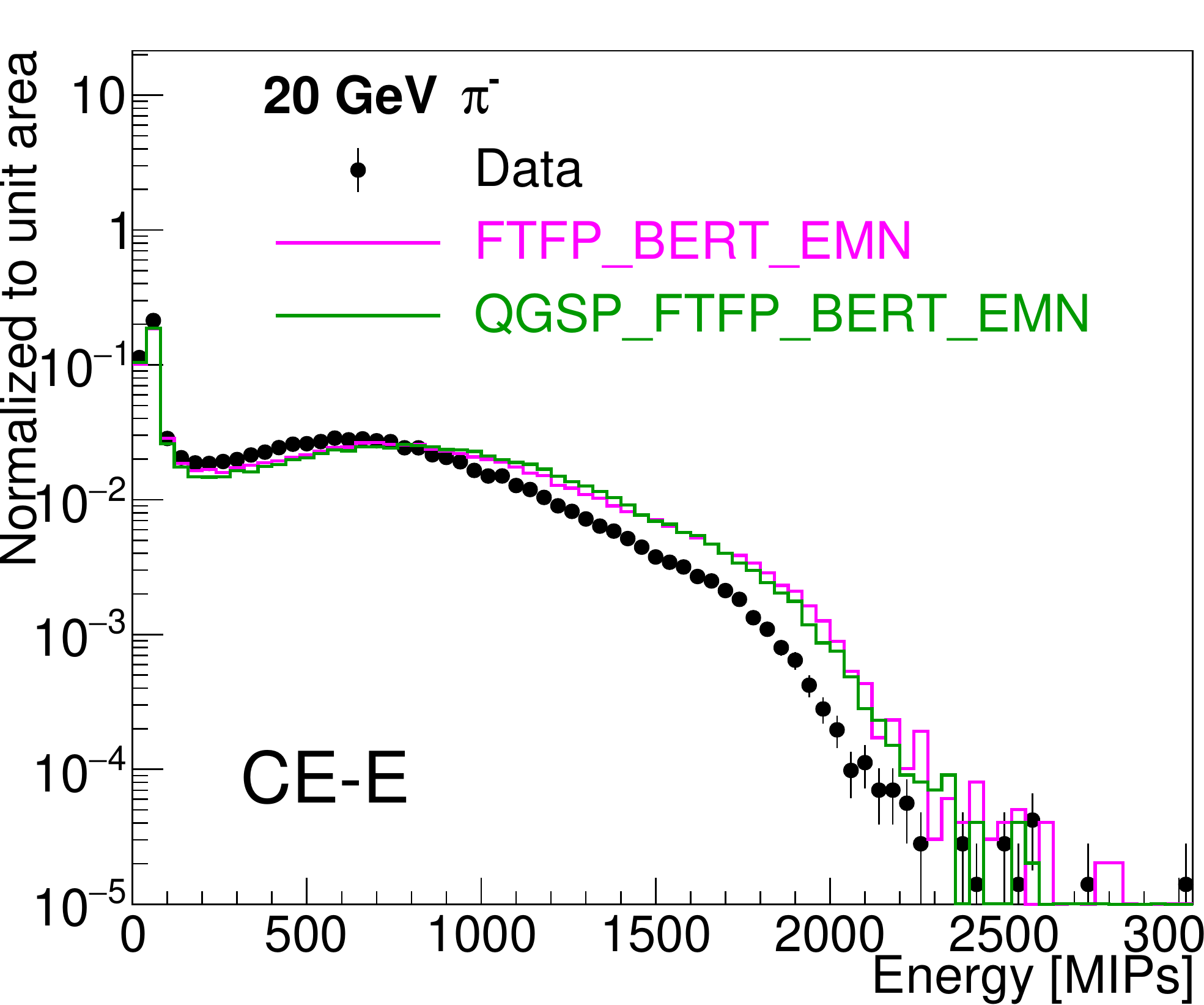}
  \includegraphics[width=0.32\linewidth]{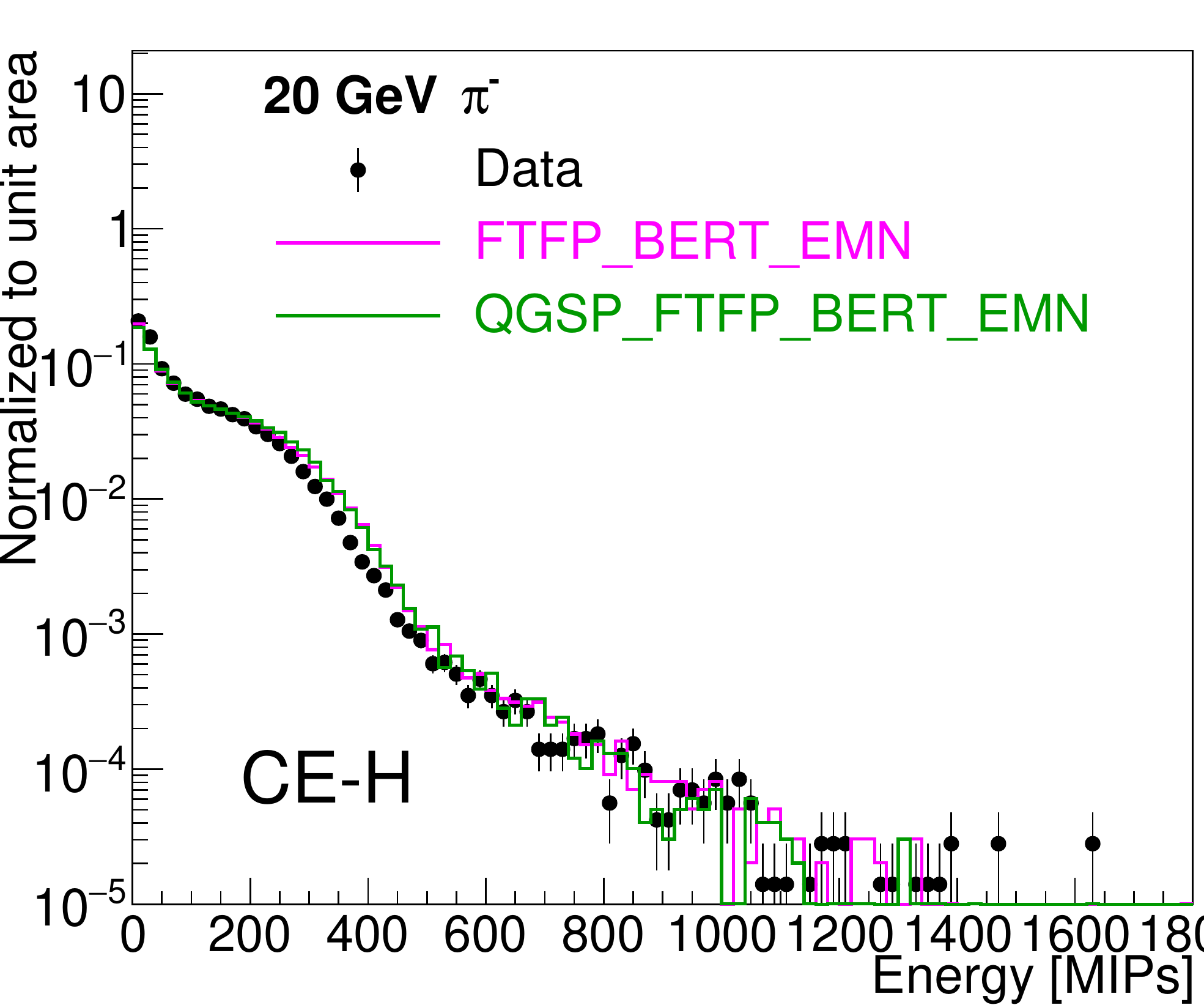}
  \includegraphics[width=0.32\linewidth]{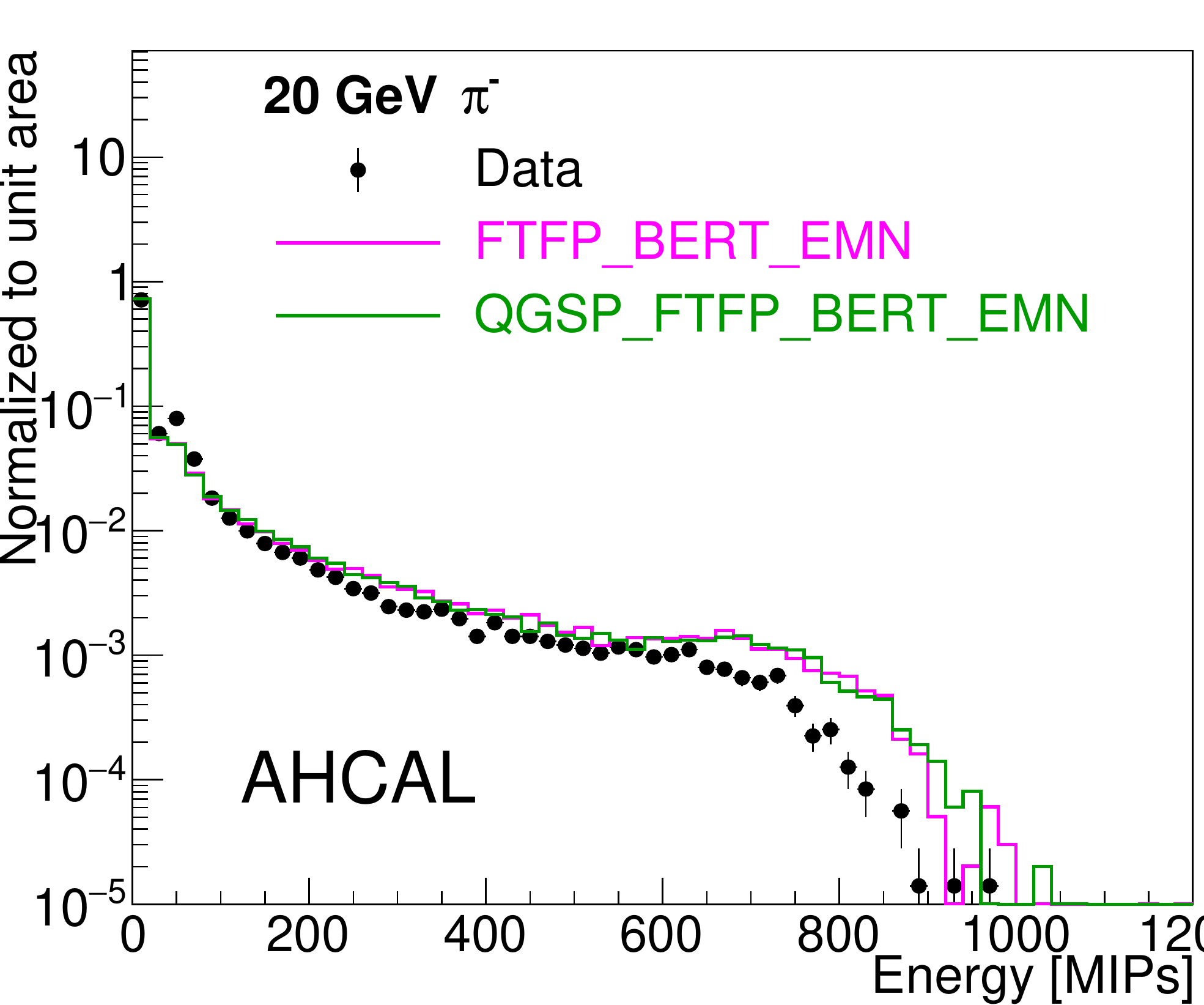}
  \includegraphics[width=0.32\linewidth]{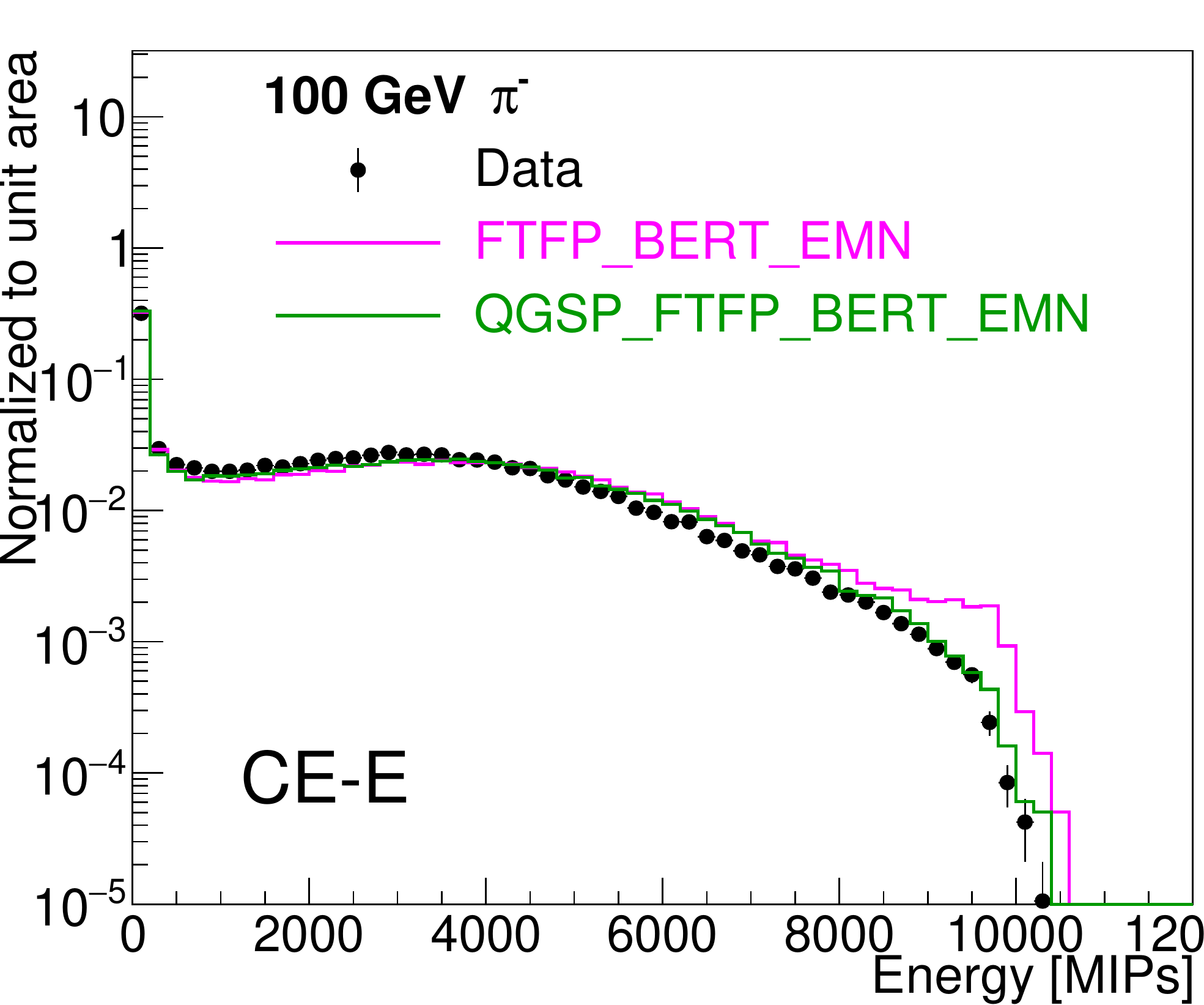}
  \includegraphics[width=0.32\linewidth]{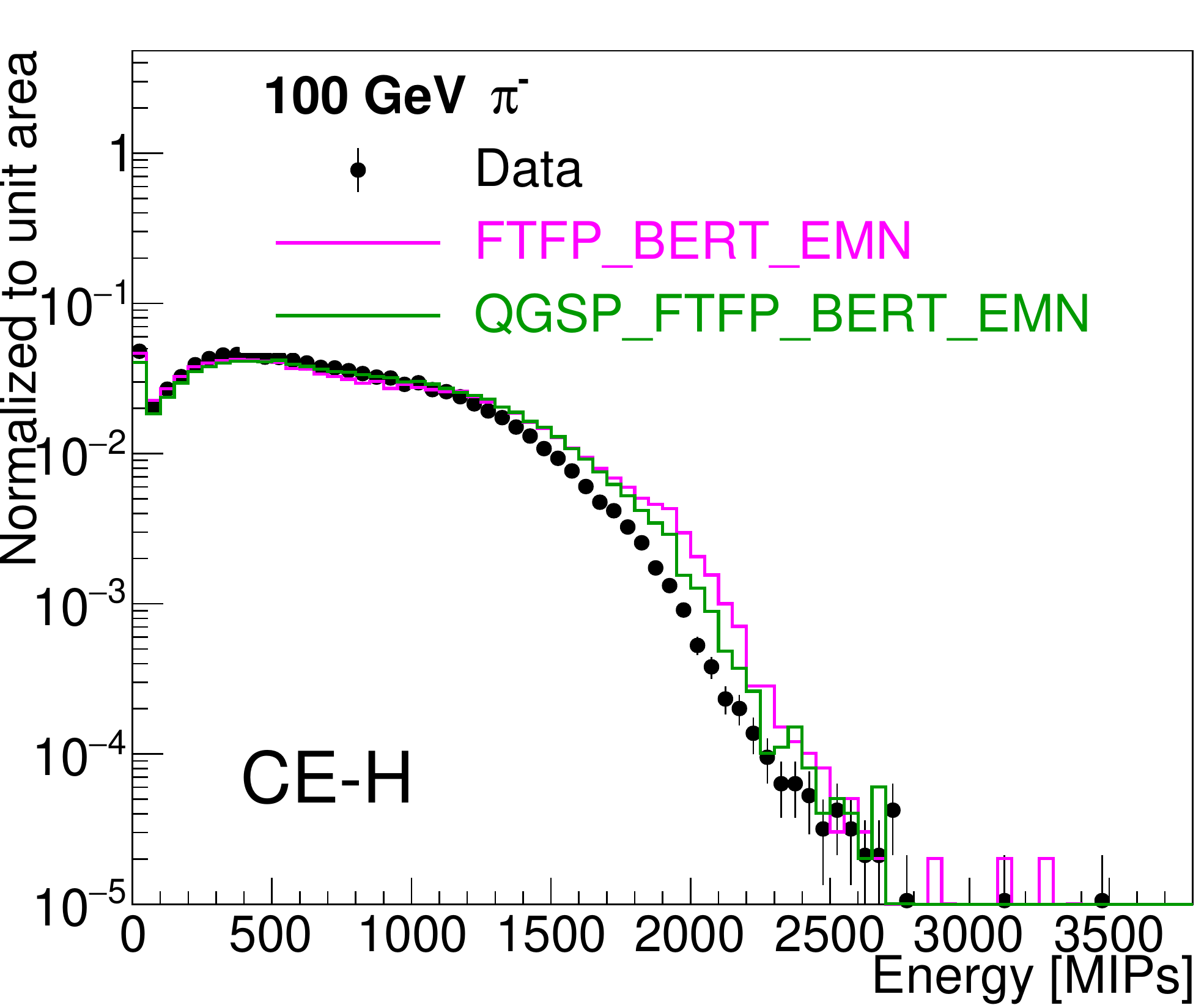}
  \includegraphics[width=0.32\linewidth]{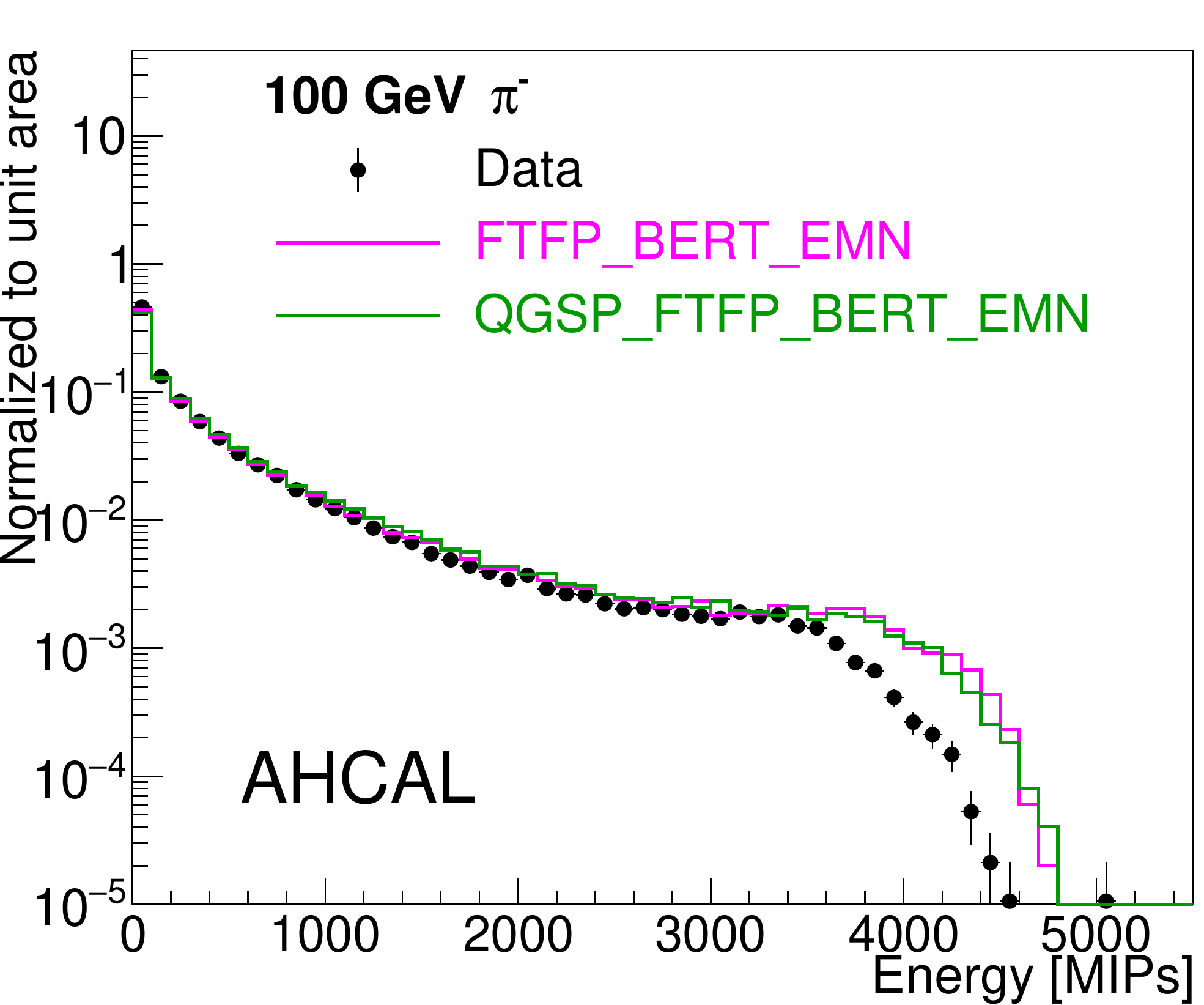}
  \caption{\label{fig:emips-datasim-EEFHAH}{Energy in units of MIPs deposited by 20 GeV (upper row) and 100 GeV (lower row) pions in data and simulation in the CE-E (left), CE-H (middle), and AHCAL (right) prototypes.}}
\end{figure}

\section{Depth of the first hadronic interaction} \label{sec:showerstart}

The fine transverse and longitudinal granularity of the electromagnetic and hadronic calorimeter sections of the HGCAL make it possible to determine the depth at which the pion shower starts (\zss), that is, the depth at which a pion underwent the first hadronic interaction. As shown in the event display presented in Figure \ref{fig:event-display-pi300}, a pion continues through the detector as a minimum ionizing particle until it initiates a hadronic shower in the later layers of the CE-E, which results in particle multiplication and the subsequent development of a shower extending into the CE-H and the AHCAL sections. An algorithm is developed to identify the position at which the shower started along the pion trajectory using the increase in particle multiplicity and transverse spread of the energy deposited as the shower develops longitudinally. Muons are used as a reference to study the various observables.

\begin{figure}[h]
  \centering
  \includegraphics[width=0.95\linewidth]{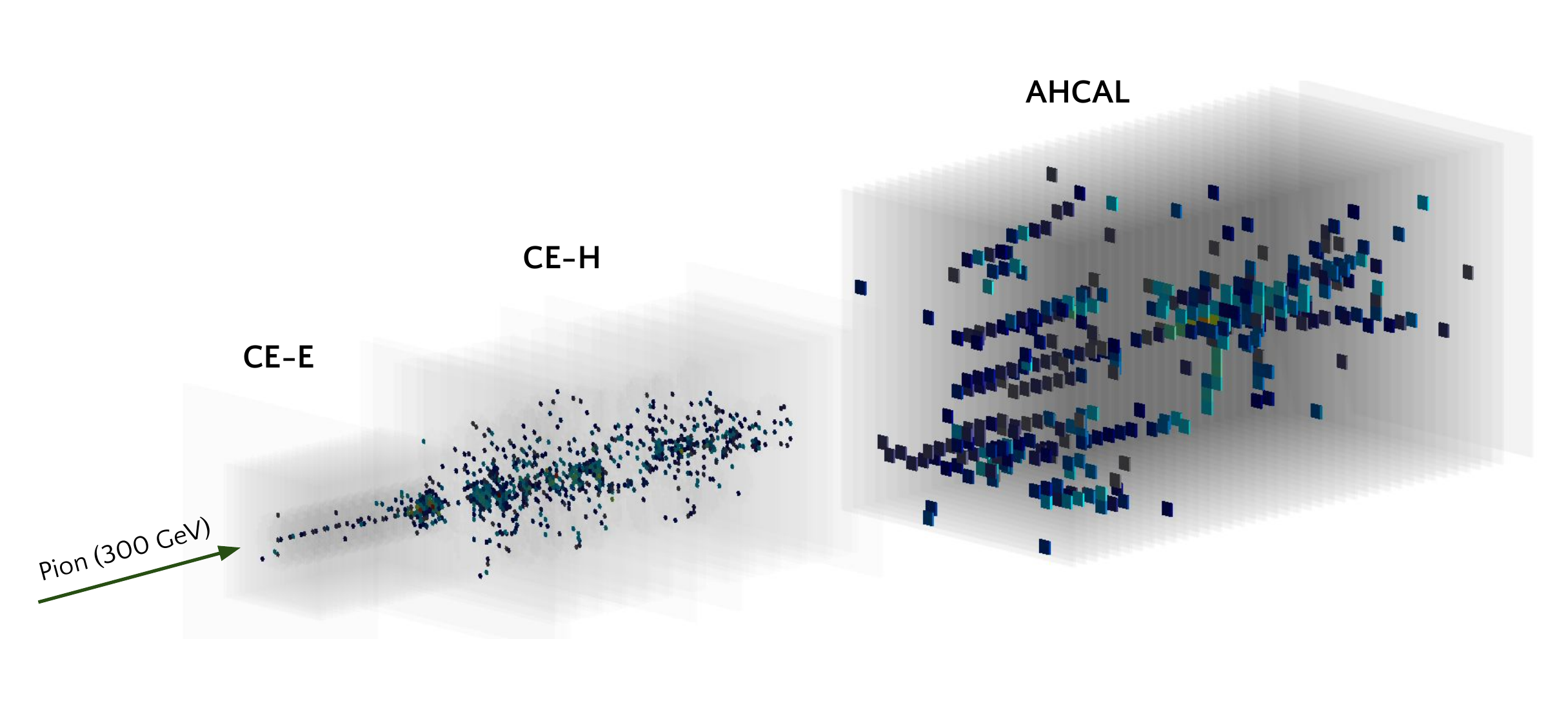}
  \caption{\label{fig:event-display-pi300} An event display illustrating the development of a hadronic shower initiated by a pion of 300 GeV energy starting in the last layers of the CE-E and depositing energy throughout CE-H and AHCAL.}
\end{figure}

The truth information available from the GEANT4 simulation such as the interaction processes, positions, and energy deposited as a particle traverses the detector, is used for optimizing the algorithm and assessing its performance. In the simulation, each particle is tracked and propagated to the next step with a predefined probability for potential interactions that it can undergo. If, at a given step, the primary track (incident pion) encounters a hadronic interaction, the coordinates of the step and the number of secondary particles produced, as well as their particle type, charge, position, and 4-momenta are saved. Soft hadronic interactions do not initiate a hadronic shower, and such events are rejected from the training dataset by requiring the total kinetic energy of the secondary particles other than the leading hadron to be more than 40\% of the incident particle's energy. The true z-position (\zsstrue) is a continous distribution but for the actual experiment, only the layer number is known. Hence, the layer number of the closest active layer following the \zsstrue{} is assigned as the true layer where the shower started (\layi{}) to perform a comparison with the corresponding layer identified using the reconstructed observables (\layireco{}) in data and simulation.

For each pion, the layer \layi{} corresponding to \zsstrue{} is determined. The energy pattern deposited in layers before \layi{} is expected to be similar to the one for muons, and in layers after \layi{} is expected to be more populated. The increase in particle multiplicity is examined by counting the number of cells  with energy deposited above the noise thresholds (as explained in Section \ref{sec:evtsel}) for each layer. The position of the center-of-gravity (COG) is calculated as energy weighted average of the position of all the cells in a given layer. The increase in energy deposited as the shower develops is examined using energy measured in a radius of 10 cm around the COG. The transverse energy spread is defined by the ratio of the energy deposited in a radius of 2 cm around the layer COG to that in a radius of 10 cm (\ratioavg), in a moving window of three layers. Hence, the \ratioavg{} for layer {\it{i}} is defined as:

\begin{equation}
  R_{i} =  \frac{\sum_{layer=i}^{i+2}~ E_{2\,cm}^{layer} }{\sum_{layer=i}^{i+2}~ E_{10\,cm}^{layer}}
\end{equation}
The summation is performed only up to the next layer for the penultimate layer of the CE-H. The value of \ratioavg{} is expected to be close to one for the pions before the first hadronic interaction and for muons since the energy is deposited only along the track of the particles via ionization processes. In the layers following \layi{},  \ratioavg{} is expected to be smaller.

Representative distributions of the number of cells in the CE-H prototype, of the energy deposited in a radius of 10 cm around the COG in CE-E and of \ratioavg{} in the layers \layi{} and \layi{}\,-\,2 are shown in Figure \ref{fig:ss-algo-var} for 100 GeV pions, together with the corresponding distributions for muons. Distributions of \ratioavg{} for pions in the layers \layi{} and \layi{}\,-\,2 and for muons are shown in the Figure \ref{fig:ss-algo-var} (right). To identify the position of the shower start, the layer should have at least three cells with energy deposited above the noise thresholds. The total energy deposited in a radius of 10 cm around the COG in \layi{} is required to be greater than 12 MIPs (40 MIPs) for pions of beam energy 20 GeV (200 GeV). We also require \ratioavg{}$<0.96$. The first layer fulfilling all the conditions is assigned to be the \Lss{} in a given event. 

\begin{figure}[h]
  \centering
  \includegraphics[width=0.3\linewidth]{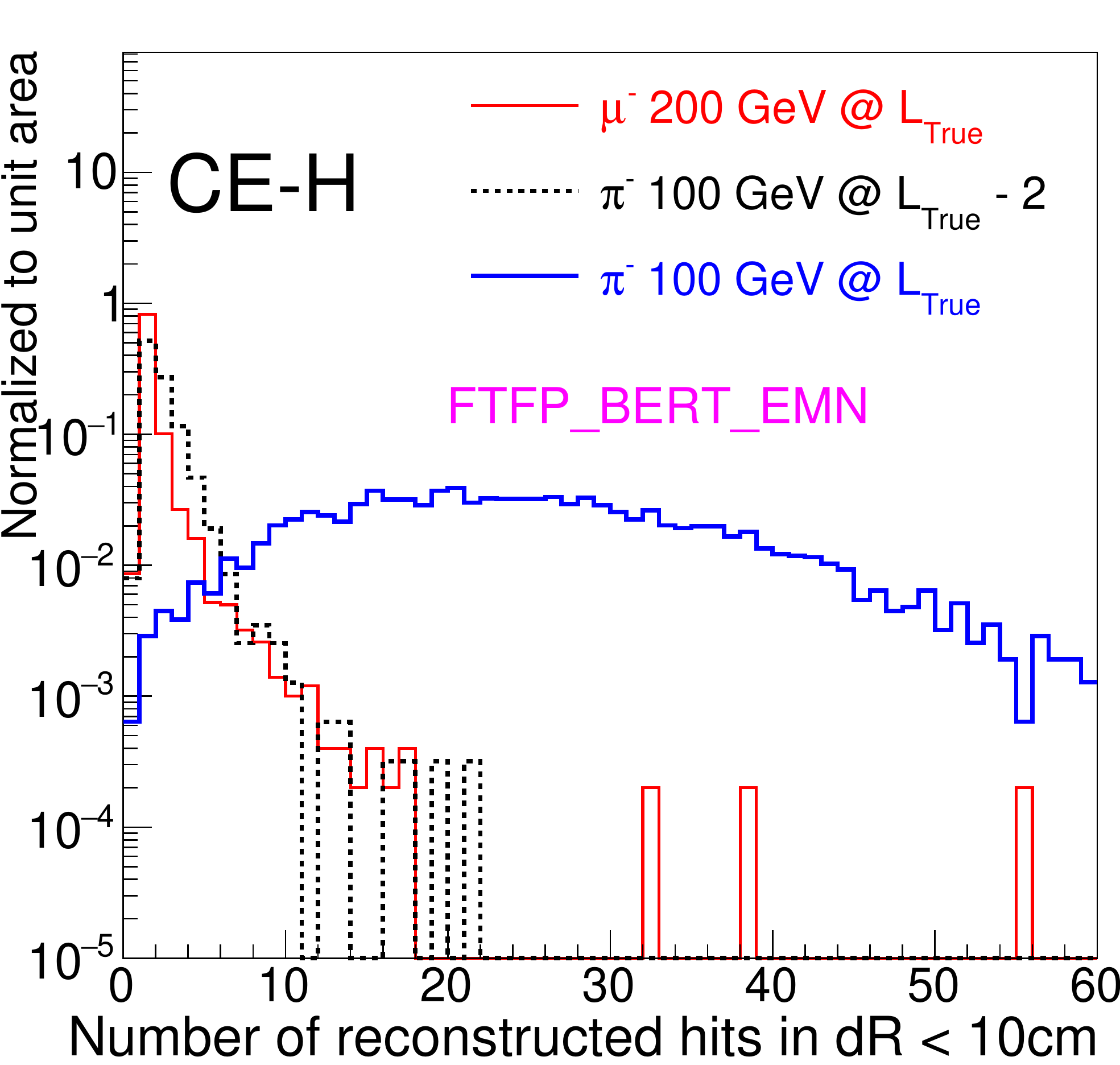}
  \includegraphics[width=0.3\linewidth]{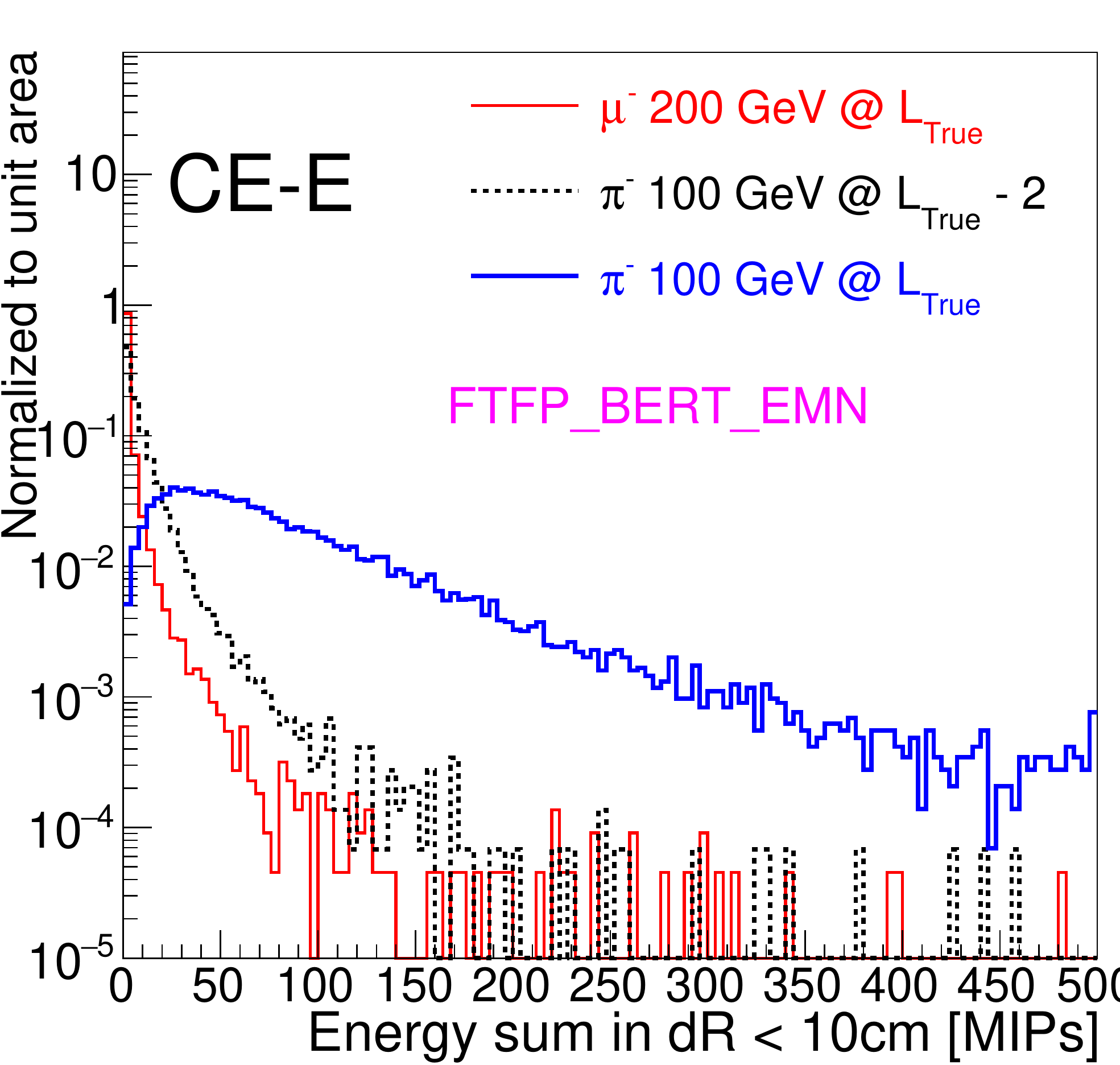}
  \includegraphics[width=0.3\linewidth]{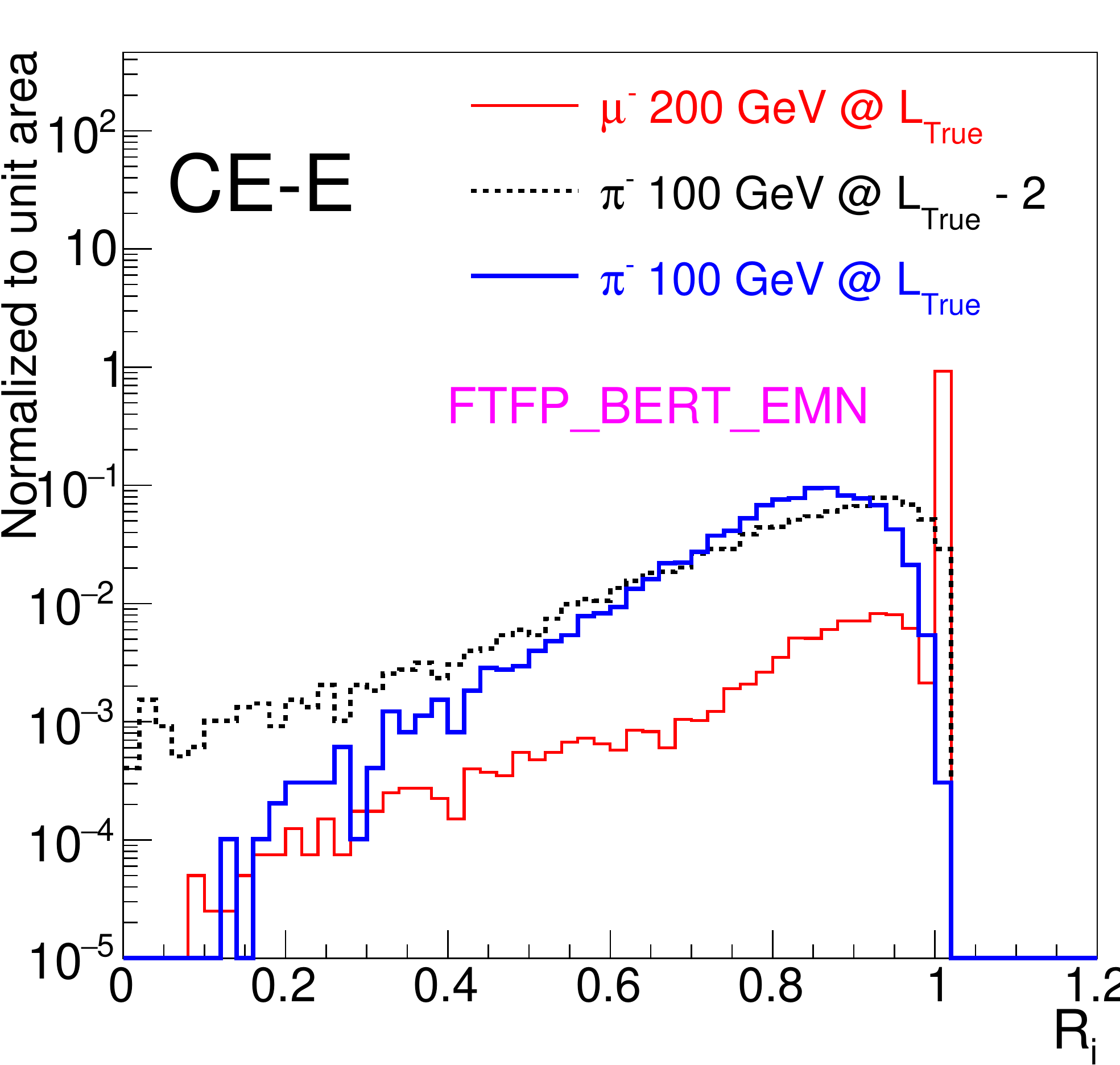}
  \caption{\label{fig:ss-algo-var} Variables used in the shower start algorithm for simulated pions in the layer closest to the first hadronic interaction layer, \layi{} (blue line) and in the layer \layi{}\,-\,2 (black dotted) compared to muons (red line): number of reconstructed hits in CE-H layers (left), energy deposited in  CE-E layers (middle), and ratio of average energy deposited in three layers of CE-E in a radius of 2 cm to that in 10 cm around the COG (right).}
\end{figure}

For the particles which start showering in \layi{} as determined from the GEANT4 truth information, the efficiency of the algorithm is defined as the fraction of events for which the \Lss{} is reconstructed within \layi{} $\pm$ {\it n}, where {\it n} $=$ 1 or 2. For almost 95\% of pions showering in HGCAL, the algorithm is able to correctly find \Lss{} within \layi{} $\pm$ 2 in CE-E and within \layi{} $\pm$ 1 in CE-H, with slightly lower efficiency for 20 GeV pions. A summary of the performance of the algorithm to find the depth of first hadron interaction for pions simulated with the \qgsp{} and \ftfp{} physics lists is are shown in Figure~\ref{fig:eff-ss-npi} (left) for all the beam energies used in this publication.

\begin{figure}[h]
  \centering
  \includegraphics[width=0.48\linewidth]{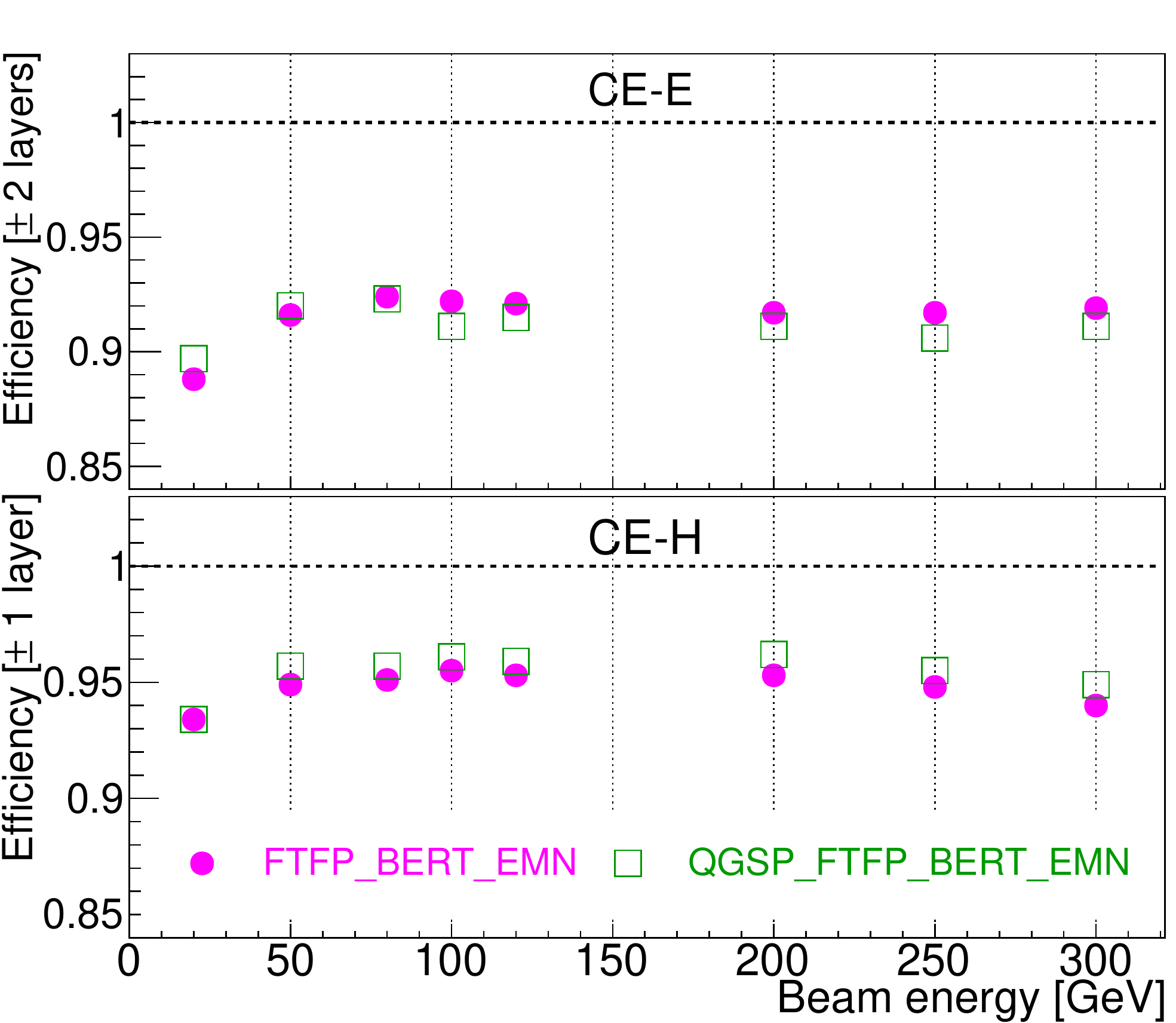}
  \includegraphics[width=0.46\linewidth]{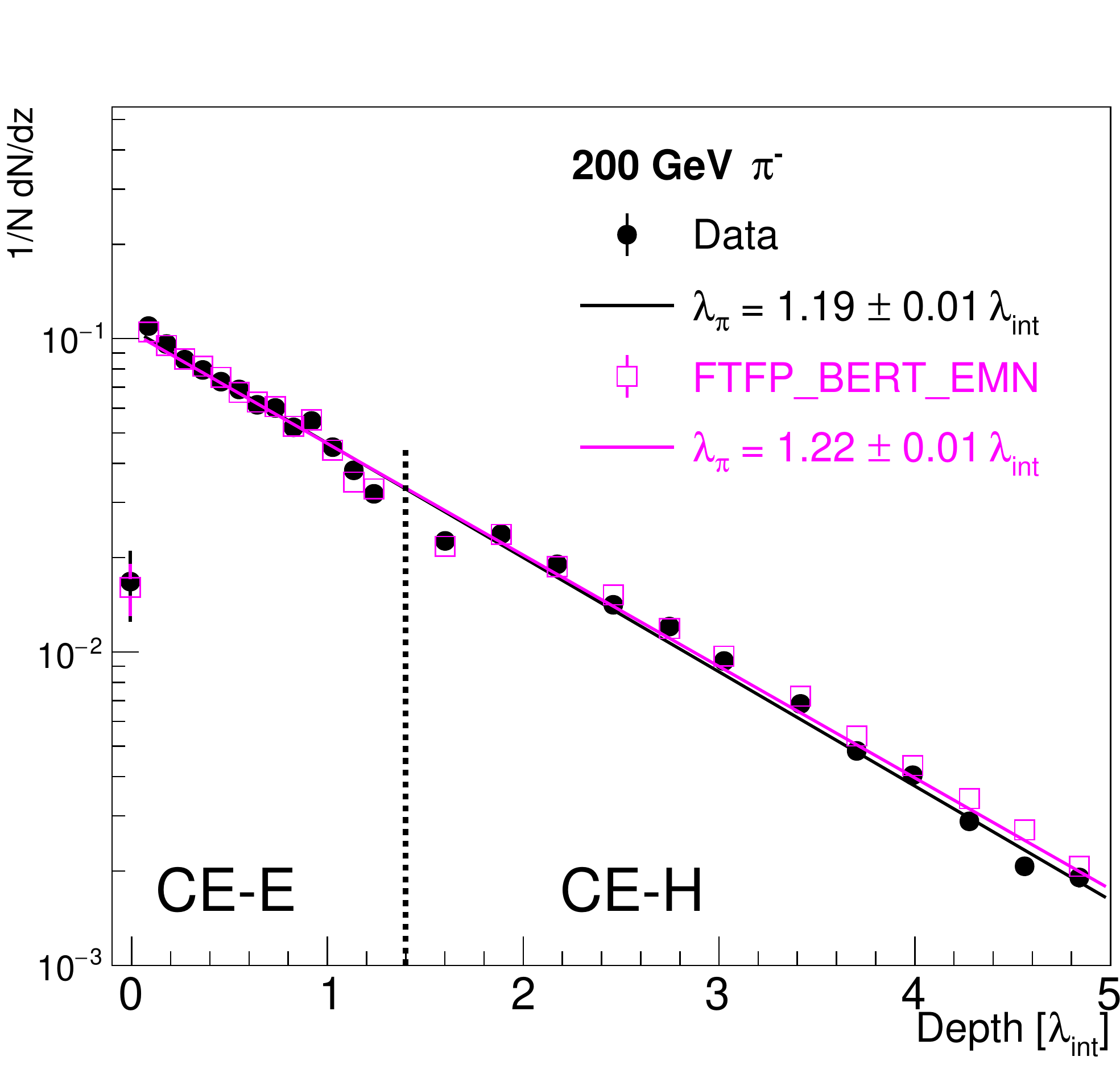}
\caption{\label{fig:eff-ss-npi} Efficiency of finding the shower start position in the CE-E and CE-H prototypes (left), and number of pions not undergoing a hadronic interaction as a function of calorimeter depth for 200~GeV pions in simulation (right). The distribution is fitted with an exponential function (solid lines).}
\end{figure}

The optimized algorithm to find \Lss{} is used to measure the position of the first hadronic interaction in the simulated and beam test pion datasets. The mean free path of hadrons in the detector material is characterized by \intL{}. The distribution of the depth of the shower starting point follows an exponential distribution given as
\begin{equation}
  z^{shower-start} = C~{\it e}^{\frac{-z}{\lambda_{int}}}
\end{equation}

Figure~\ref{fig:eff-ss-npi} (right) shows the distribution of the number of pions of 200 GeV that survive without a hadronic interaction as a function of depth in calorimeter in the units of \intL{}, which is taken as the depth corresponding to \Lss{} identified by the algorithm. The figure also shows the distribution obtained from the simulated pion sample using the \ftfp{} physics list, which agrees very well with the distribution measured in the data. Similar agreement between data and simulation is observed for the pions of other beam energies, and with those simulated with the \qgsp{} physics list. These distributions are fitted with an exponential function. The exponent is slightly higher than unity, indicating that the pion interaction length (\piL) is different from the interaction length obtained using inputs from {\tt PDG}, corresponding to 200 GeV neutrons or protons \cite{bib.pdg}. This is expected as the interaction cross-section of pions with protons or neutrons is smaller than that of protons. The fitted simulation and data distributions give \piL $\approx$ 1.2\,\intL{} for high energy pions which is in agreement with cross section measurements on iron \cite{bib.cosmic-p-pi-xsec,bib.ahcal-p-pi}.

\section{Pion energy reconstruction}
\label{sec:enereco}

The HGCAL calorimeter is designed to fully contain the showers initiated by hadrons to measure their energies and positions. The CE-E, optimized for measuring electromagnetic showers, corresponds to a depth of 1.4 \intL{}. Approximately 70\% of charged pions entering the detector are expected to undergo a first hadronic interaction in the CE-E. Since their hadronic showers continue to develop much deeper into the detector, the energy deposited is shared among the CE-E, CE-H and AHCAL sections. Figure~\ref{fig:eneEE-vs-eneFHAhcal} shows the correlation between the energies measured in units of MIPs in active layers of the CE-E and CE-H $+$ AHCAL sections for pions with an energy of 50~GeV and 200~GeV. The pions which start showering in CE-E are referred to as {\it{``CE-E pions''}}, and those that behave like MIPs in CE-E are referred to as {\it{``CE-H pions''}} in the following. 

The absorbers used for the electromagnetic (Pb and CuW/Cu) and hadronic (stainless steel) compartments have a different evolution of showers generated by \epm/$\gamma$ ({\it{e}}) and hadrons ({\it{h}}). The sampling fractions of the three prototype calorimeter sections are also different. Hence, the energies measured in units of MIPs in the different compartments need to be converted, with different factors for each section, to GeV to reconstruct the total energy of pions. In Section~\ref{sec:enereco-det}, we reconstruct the pion energy using the MIP-to-GeV conversion scale obtained using 50~GeV positrons and 50~GeV pions for the electromagnetic and hadronic compartments, respectively. This allows for a direct comparison of data and simulation for CE-E pions and CE-H pions, and is used to correct the scales in simulation to match those measured in the data. A further optimization of the energy measurement using calibration factors dependent on beam energy is presented in Section~\ref{sec:enereco-optimize}.
\begin{figure}[ht]
  \centering
  \includegraphics[width=0.48\linewidth]{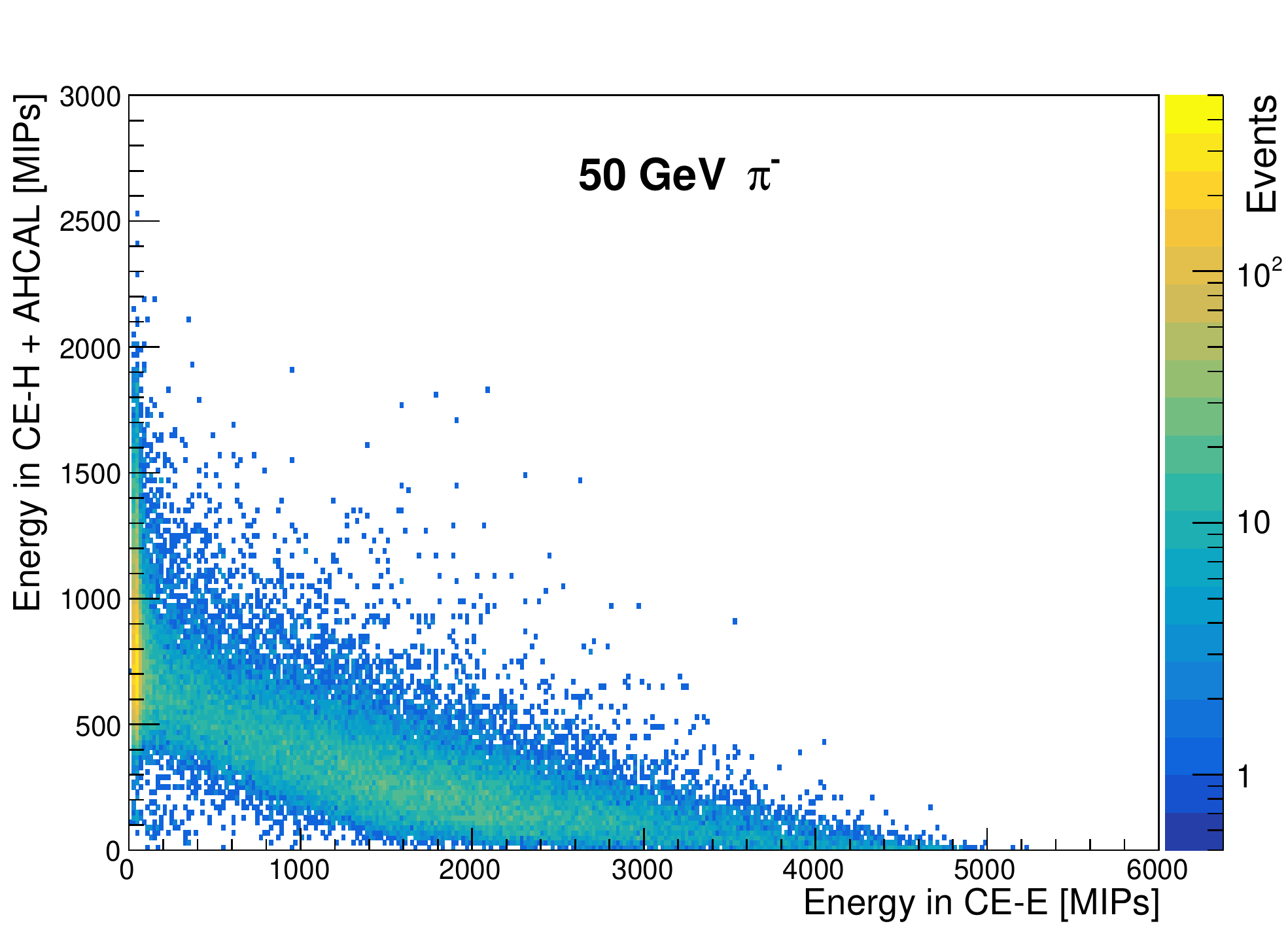}
  \includegraphics[width=0.48\linewidth]{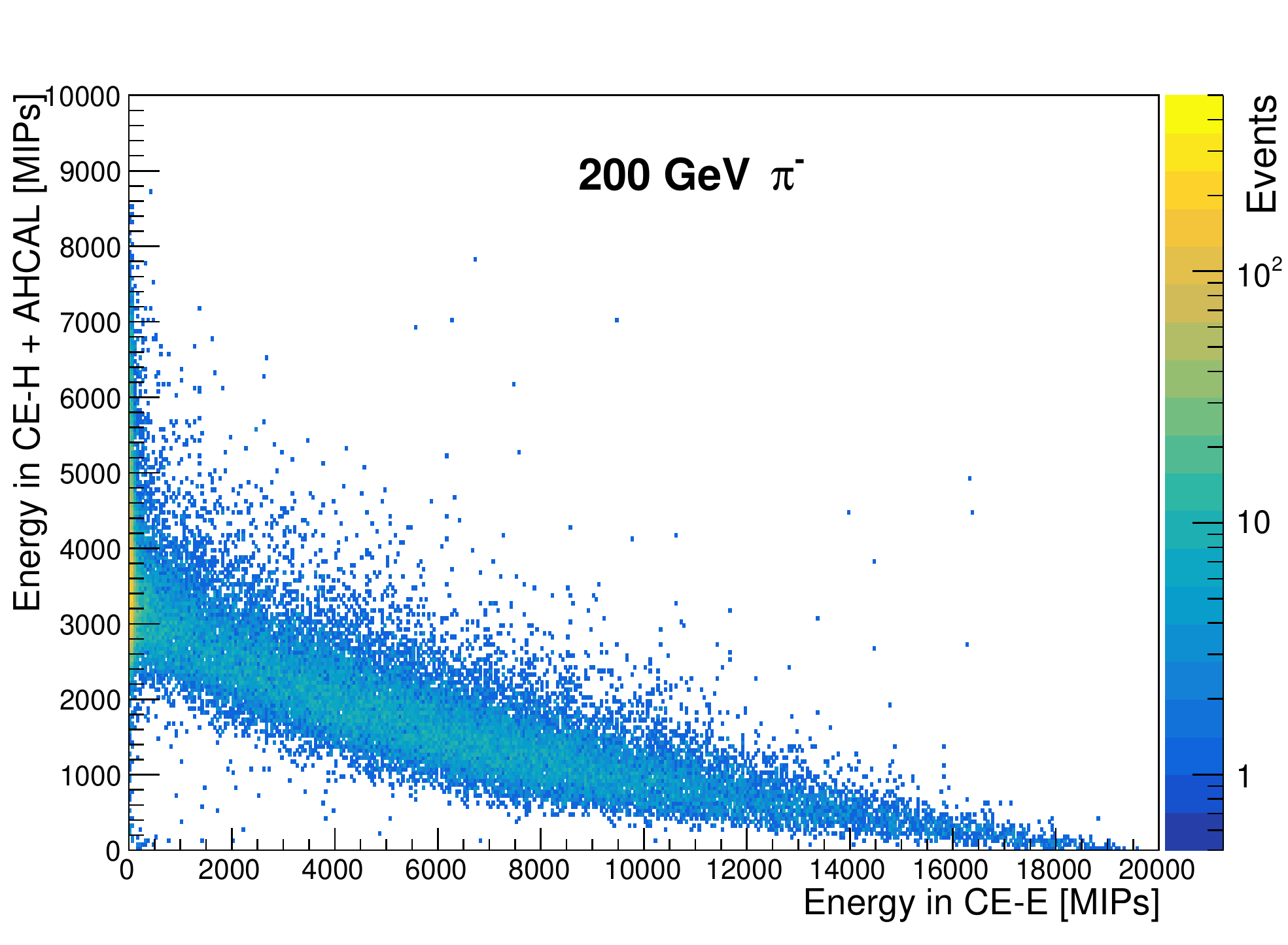}
  \caption{\label{fig:eneEE-vs-eneFHAhcal} Energy sharing (MIP units) between the CE-E and CE-H $+$ AHCAL prototypes for pions of (left) 50 and (right) 200~GeV, respectively.}
\end{figure}

\subsection{Energy reconstruction of pions} \label{sec:enereco-det}

Starting with the total energies measured in units of MIPs in the different subsections of the calorimeter, denoted $E^{CE-E}_{MIPs}$, $E^{CE-H}_{MIPs}$ and $E^{AHCAL}_{MIPs}$, the reconstructed pion energy is given by 
\begin{equation}
    E\,(GeV) = \alpha^{fix} \times E^{CE-E}_{MIPs} ~+~ \beta^{fix} \times (E^{CE-H}_{MIPs} + \delta^{fix} \times E^{AHCAL}_{MIPs}).
   \label{eq:reco:ene-fix}
\end{equation}
where $\alpha^{fix}$ and $\beta^{fix}$ are MIP-to-GeV conversion factors for the electromagnetic and hadronic sections respectively, and $\delta^{fix}$ is a relative weight factor between the energy deposited in CE-H and AHCAL. 

The $\alpha^{fix}$ sets the so-called electromagnetic scale, and it can be determined independently using \epm{} or $\gamma$ since these showers are fully contained in CE-E by design. For the studies presented here, it is determined using 50 GeV positron beam test data, and is found to be approximately 10.5~MeV per MIP. The CE-E response to positrons of the energy range 20--300~GeV measured in data is linear, and has been described in detail in \cite{bib.hgcal-2018-positrons}.

The 50 GeV pion data are used to determine the $\beta^{fix}$ since the positron showers do not extend beyond the CE-E. To avoid complications due to the different ${\it{e}}/\pi$ response from lead and steel absorbers, a sample of CE-H pions is selected using the shower start algorithm described in the previous section. The value of $\delta^{fix}$ = 0.4 combines the $E^\mathrm{CE-H}_\mathrm{MIPs}$ and $E^\mathrm{AHCAL}_\mathrm{MIPs}$ for 50 GeV CE-H pions in eqn. \ref{eq:reco:ene-fix} such that it minimizes the relative resolution of the reconstructed energy which is determined, from the mean ($\mu$) and width ($\sigma$) of a Gaussian fit, as $\sigma/\mu$. A similar value of $\delta^{fix}$ is obtained for the other beam energies. After fixing $\delta^{fix}$, $\beta^{fix}$ is obtained by fitting the combined energy of 50 GeV CE-H pions in data with a Gaussian function such that the mean value reproduces the correct energy, and is measured to be approximately 80~MeV per MIP.

The distribution of the energy determined for 50~GeV pions using the CE-E calibrated to 50 GeV positrons and CE-H + AHCAL calibrated to 50~GeV pions in the beam test data is shown in Figure~\ref{fig:eneReco-fixedWt} (left). The CE-H pions peak at 50~GeV as the factors $\beta^{fix}$ and $\delta^{fix}$ are obtained from the same sample. The average energy of CE-E pions is less than 50~GeV because hadronic showers deposit less visible energy than the electromagnetic showers in our noncompensating calorimeter configuration. In simulated pion samples, the energy recorded in the different calorimeter sections is combined using the same values of $\alpha^{fix}$, $\beta^{fix}$, and $\delta^{fix}$ as obtained from the data. A comparison of the reconstructed energy in data and simulation for 50~GeV CE-H pions is presented in Figure~\ref{fig:eneReco-fixedWt} (right). Both the \qgsp{} and \ftfp{} simulation physics lists over-predict the energy scale in the hadronic sections of the calorimeter prototype.

\begin{figure}[ht]
  \centering
  \includegraphics[width=0.45\linewidth]{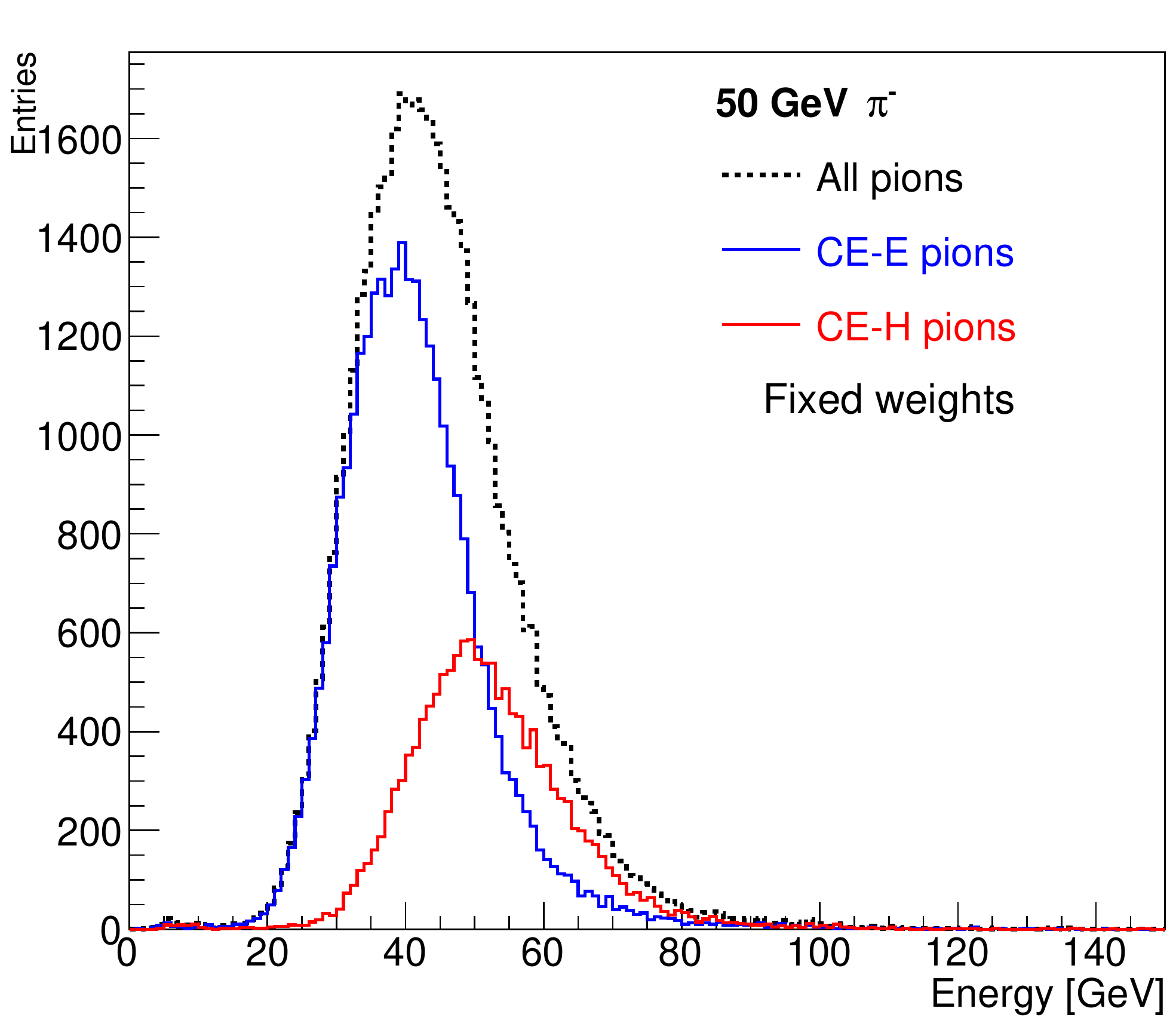}
  \includegraphics[width=0.45\linewidth]{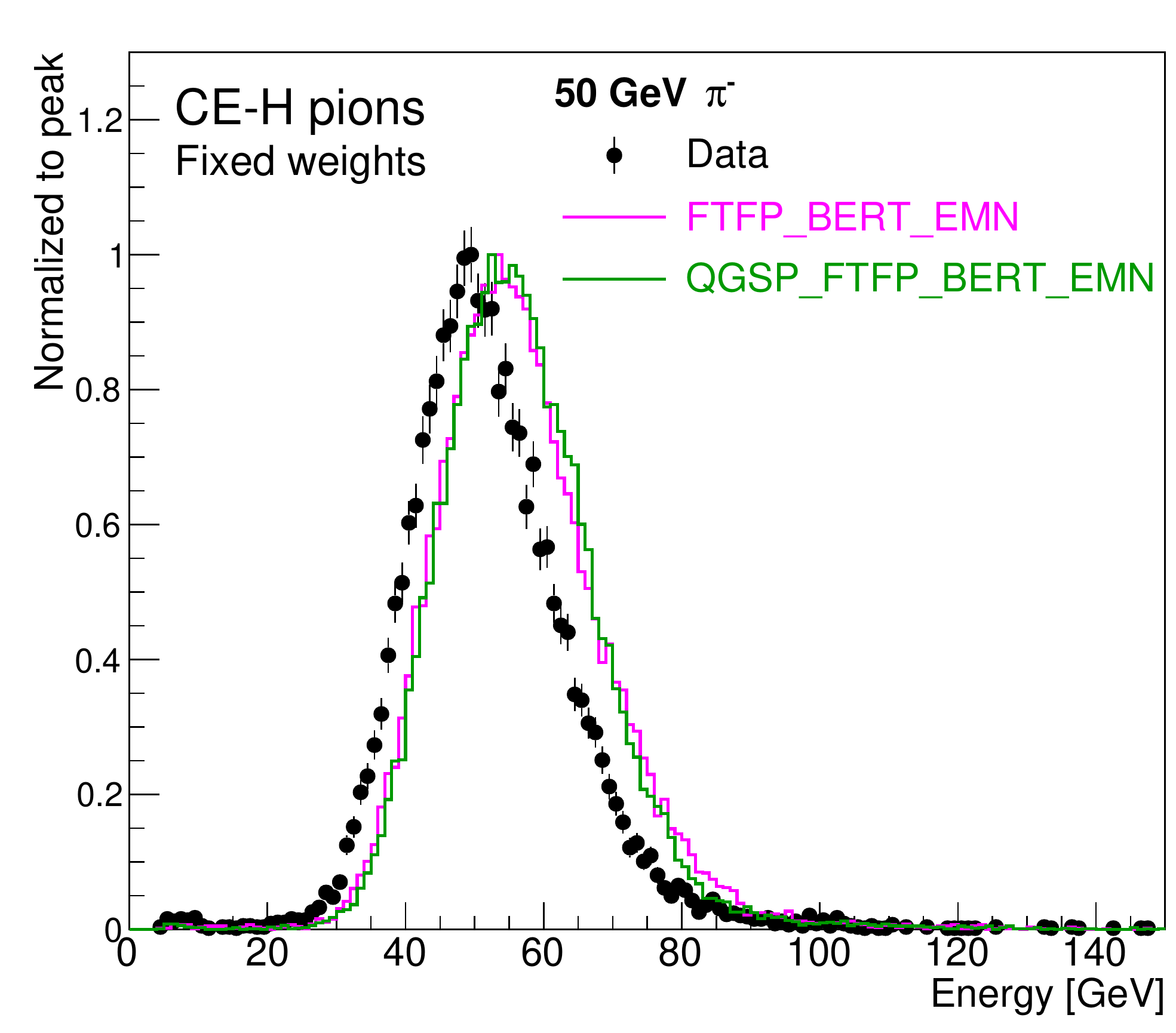}
  \caption{\label{fig:eneReco-fixedWt} Energy measured for 50~GeV CE-E pions, CE-H pions, and all pions (left), and energy measured for 50 GeV CE-H pions and simulation (right).}
\end{figure}

Due to the non-compensating nature of calorimeters and the fact that the $\pi^0$ component produced in hadronic showers depends on the incident beam energy, a nonlinearity in response is expected. Here, the response is defined as the average of the energy measured, taken as the $\mu$ of the Gaussian function used to fit the measured energy spectrum, normalized to the incident beam energy. The measured response as a function of beam energy is shown in Figure~\ref{fig:response-fixedWeight} (left) for CE-H pions. The response is unity for the 50 GeV pions in data, by construction. For the incident energy of pions in the range 20$-$300 GeV, the response is nonlinear by $\pm$10\% with respect to the energy scale fixed using 50 GeV pions. The simulated response is consistently over-predicted by $\sim$10\% for all energies in hadronic compartments. However, the nonlinearity of the response is reproduced by the simulation.

\begin{figure}[ht]
  \centering
  \includegraphics[width=0.45\linewidth]{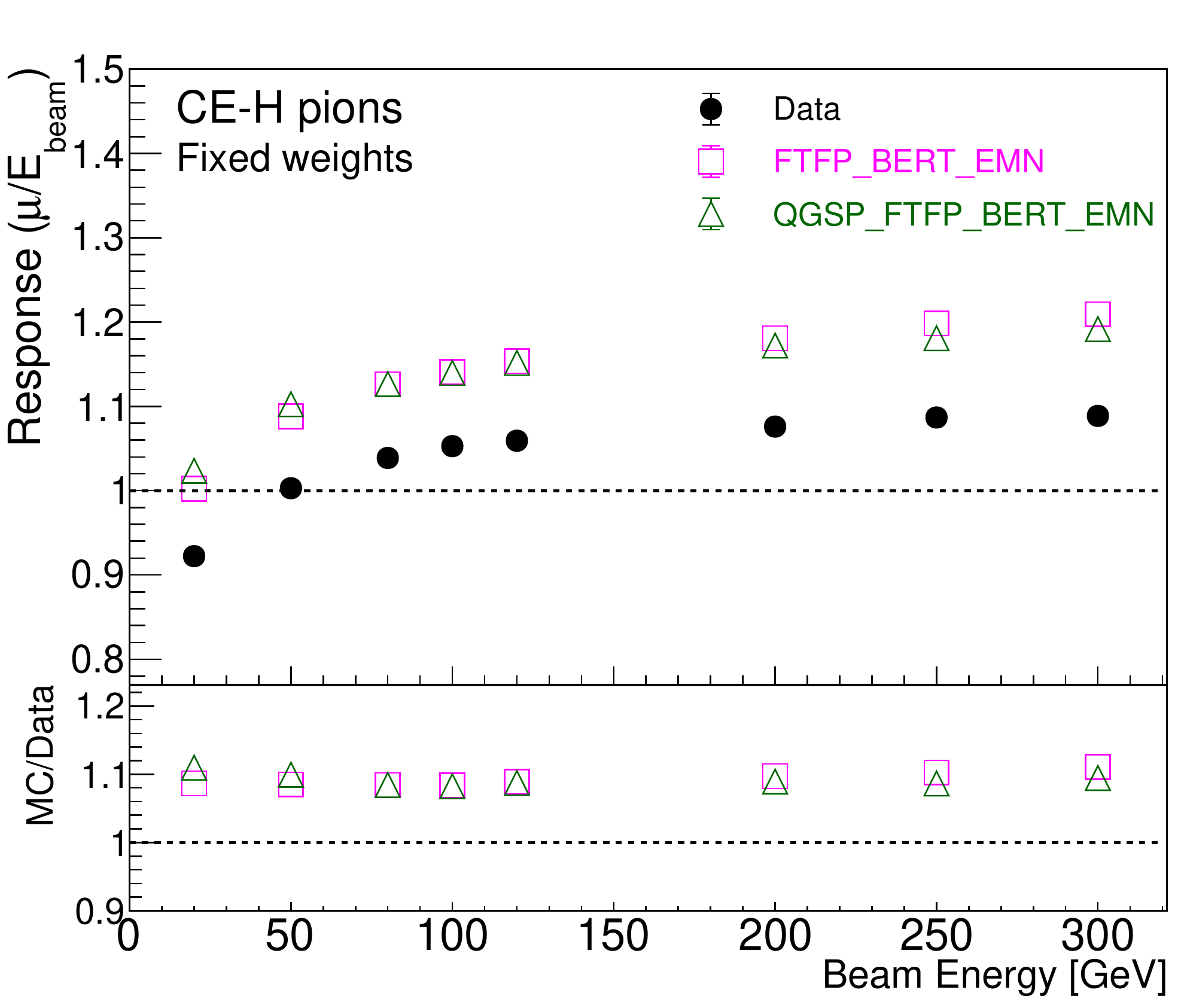}
  \includegraphics[width=0.45\linewidth]{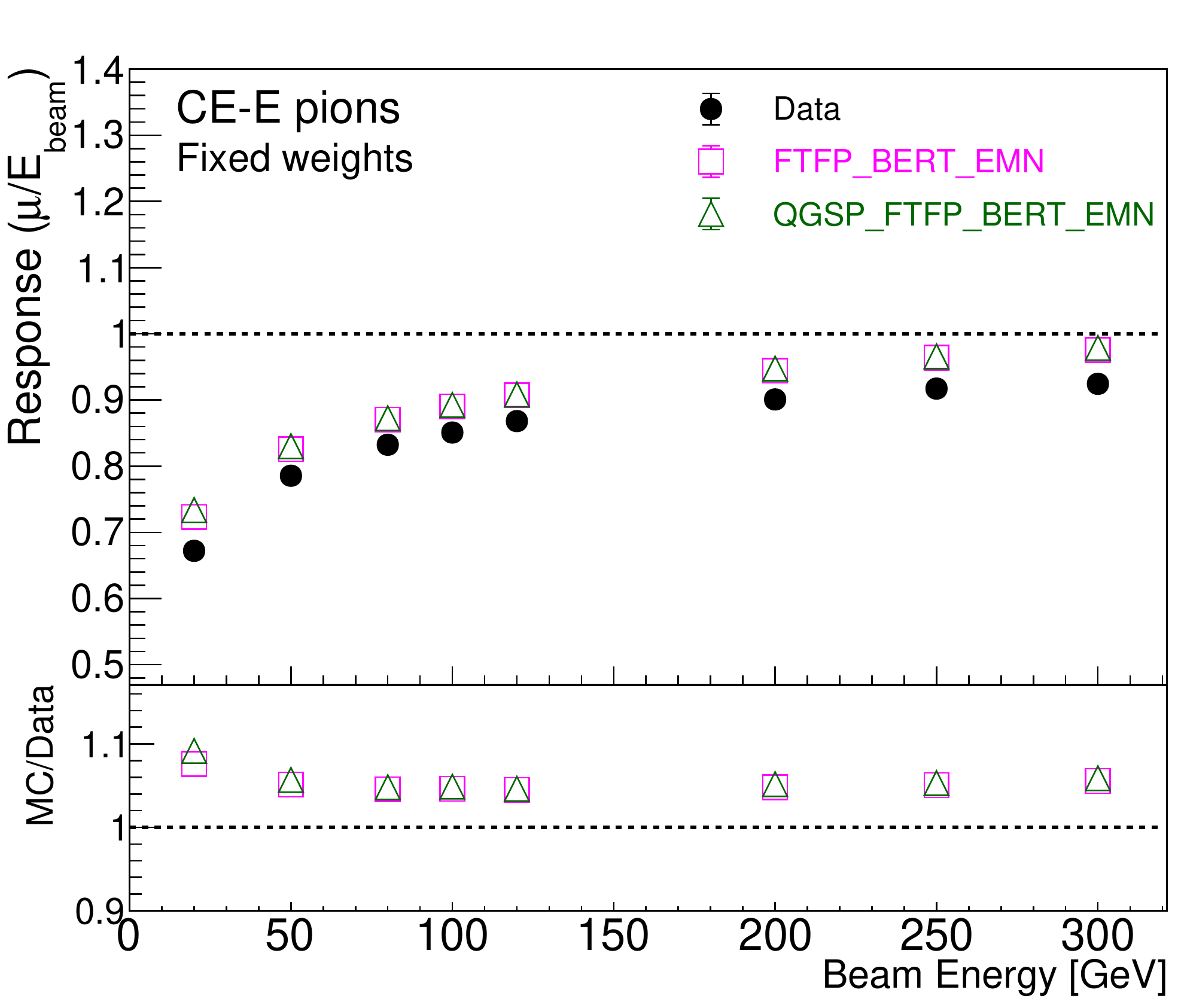}
  \includegraphics[width=0.45\linewidth]{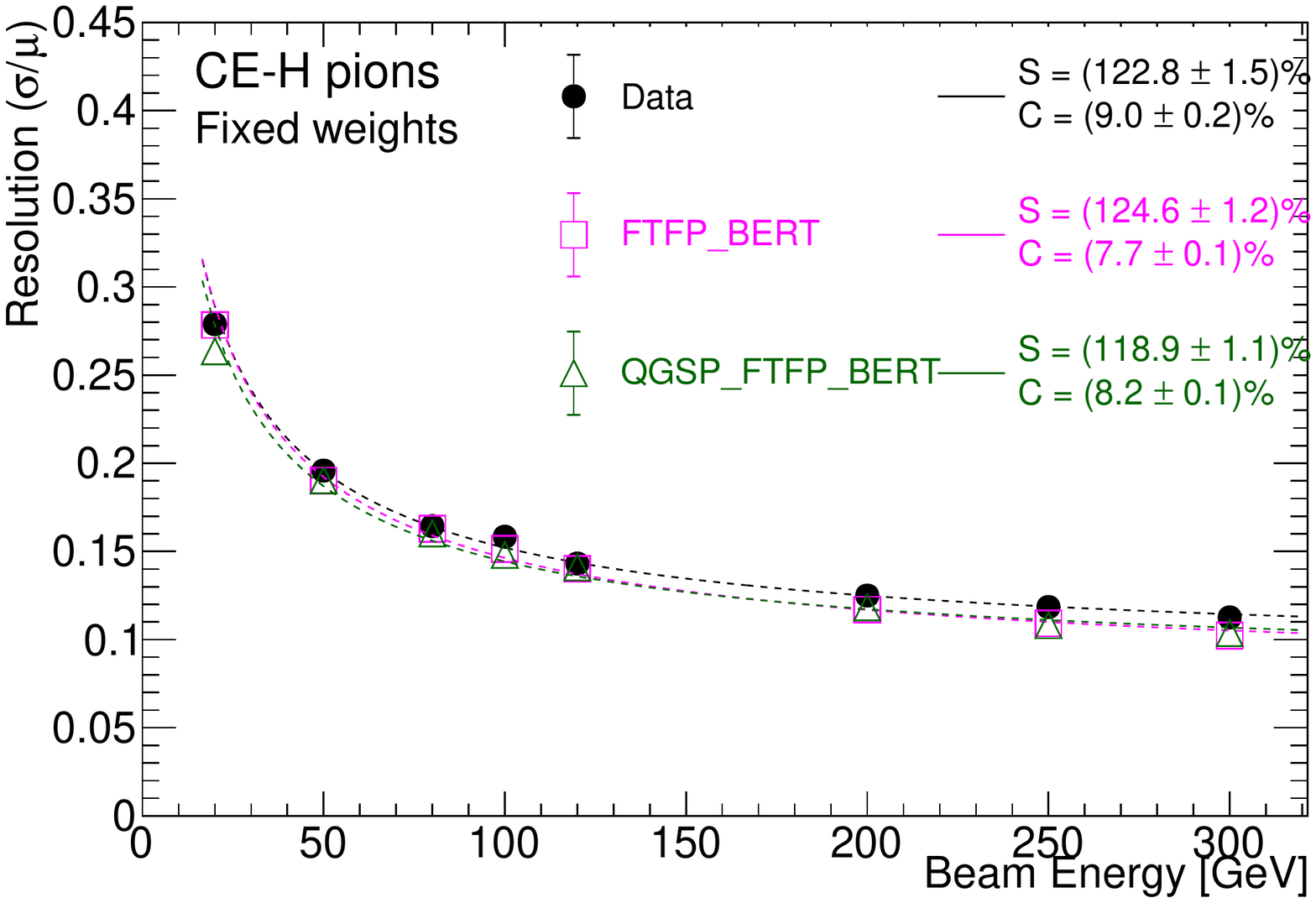}
  \includegraphics[width=0.45\linewidth]{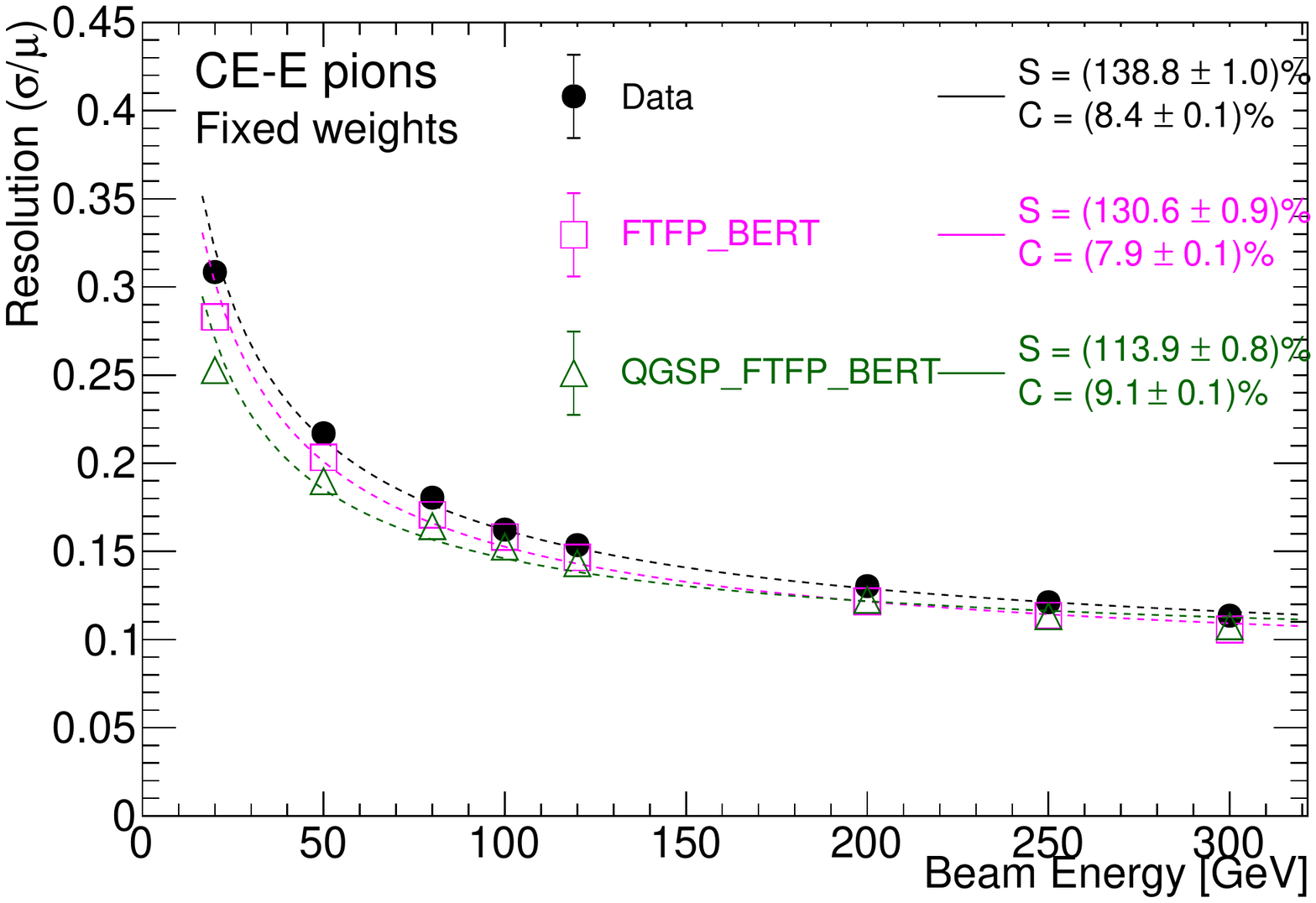}
  \caption{\label{fig:response-fixedWeight} Energy response and resolution as a function of beam energy for CE-H pions (left) and CE-E pions (right) where the energy scale of the CE-E is fixed to 50~GeV positrons and that of CE-H + AHCAL is fixed to 50~GeV using CE-H pions. The {\it{S}} and {\it{C}} are the fit parameters of the function $\sqrt{\it{S^2/E + C^2}}$ used to fit resolution as a function of energy.}
\end{figure}

The response of CE-E pions from simulation is higher than that measured from data by $\sim$5\% across all beam energies as shown in Fig.~\ref{fig:response-fixedWeight} (right). Detailed studies of the CE-E response to positrons concluded that the simulated response is higher by $\sim$3.5\% than the measured response \cite{bib.hgcal-2018-positrons}. The simulation reproduces the  observed nonlinearity for CE-E pions. The response is lower than one because the CE-E is calibrated to 50~GeV positrons and {\it e/h} for our calorimeter setup is less than one.

Based on these studies, we correct the energy scale in simulation by 9.5\% for CE-H and AHCAL using the 50 GeV pion response, and by 3.5\% for CE-E using positron response. The resolution as a function of beam energy is shown in Figure~\ref{fig:response-fixedWeight} (bottom) for the CE-H pions (left) and CE-E pions (right) in data and simulation. The resolution is fitted using the functional form $\sqrt{\it{S^2/E + C^2}}$ where {\it{S}} and {\it{C}} are the fit parameters, and are referred to as stochastic and constant terms of the resolution, respectively. The noise term of the resolution is found to be negligible and is omitted from the discussion of the results. The resolution is observed to scale inversely with $\sqrt{\it{E}}$. We obtain a stochastic term of $\sim$123\% ($\sim$139\%) and a constant term of 9.0\% (8.4)\% for CE-H pions (CE-E pions). Using the same weights as used for the data, the \qgsp{} and \ftfp{} predicts a slightly better resolution for most of the beam energies. The values of the {\it{S}} and {\it{C}} parameters obtained using the two physics lists match within 7$-$10\% with those obtained in data.

\subsection{Optimization of energy reconstruction of pions}
\label{sec:enereco-optimize}
As discussed in the previous section, owing to the nature of hadronic showers and the different response of calorimeters to electromagnetic and hadronic shower components, the response of the detector to pions is not linear as a function of incident energy. In addition, these aspects also result in an overall degradation of resolution when the electromagnetic and hadronic sections are calibrated to a fixed energy scale, see Fig.~\ref{fig:eneReco-fixedWt} (left). Even if one uses an overall factor to correct for the mean response of CE-E pions to one, the nonlinearity cannot be fixed. Following the approach used for the calibration of particle flow hadrons in CMS \cite{bib.cmsPFPaper}, we optimize the MIP-to-GeV scale factors as a function of the incident beam energy. Hence, the total energy measured is given as
\begin{equation}
    E(GeV) = \alpha_1\,(E^{beam}) \times \left(E^{CE-E}_{fix}\right) ~+~ \beta_1\,(E^{beam}) \times \left(E^{CE-H}_{fix}\right) + \gamma_1\,(E^{beam}) \times \left(E^{AHCAL}_{fix}\right), \\
  \label{eq:recoEne-CEEShow}
\end{equation} 
for CE-E pions and

\begin{equation}
    E(GeV)  = E^{CE-E}_{fix} ~+~ \beta_2\,(E^{beam}) \times \left(E^{CE-H}_{fix}\right) + \gamma_2\,(E^{beam}) \times \left(E^{AHCAL}_{fix}\right),
  \label{eq:recoEne-CEEMIPs}
\end{equation}
for CE-H pions. Here $\alpha_1$, $\beta_{1,2}$, and $\gamma_{1,2}$, also referred to as weights, are the parameters optimized for each beam energy using a $\chi^2$ minimization procedure described in the following. The $E^{CE-E}_{fix}$, $E^{CE-H}_{fix}$, and $E^{AHCAL}_{fix}$ refer to the energy measured in respective compartments using the MIP-to-GeV scale fixed to 50~GeV positrons and pions beam data as described in Section~\ref{sec:enereco-det}. The respective energy scales in simulation have been corrected to account for the differences with the data. For CE-H pions, the energy of the pion track measured in CE-E is added to the total energy of the pions and is represented by the first term of equation~\ref{eq:recoEne-CEEMIPs}.

The $\chi^2$ used in the optimization of these weights is defined as:
\begin{equation}
  \chi^2 = \sum_{i} \frac{(E^{beam} - E^i)^2}{\sigma^2(E^{i}_{fix})},
  \label{eq:reco-chi2}
\end{equation}
where the sum runs over all the pion events of a given beam energy, $E^i$ is the energy measured using Equations~\ref{eq:recoEne-CEEShow} and \ref{eq:recoEne-CEEMIPs} for the ${\it{i}^{th}}$ event separately, and $\sigma(E^i_{fix})$ in the denominator is a preliminary estimate of the resolution corresponding to the energies measured using fixed weights as described in Section~\ref{sec:enereco-det}. The $\chi^2$ values corresponding to equations~\ref{eq:recoEne-CEEShow} and \ref{eq:recoEne-CEEMIPs} are separately minimized by zeroing the first derivative with respect to the $\alpha$, $\beta$ and $\gamma$ parameters. The simultaneous equations obtained in this way corresponding to CE-E pions or CE-H pions are solved using a matrix formulation. The weights obtained by this procedure for different beam energies are summarized in Figure~\ref{fig:chi2-weights}. The weights are also fitted as a function of beam energy using a two parameter function, namely $p_0 + p_1/\sqrt{E^{beam}}$. The fitted functions extrapolated down to $E^{beam} =$ 5 GeV along with the values of $p_0$ and $p_1$ are shown in the same figure.

\begin{figure}[htbp]
  \centering
  \includegraphics[width=0.45\textwidth]{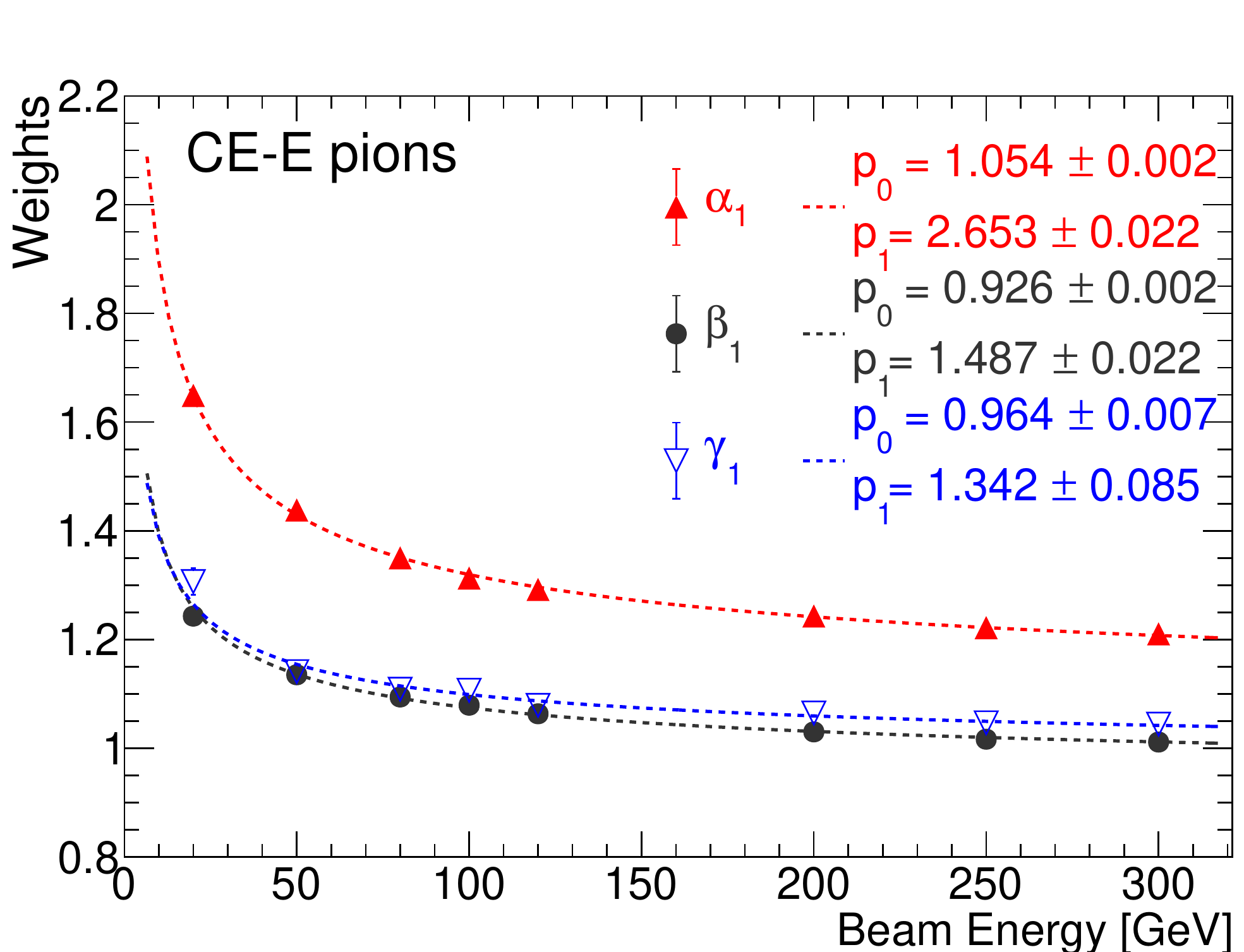}
  \includegraphics[width=0.45\textwidth]{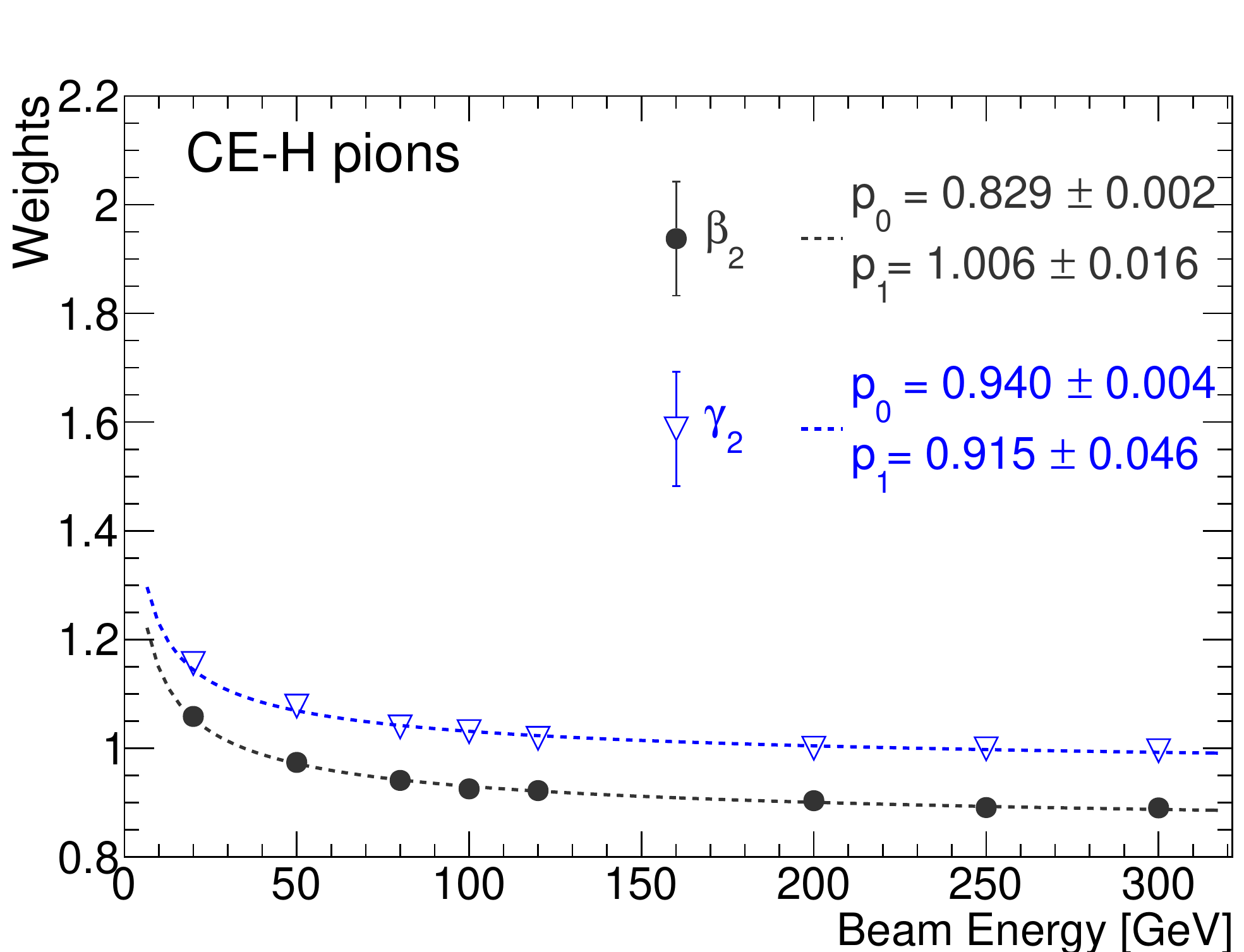}
  \caption{\label{fig:chi2-weights} Weights $\alpha_{1}$, $\beta_{1,2}$, and $\gamma_{1,2}$ for different beam energies for CE-E pions (left) and CE-H pions (right).}
\end{figure}

The energy measured using the weights $\alpha_{1}$, $\beta_{1}$, and $\gamma_{1}$ for CE-E pions, and $\beta_{2}$, and $\gamma_{2}$ for CE-H pions are shown in Figure~\ref{fig:ene-ch2Weight} for 50 and 200~GeV incident energies. The same calibration factors, obtained from data, are also used to reconstruct the energy of pions simulated using \qgsp{} and \ftfp{} physics lists. The energy distributions predicted by the simulation agree very well with the data as shown in Figure~\ref{fig:ene-ch2Weight} for CE-E pions (left) and CE-H pions (right) for 50 GeV and 200~GeV beam energies.

\begin{figure}[ht]
  \centering
  \includegraphics[width=0.40\linewidth]{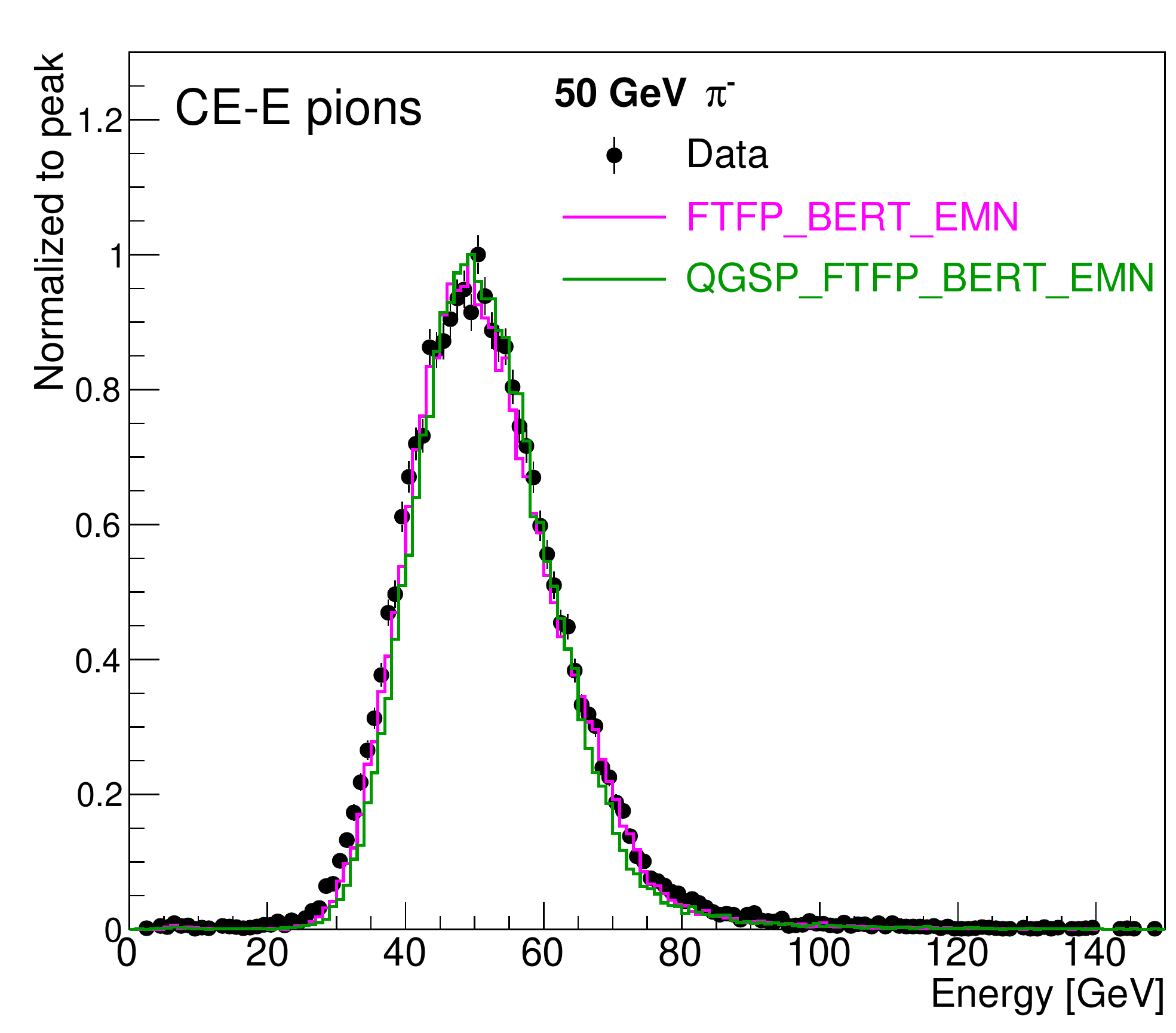}
  \includegraphics[width=0.40\linewidth]{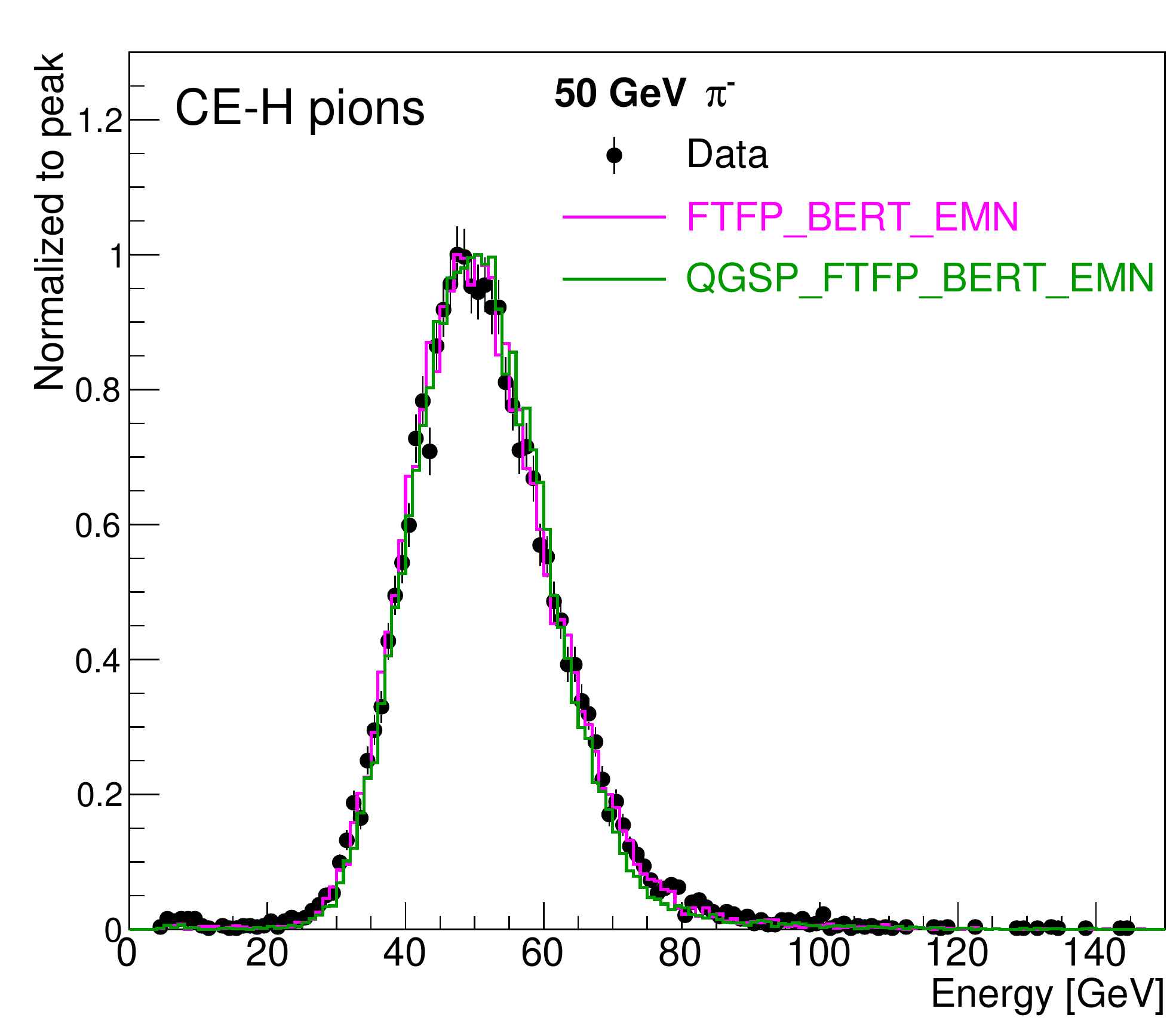}
  \includegraphics[width=0.40\linewidth]{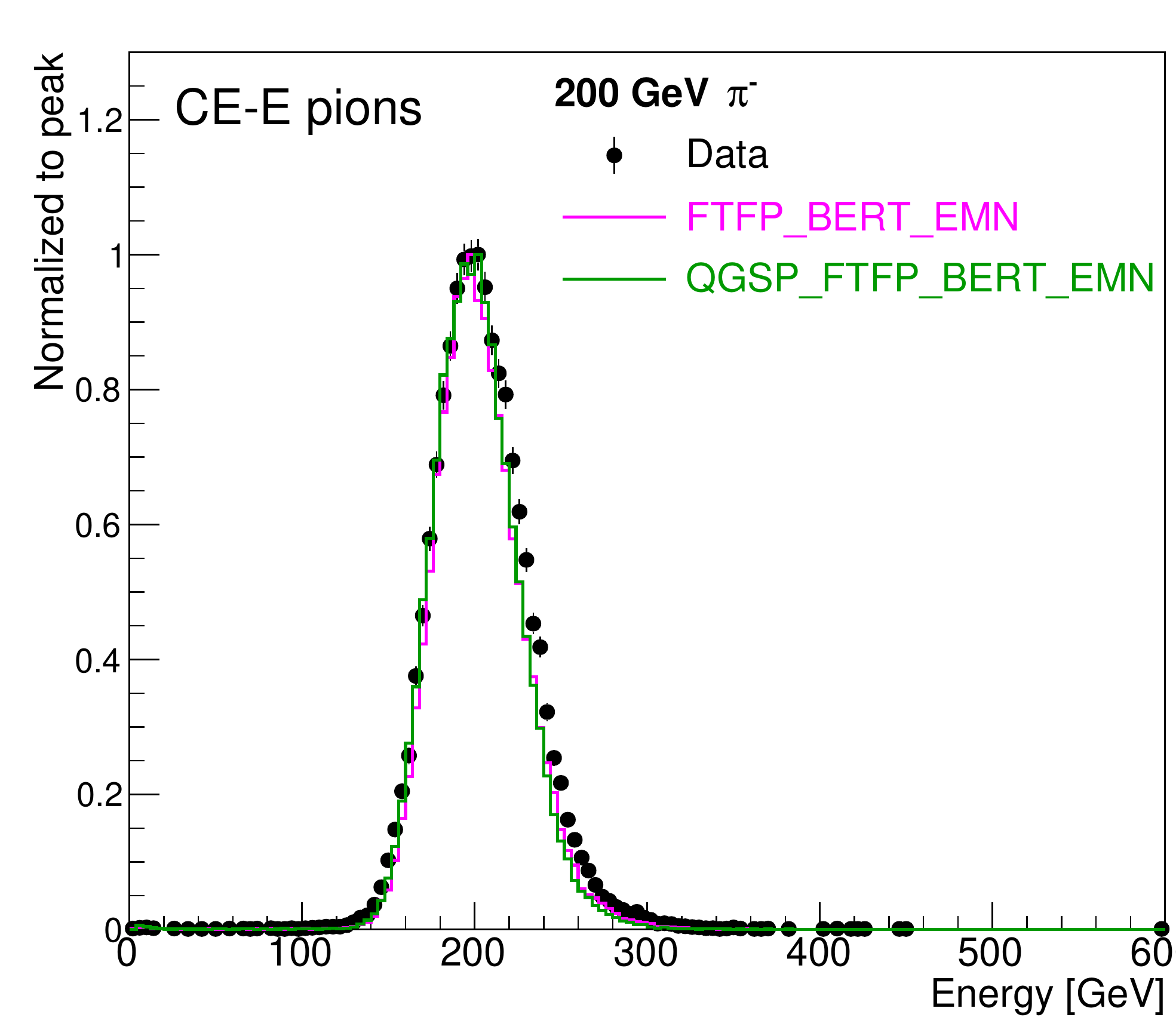}
  \includegraphics[width=0.40\linewidth]{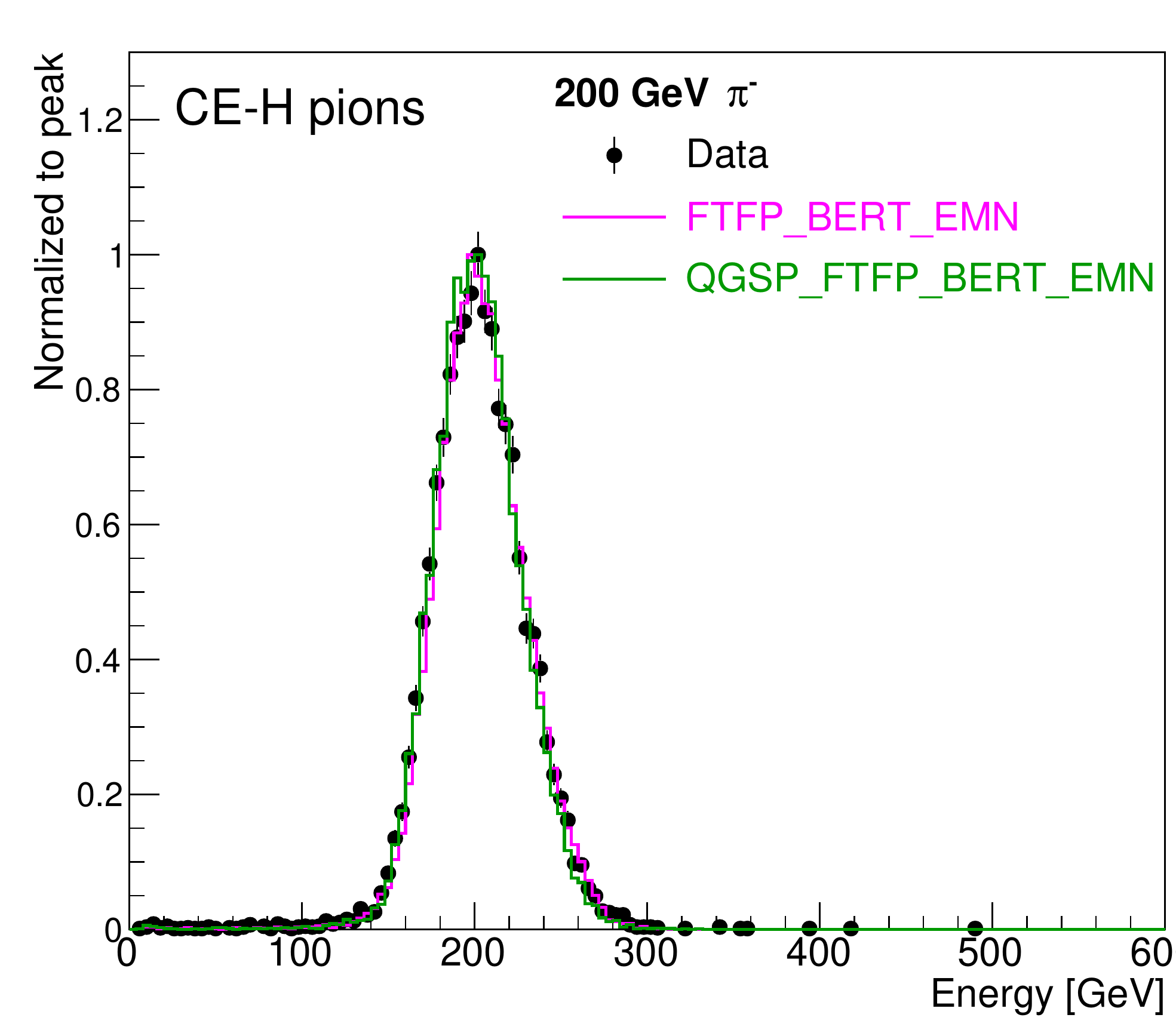}
  \caption{\label{fig:ene-ch2Weight} Energy measured for 50 GeV (upper) and 200~GeV (lower) CE-E pions (left) and CE-H pions (right) where energies measured in CE-E, CE-H and AHCAL are combined using a $\chi^2$ minimization. The distributions measured in data are compared to those predicted by simulation using the same weights as obtained from data.}
\end{figure}

These reconstructed energy distributions are fitted with a Gaussian function iteratively, updating the range of fit to be $\mu\,\pm$1.5$\sigma$ at each successive iteration. The response is defined as the fitted parameter $\mu$ normalized to the beam energy, and resolution is defined as the $\sigma$/$\mu$ for each beam energy, and these are summarized as a function of beam energy in Figure~\ref{fig:resp-resol-chi2Weight} (left) for CE-E pions and in Figure~\ref{fig:resp-resol-chi2Weight} (right) for CE-H pions. The response of pions in simulation is also linear within 2--3\% using the same set of energy-dependent weights determined from the data. The weights obtained using dataset corresponding to the given beam energy are used in the results presented here. The \qgsp{} physics list predicts slightly better resolution for CE-E pions, especially at energies below 100~GeV (lower stochastic term) while \ftfp{} predicts a smaller constant term resulting in slightly better resolution at high energies. For CE-H pions, the resolution predicted by simulation matches well with that measured in data.

\begin{figure}[ht]
  \centering
  \includegraphics[width=0.48\linewidth]{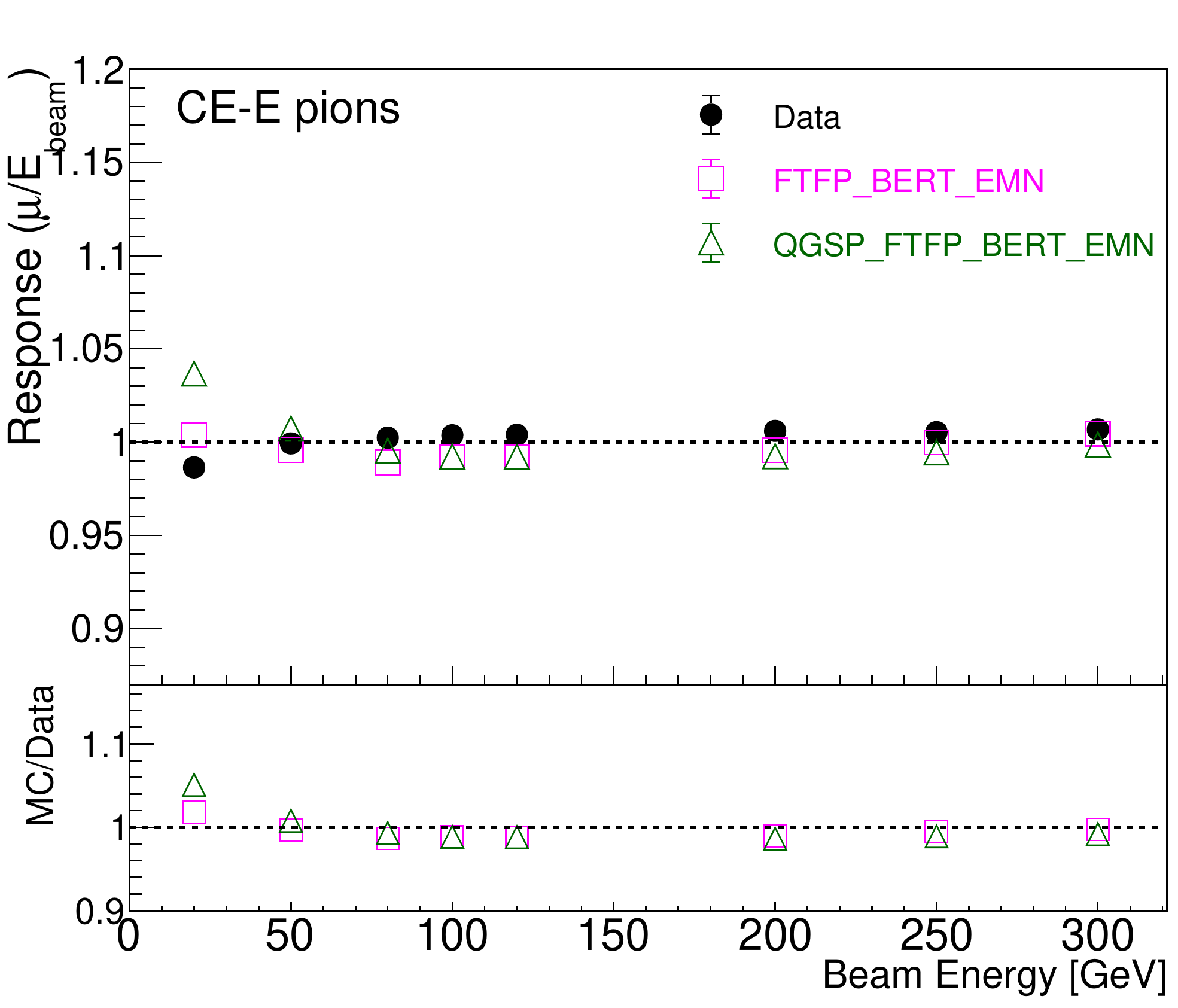}
  \includegraphics[width=0.48\linewidth]{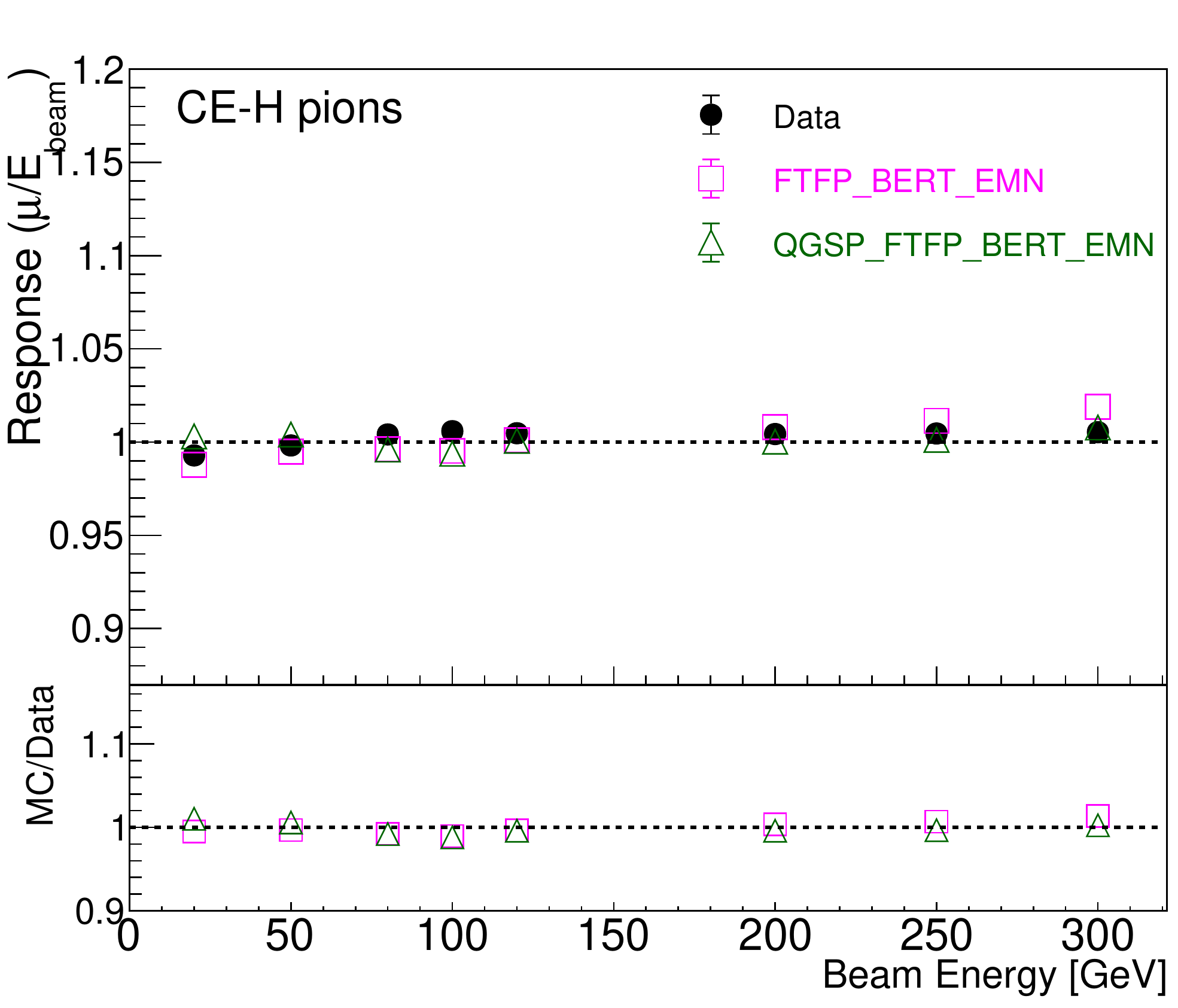}
  \includegraphics[width=0.48\linewidth]{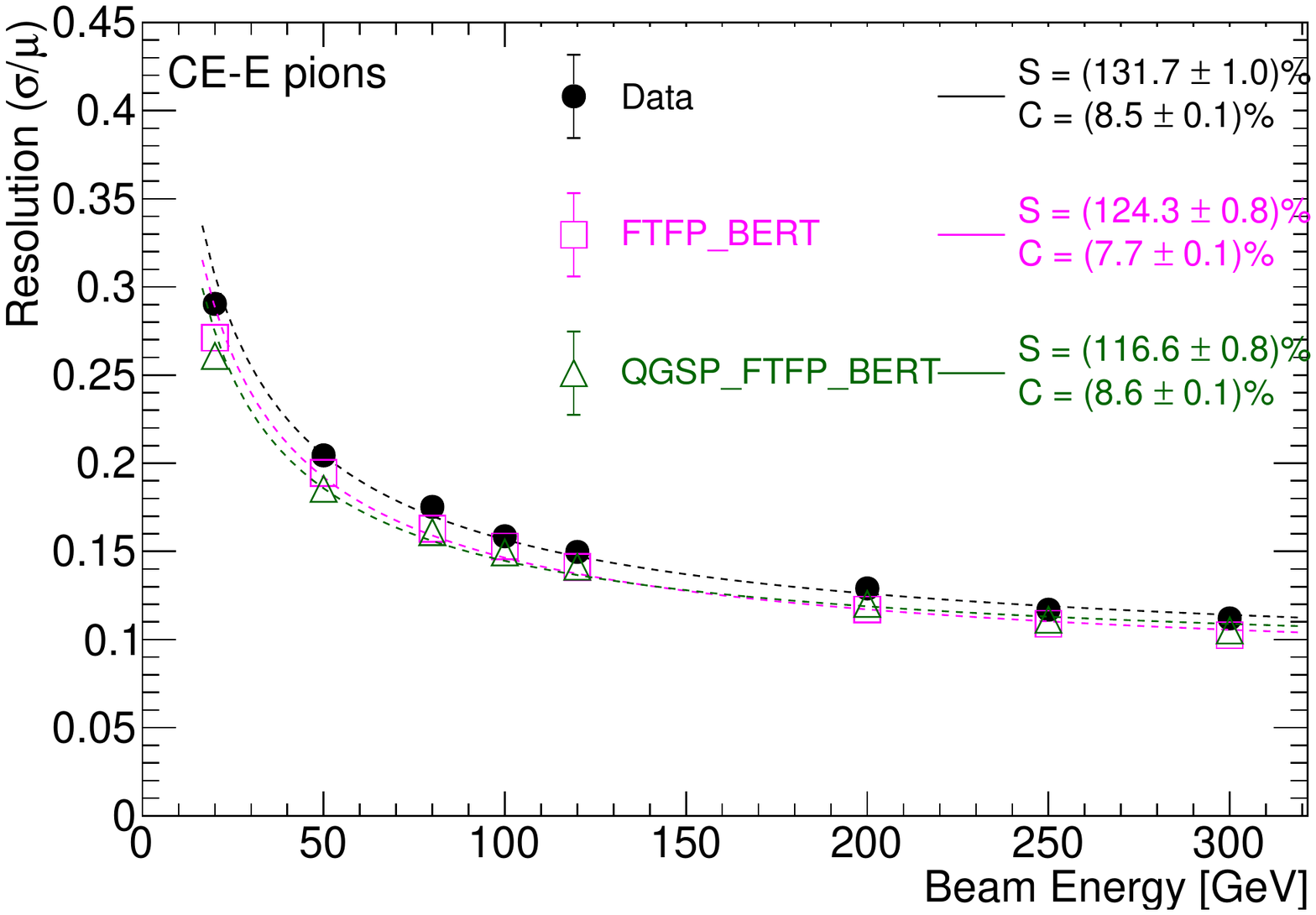}
  \includegraphics[width=0.48\linewidth]{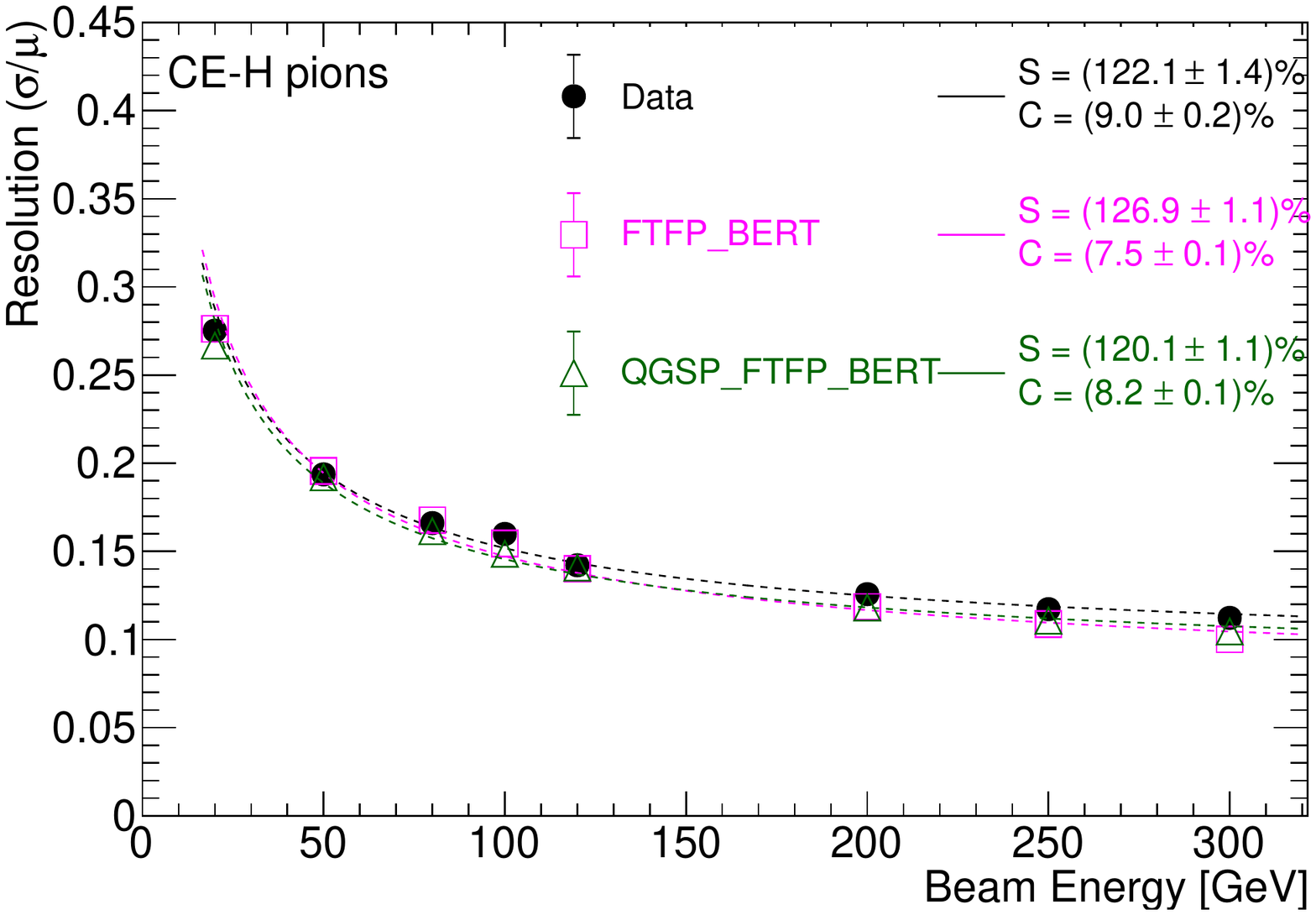}
  \caption{\label{fig:resp-resol-chi2Weight} Energy response and resolution as a function of beam energy for CE-E pions (left) and CE-H pions (right) for which the energy scale of CE-E, CE-H and AHCAL is obtained using $\chi^2$ minimization.}
\end{figure}

To summarize, for the prototype setup used in this beam test experiment, the pion showers are reconstructed with a stochastic term of $\sim$122\% ($\sim$132\%) and a constant term of 9.0\% (8.5)\% for CE-H pions (CE-E pions) in data. The corresponding values measured in the pion showers simulated using \qgsp{} and \ftfp{} physics lists are consistent with the data within 10\%. Using this method of combining the energies measured in the CE-E, CE-H and AHCAL prototype sections using the method of $\chi^2$ minimization, the CE-E pions and CE-H pions show similar resolution, see Figure~\ref{fig:resol-data-erecoref} (left).

For the results presented in the Figures \ref{fig:resp-resol-chi2Weight} and \ref{fig:resol-data-erecoref} (left), we have used the $\chi^2$ weights determined for the given beam energy. In the regions of the detector which are not covered by tracking elements, a precise proxy to true particle energy is not available as a reference. In that case, one can use the preliminary energy reconstructed using fixed energy scales (equation \ref{eq:reco:ene-fix}) as a reference to assign $\alpha_1$, $\beta_{1,2}$, and $\gamma_{1,2}$ from the fitted functions presented in Figure~\ref{fig:chi2-weights} for a given hadron. The resolution so obtained is compared with that obtained using the beam energy as reference in Figure~\ref{fig:resol-data-erecoref} (right). A slight improvement is observed in resolution when dynamically assigning $\chi^2$ weights for energy measurements as these weights are increasing with smaller energies, thus bringing lower reconstructed energies closer to the central values. In addition to this overall decrease in spread of the energy distribution, the response is also overestimated by up to 5\% for CE-E pions below 50~GeV as shown in Figure~\ref{fig:resp-data-erecoref}, contributing to a decrease in the value of $\sigma/\mu$. However, an implementation of this method needs to be explicitly validated using a realistic momentum spectrum of hadrons expected in proton proton collision events in the presence of pileup.

\begin{figure}[ht]
  \centering
  \includegraphics[width=0.45\linewidth]{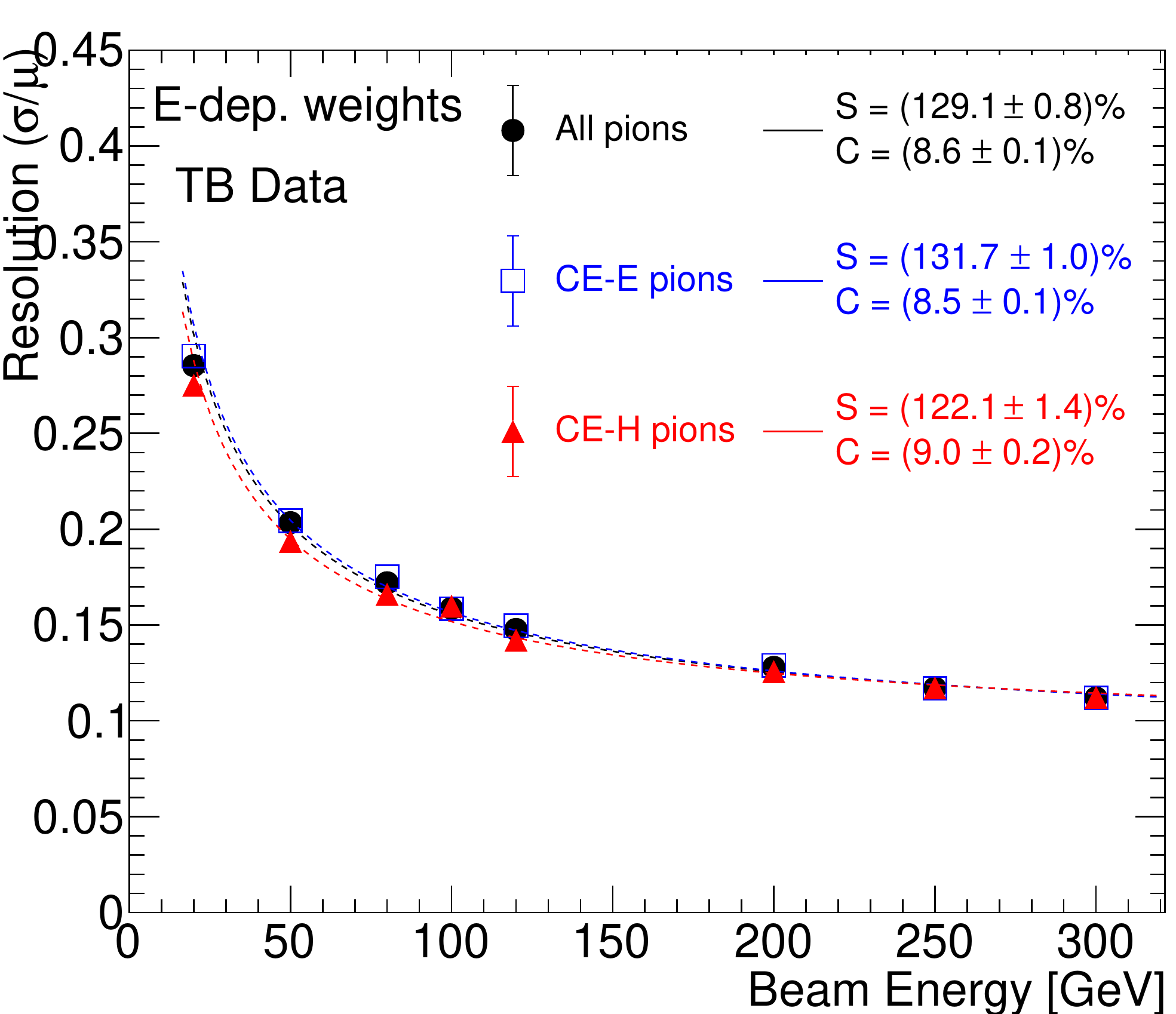}
  \includegraphics[width=0.45\linewidth]{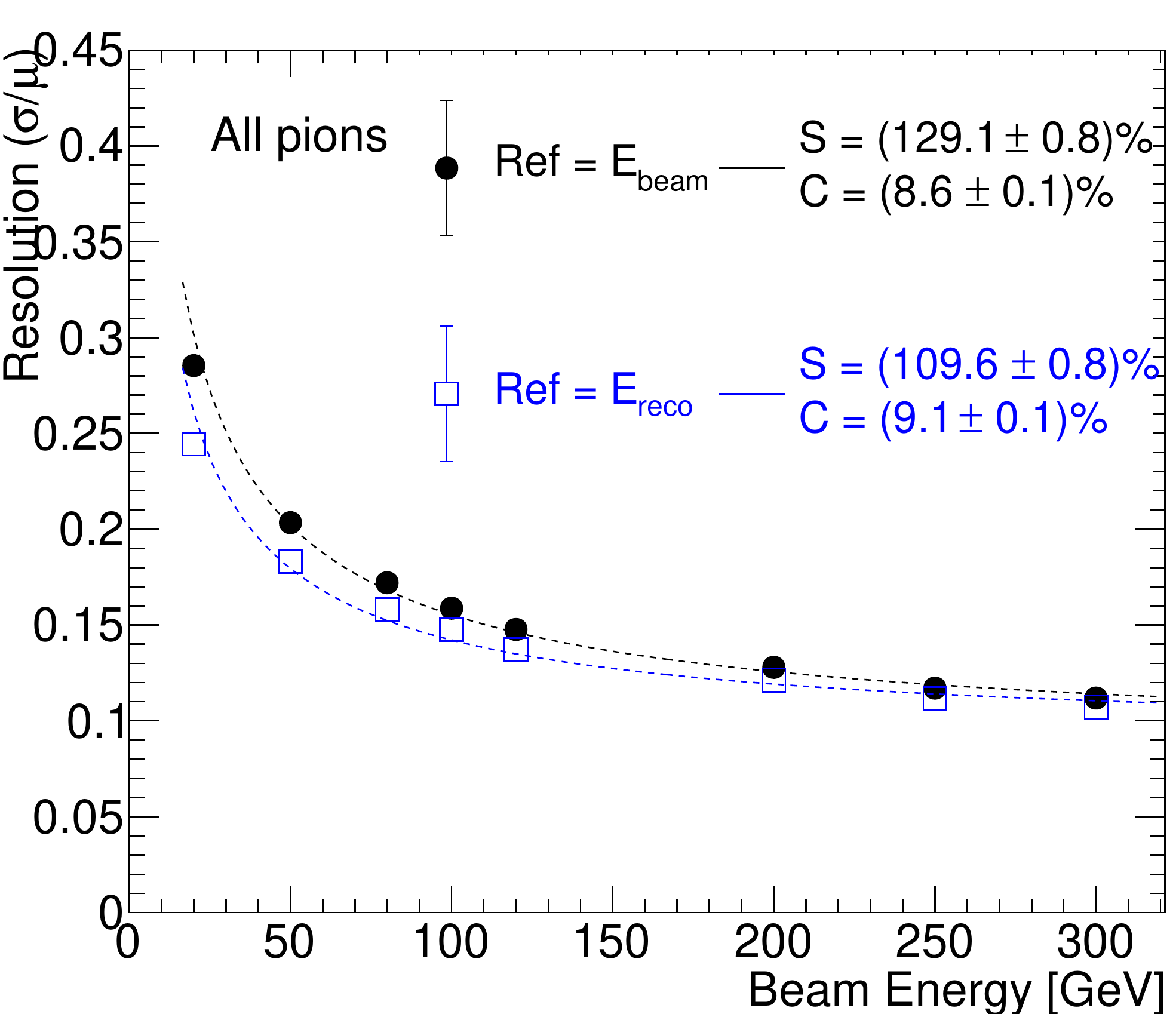}
  \caption{\label{fig:resol-data-erecoref} {Comparison of the resolution using $\chi^2 weights$ determined for the given beam energy for the sample of all pions with CE-H pions and CE-E pions (left), and with that obtained using the reconstructed calorimeter energy as reference to apply calibration factors $\alpha_1$, $\beta_{1,2}$, and $\gamma_{1,2}$ from the fitted functions (right).  }}

\end{figure}

\begin{figure}[ht]
  \centering
  \includegraphics[width=0.45\linewidth]{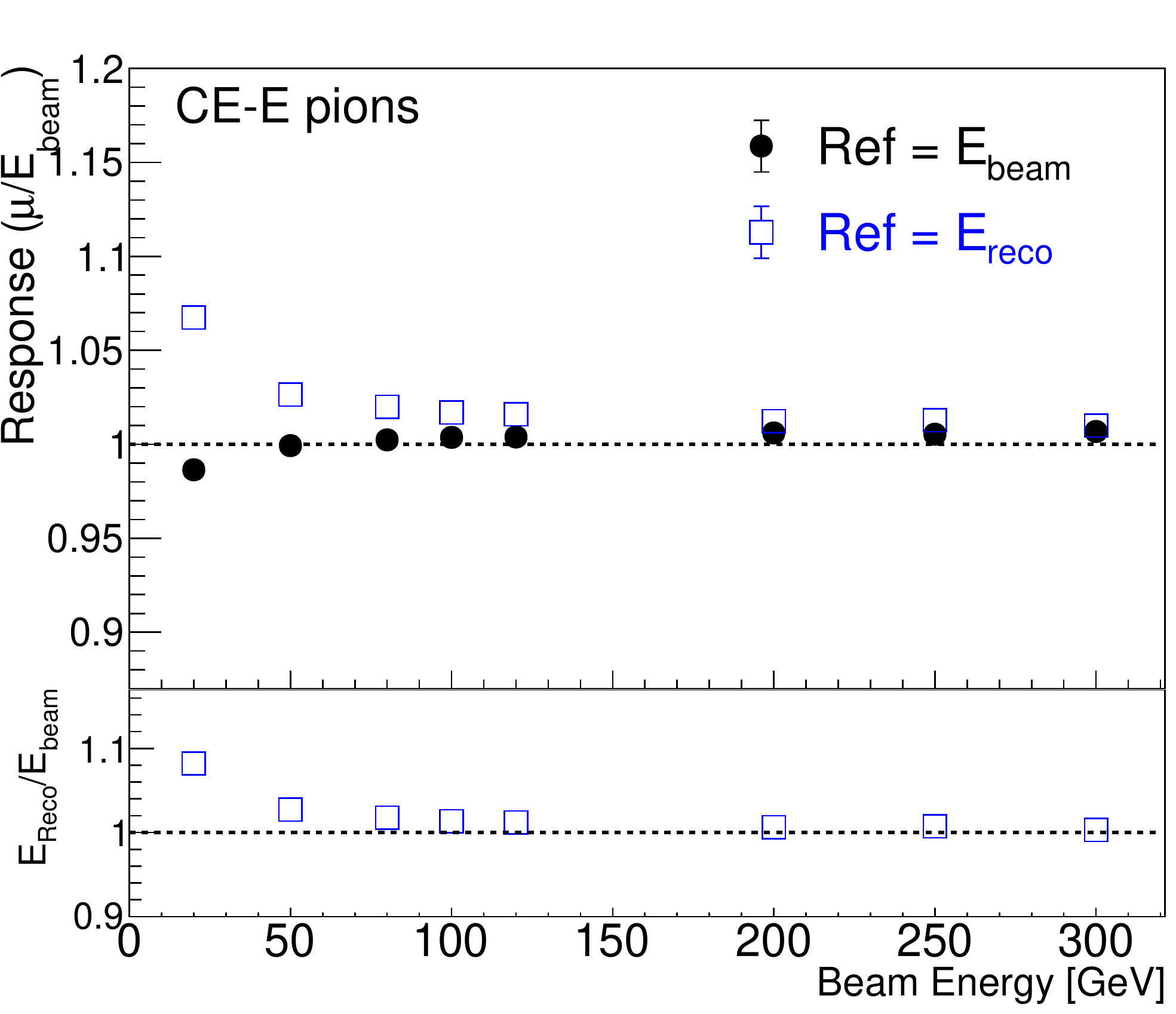}
  \includegraphics[width=0.45\linewidth]{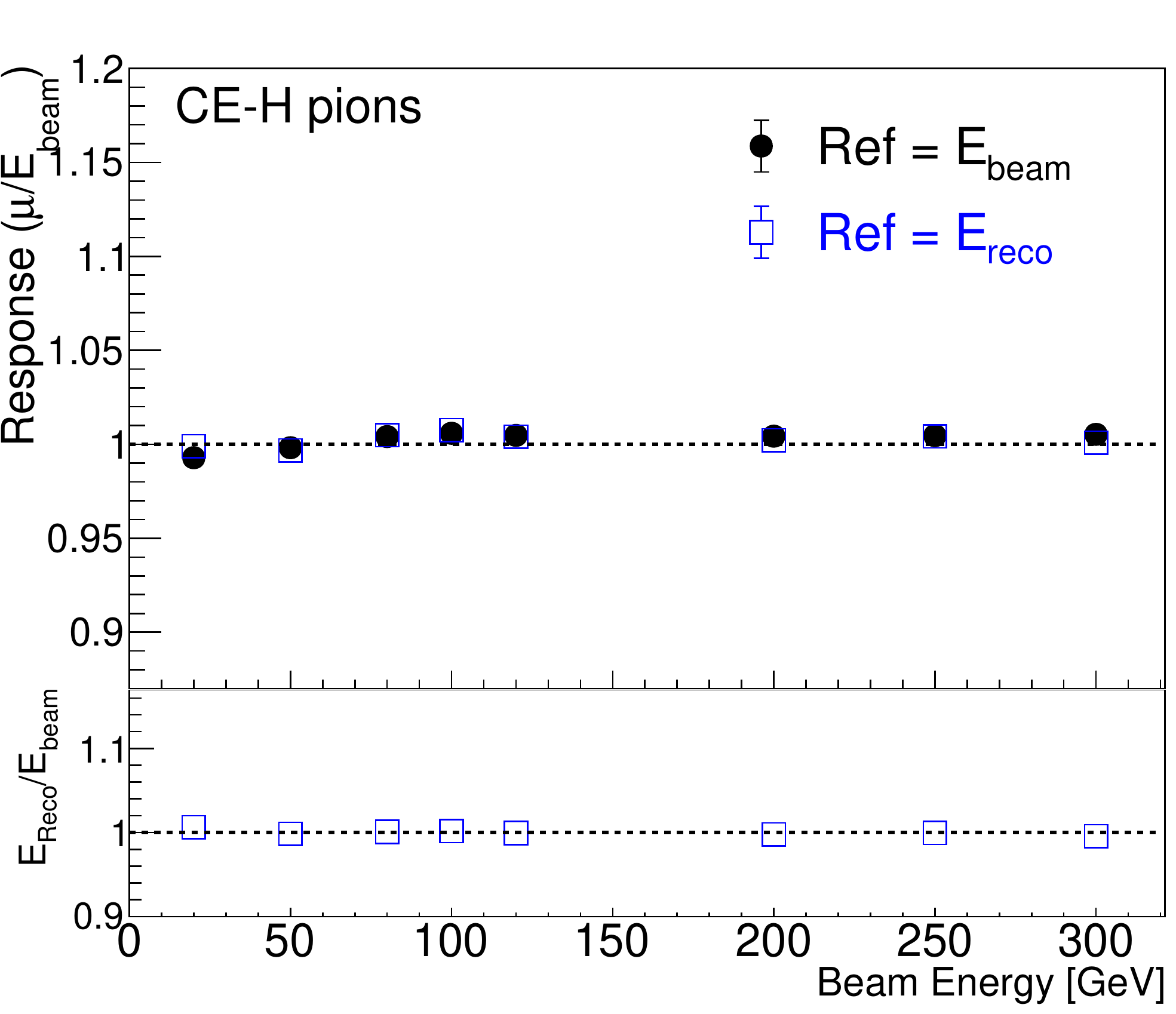}
  \caption{\label{fig:resp-data-erecoref} Response obtained using reconstructed calorimeter energy as reference to apply calibration factors $\alpha_1$, $\beta_{1,2}$, and $\gamma_{1,2}$ from the fitted functions for CE-E pions (left) and CE-H pions (right).}
\end{figure}

\section{Longitudinal and transverse shower profiles of pions}
\label{sec:long-trans-showers}

In this section, we discuss some key features of longitudinal and transverse development of the pion showers in different compartments of the prototype detector as observed in the beam test data and the simulated event sample. The energy scale in the simulation has been corrected to match the data as discussed in Section \ref{sec:enereco-det}.

\subsection{Longitudinal shower development in data and simulation} \label{sec:long-showers}

The average longitudinal profile of pions, described as the mean energy measured in units of MIPs in each active layer, is presented in Figure \ref{fig:long-data-mc} for $E_{beam} =$ 20, 100, and 300 GeV. The left column corresponds to the pions which start showering in the third layer of CE-E and the right column corresponds to those undergoing the first hadronic interaction in the second layer of CE-H prototype section. The narrow peak in the CE-E prototype section is mainly dominated by the early $\pi^0$ production which results in a compact electromagnetic shower in CE-E. The kink at the transition of CE-E and CE-H prototype sections can be attributed to a change in material thickness at the front of the CE-H layers and a larger transverse coverage due to the presence of seven modules. The lower average energies measured in the last three layers of the CE-H section are due to the presence of only one silicon module in these layers (as compared to the seven modules in its remaining layers). Both the \qgsp{} and \ftfp{} physics lists closely reproduce all the features of longitudinal shower development as observed in data in the CE-E, CE-H and AHCAL prototype sections. For showers starting in CE-E for higher beam energies, both the physics lists under-predict the energy deposited in AHCAL by 15$-$20\%, indicating shorter simulated showers (also suggested by comparisons shown in Figure \ref{fig:emips-datasim-EEFHAH}). At an intermediate energy of 100 GeV, the simulation compares well with the data. It is worth repeating here that the electronic effects have not been simulated in detail. This could affect the comparisons when energy deposits are small in a given layer. Hence, a judgement on tuning the physics lists awaits further studies including quantitative estimate of residual instrumental effects and signal digitization.

The simulation describes reasonably well the longitudinal showers measured in data for the pions which undergo their first hadronic interaction in the second layer of CE-H prototype sections, as shown in Figure \ref{fig:long-data-mc} (right). The profiles show a broad shower maximum which is also consistent with the fact that the \radL{} to \intL{} ratio is higher in steel than in the lead absorbers. As expected, the shower develops deep in the AHCAL prototype section and a substantial fraction of energy is measured here. The data and simulation mismatch observed in the last three layers of the CE-H may indicate differences in the beam profile of the particle or in the angle of their propagation in the detectors.

\begin{figure}[h]
  \centering
  \includegraphics[width=0.45\linewidth,height=0.30\linewidth]{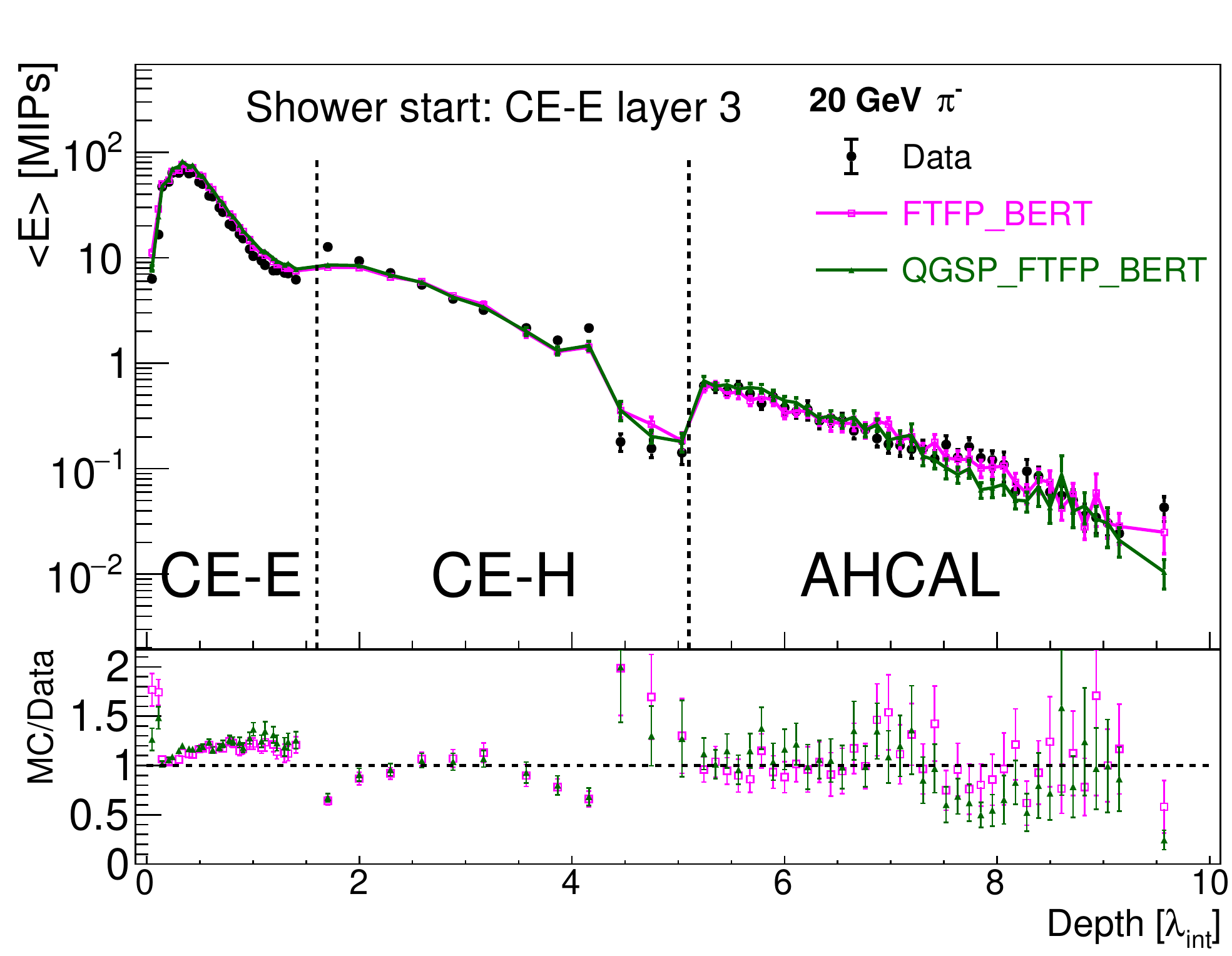}
  \includegraphics[width=0.45\linewidth,height=0.30\linewidth]{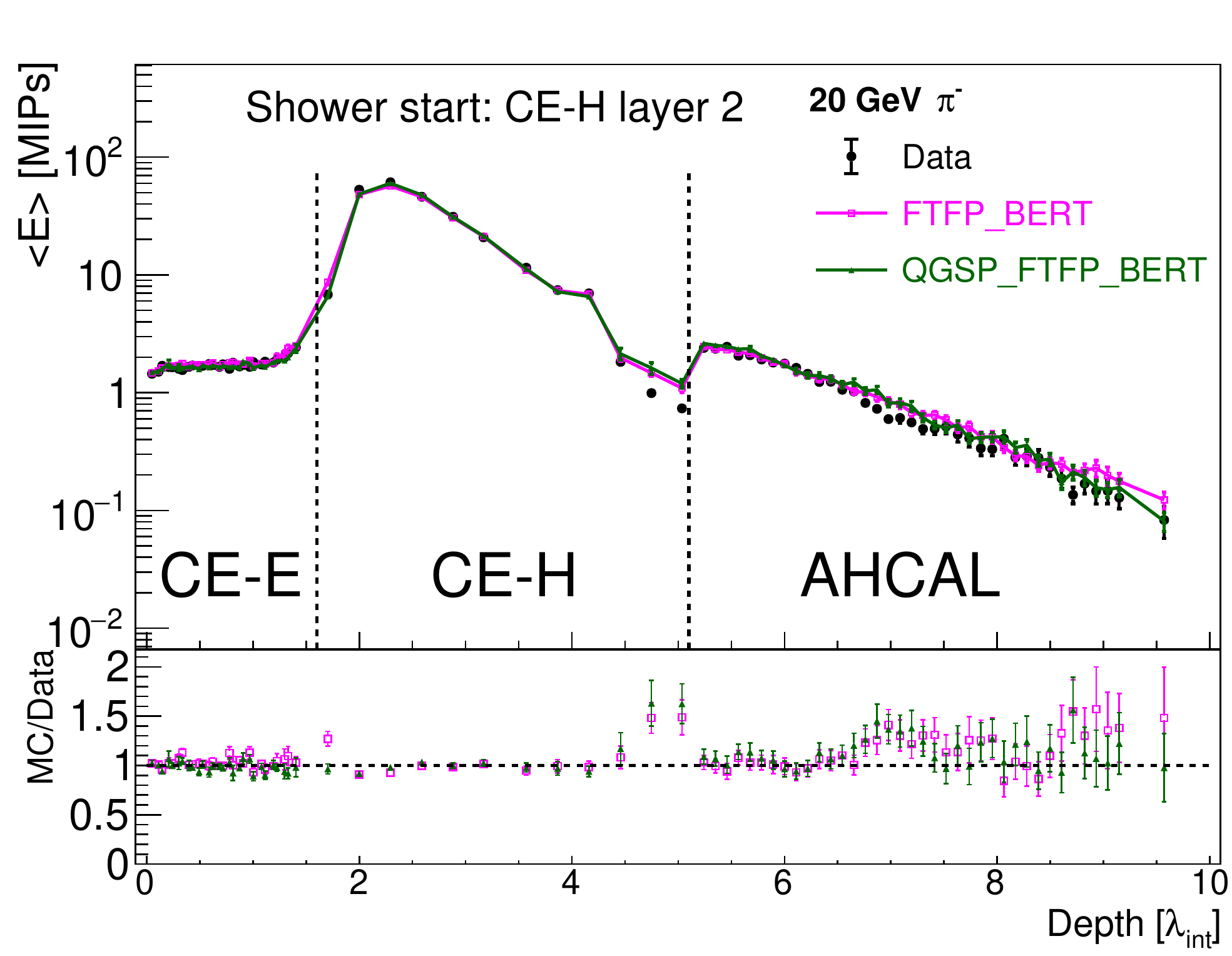}
  \includegraphics[width=0.45\linewidth,height=0.30\linewidth]{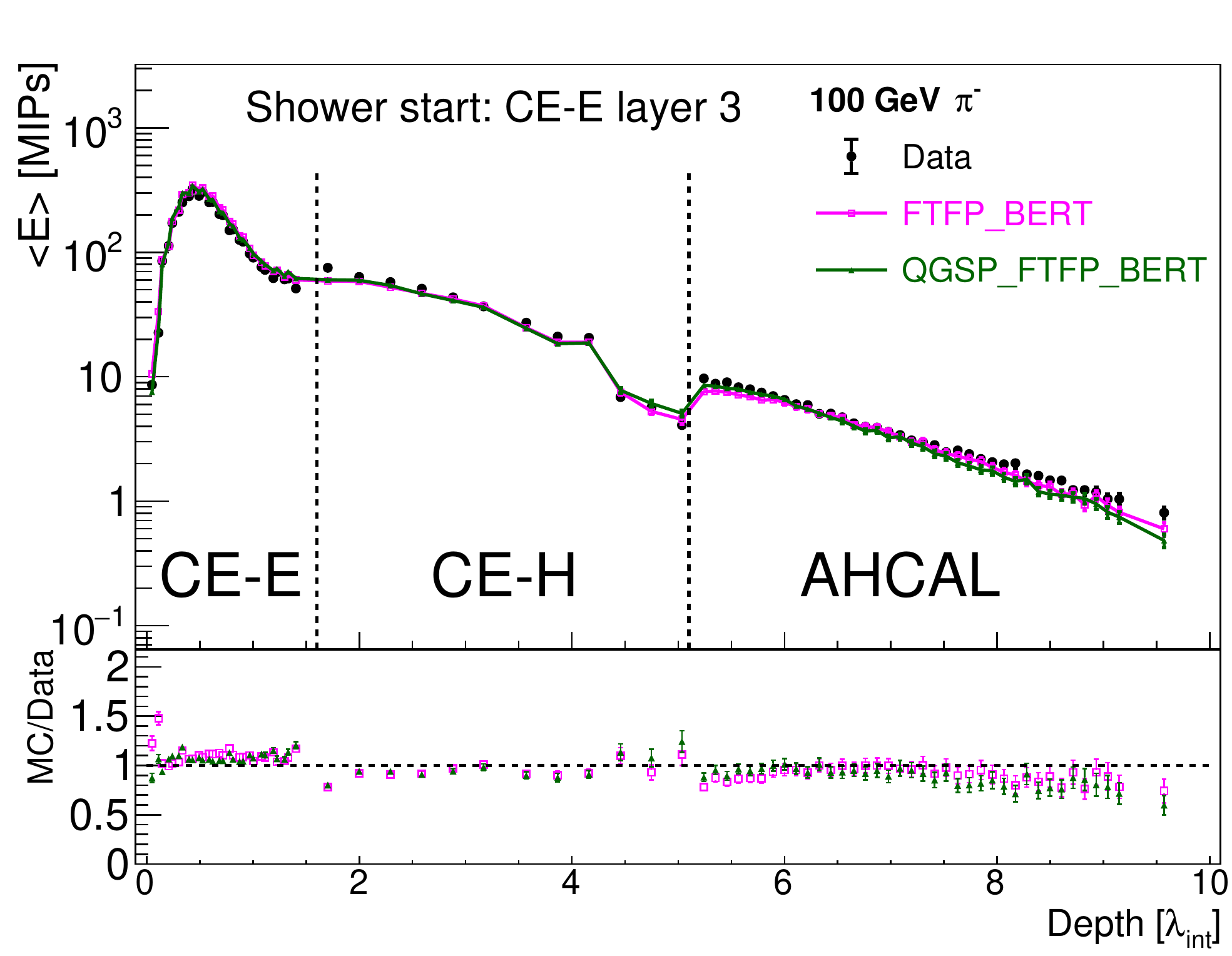}
  \includegraphics[width=0.45\linewidth,height=0.30\linewidth]{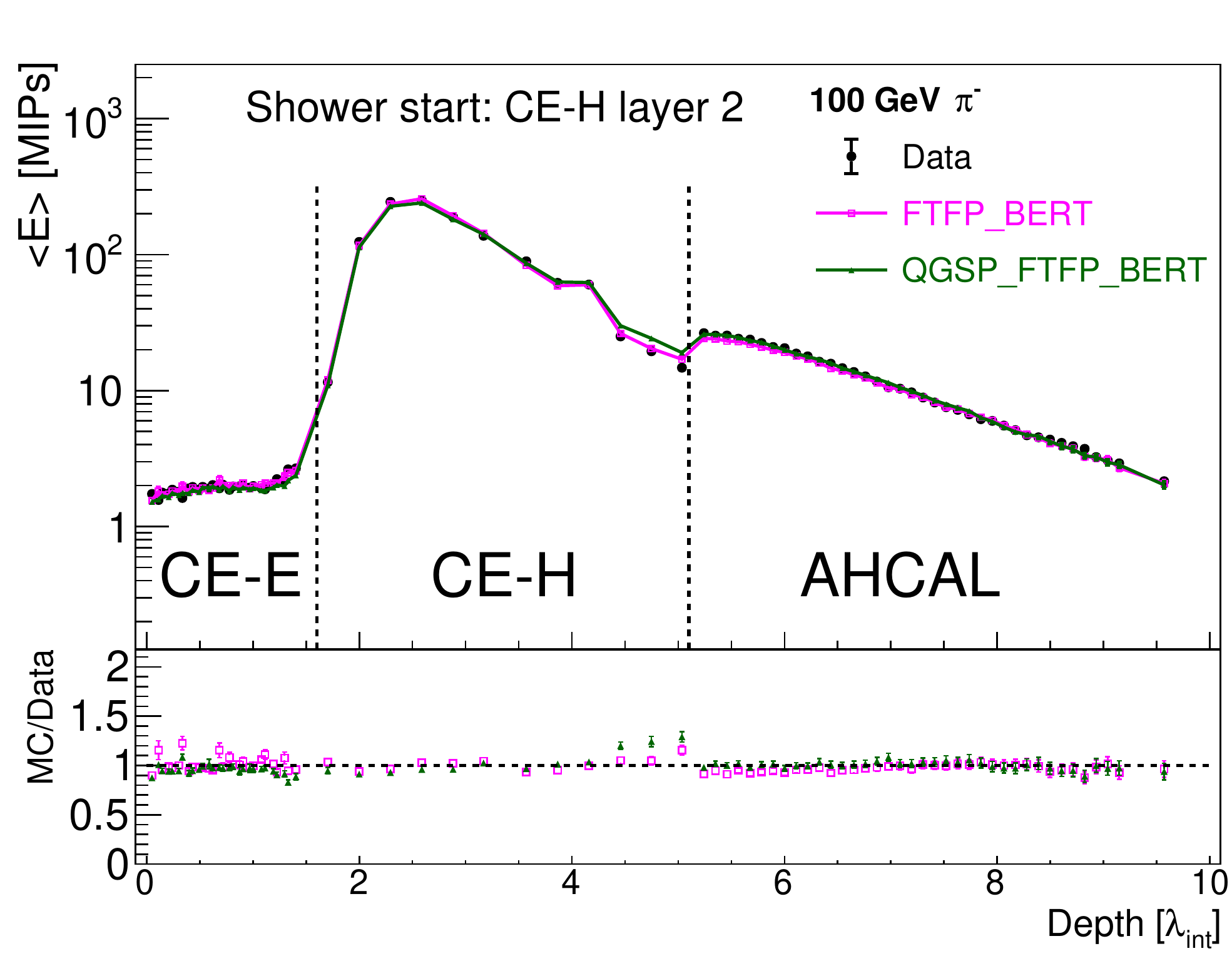}
  \includegraphics[width=0.45\linewidth,height=0.30\linewidth]{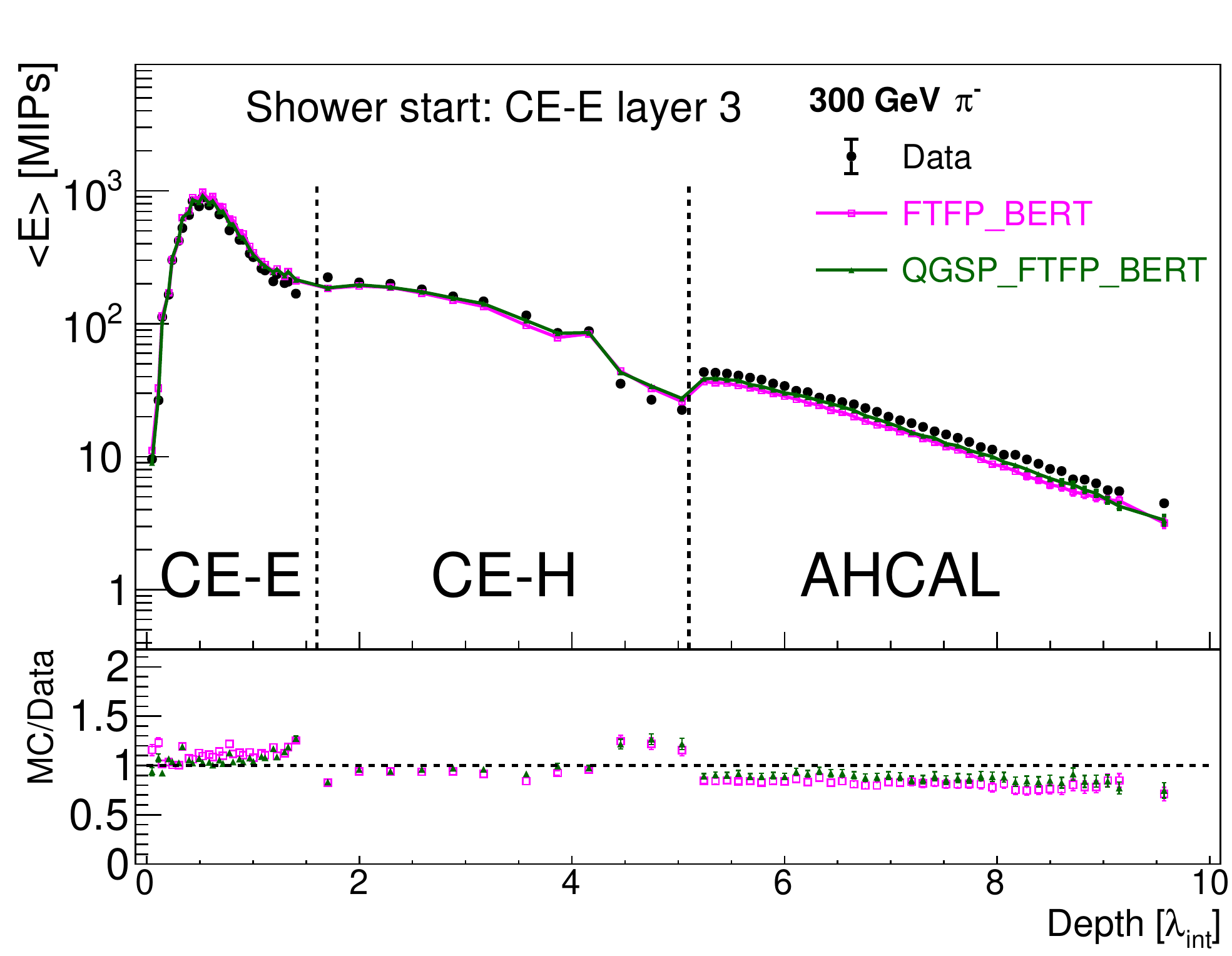}
  \includegraphics[width=0.45\linewidth,height=0.30\linewidth]{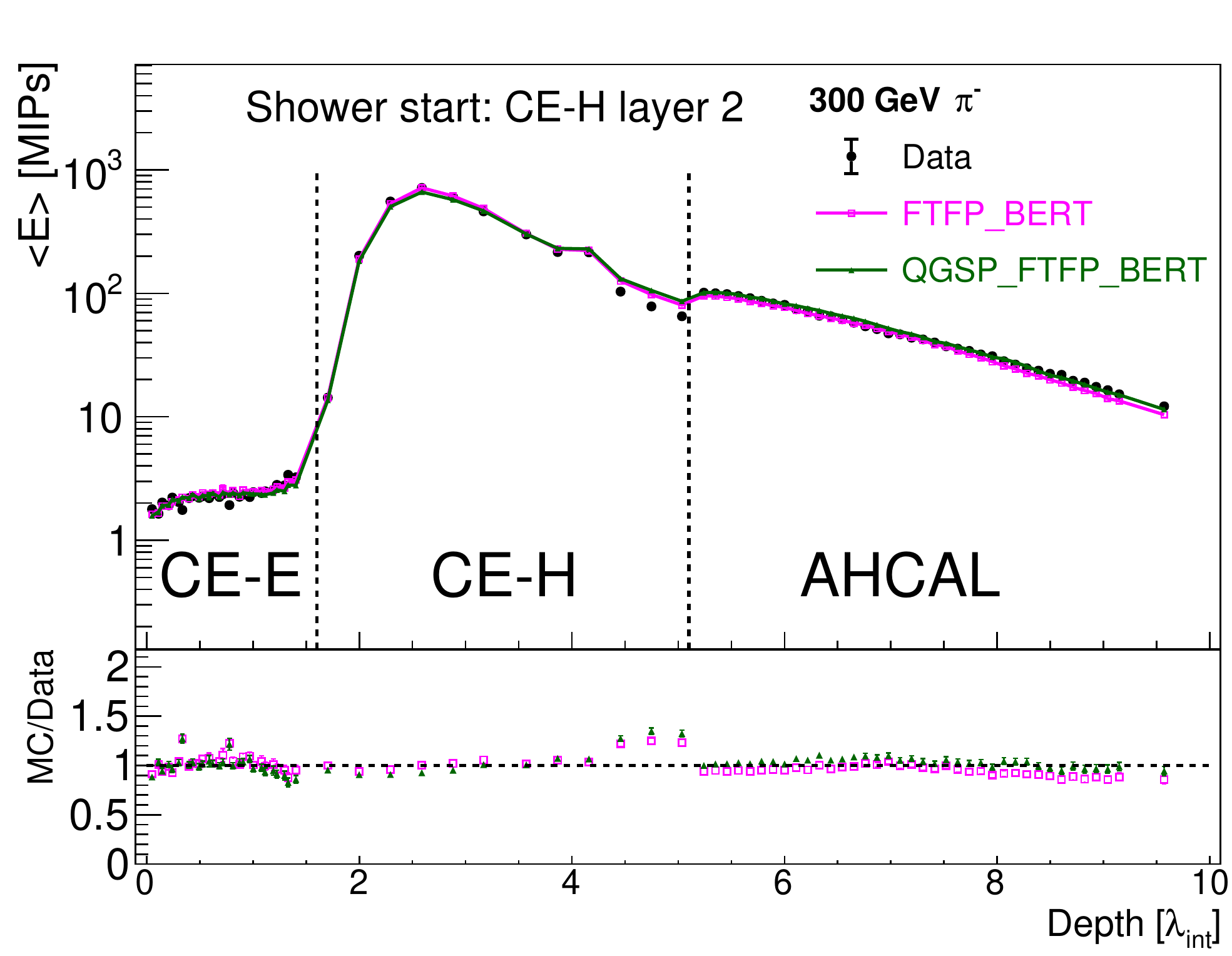}
  \caption{\label{fig:long-data-mc} Longitudinal shower profiles measured in data and predicted by the \qgsp{} and \ftfp{} physics lists as numbers of MIPs recorded in the active layers of the CE-E, CE-H and AHCAL prototypes for pions starting to shower in layer 3 of the CE-E prototype (left) and layer 2 of the CE-H prototype (right). The three rows correspond to beam energies of 20, 100 and 300 GeV (from upper to lower).}
\end{figure}

\subsection{Transverse shower profile} \label{sec:trans-showers}

The modeling of the transverse spread of pion showers as these evolve in direction of the incident particle is an important aspect required to be well simulated by the hadronic physics lists of GEANT4 for a correct assignment of energy deposits in the presence of multiple incident particles. The high longitudinal and transverse segmentation of our calorimeter sections allow us to compare the lateral profiles of energy deposited in various layers, and study the details of shower modeling. We present a brief summary of transverse spread in the CE-E, CE-H, and AHCAL compartments in this section.

The pion track reconstructed using the DWC hits (Section \ref{sec:evtsel}) is extrapolated to the last layer of the detector setup. The fraction of energy deposited in cylinders of varying radii around the track in the three calorimeter sections is presented in Figure \ref{fig:trans-data} for pions of 20 GeV and 300 GeV beam energies. The fraction of energy measured here is normalized to total energy measured in the respective calorimeter section, and does not account for potential transverse leakage. The Figure \ref{fig:trans-data} (top row) shows the transverse energy profile of pions measured in the CE-E section using the pions which start showering in layers 3$-$7 of the CE-E. The GEANT4 physics lists predict the profiles well at higher energies but show discrepancies close to the shower axis. The Figures \ref{fig:trans-data} (middle and bottom rows) show the transverse shower profile of pions in CE-H and AHCAL sections using the pions which start showering in the first layer of the CE-H. The simulated profiles show a disagreement at the core of the shower compensated by the energy distribution away from the core. These disagreements could potentially be a combined effect of differences in the beam profiles in data and simulation, and underlying differences in modeling of hadron showers in the simulation.

\begin{figure}[ht]
  \centering
  \includegraphics[width=0.45\linewidth]{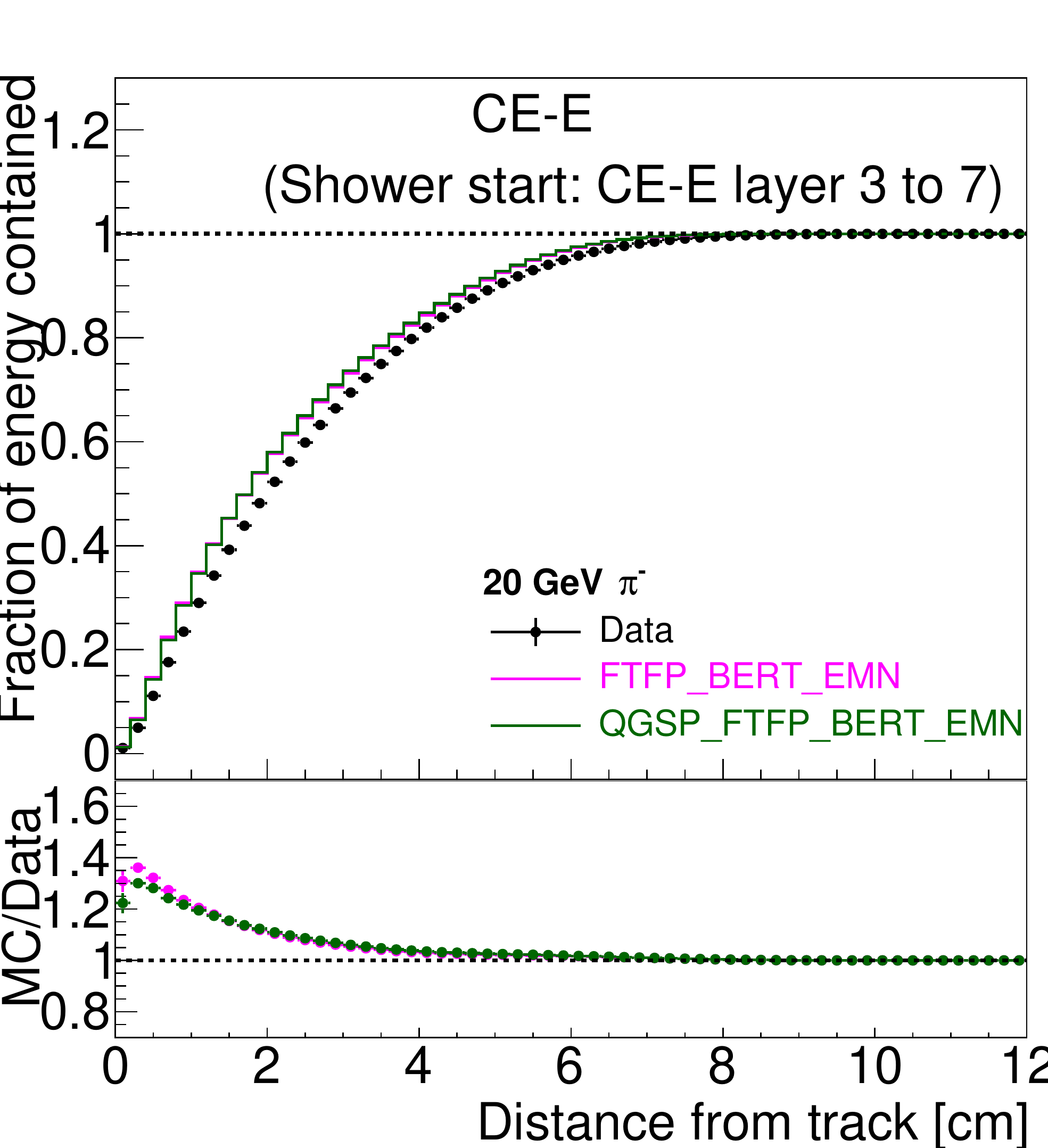}
  \includegraphics[width=0.45\linewidth]{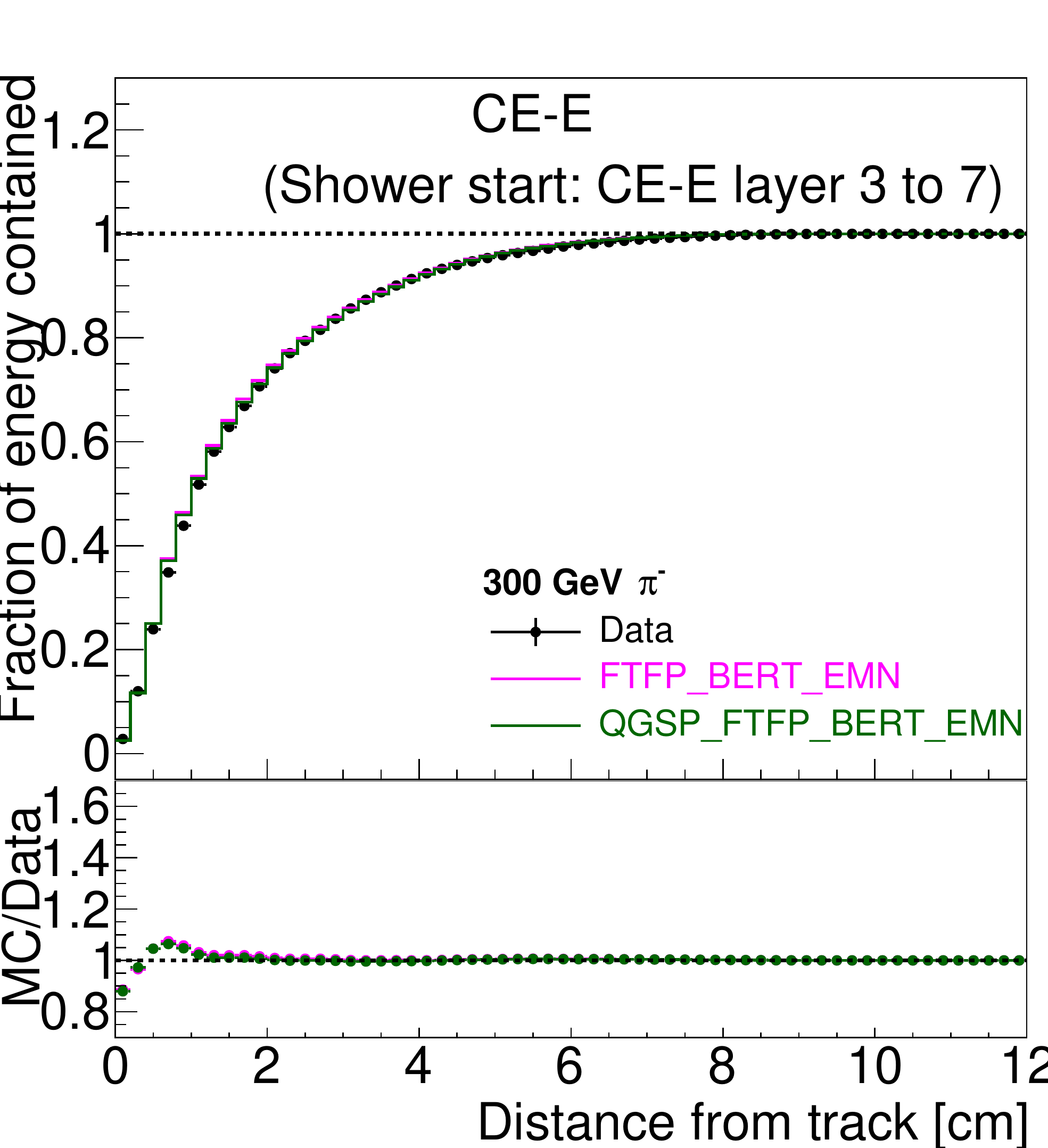}
  \includegraphics[width=0.45\linewidth]{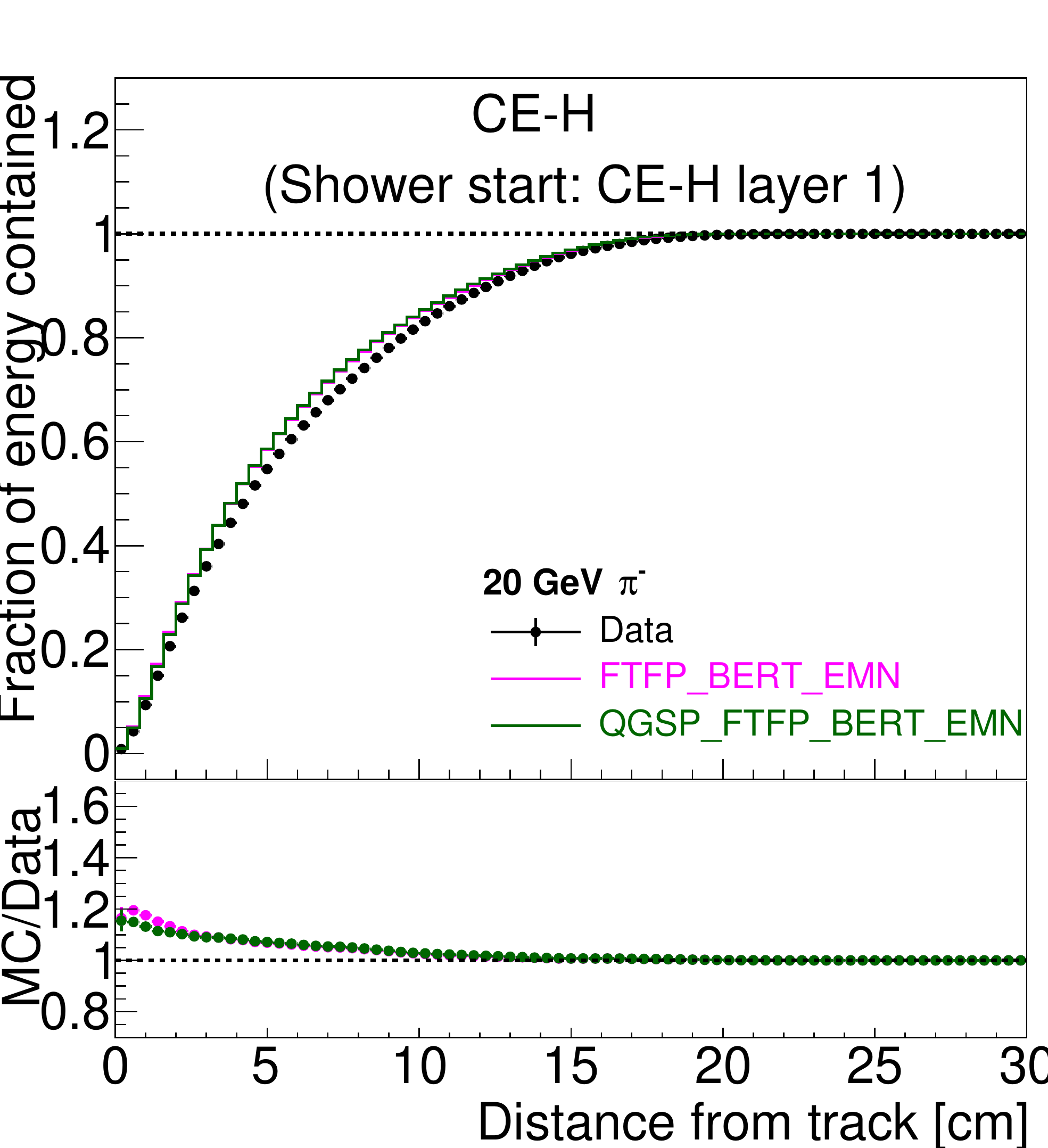}
  \includegraphics[width=0.45\linewidth]{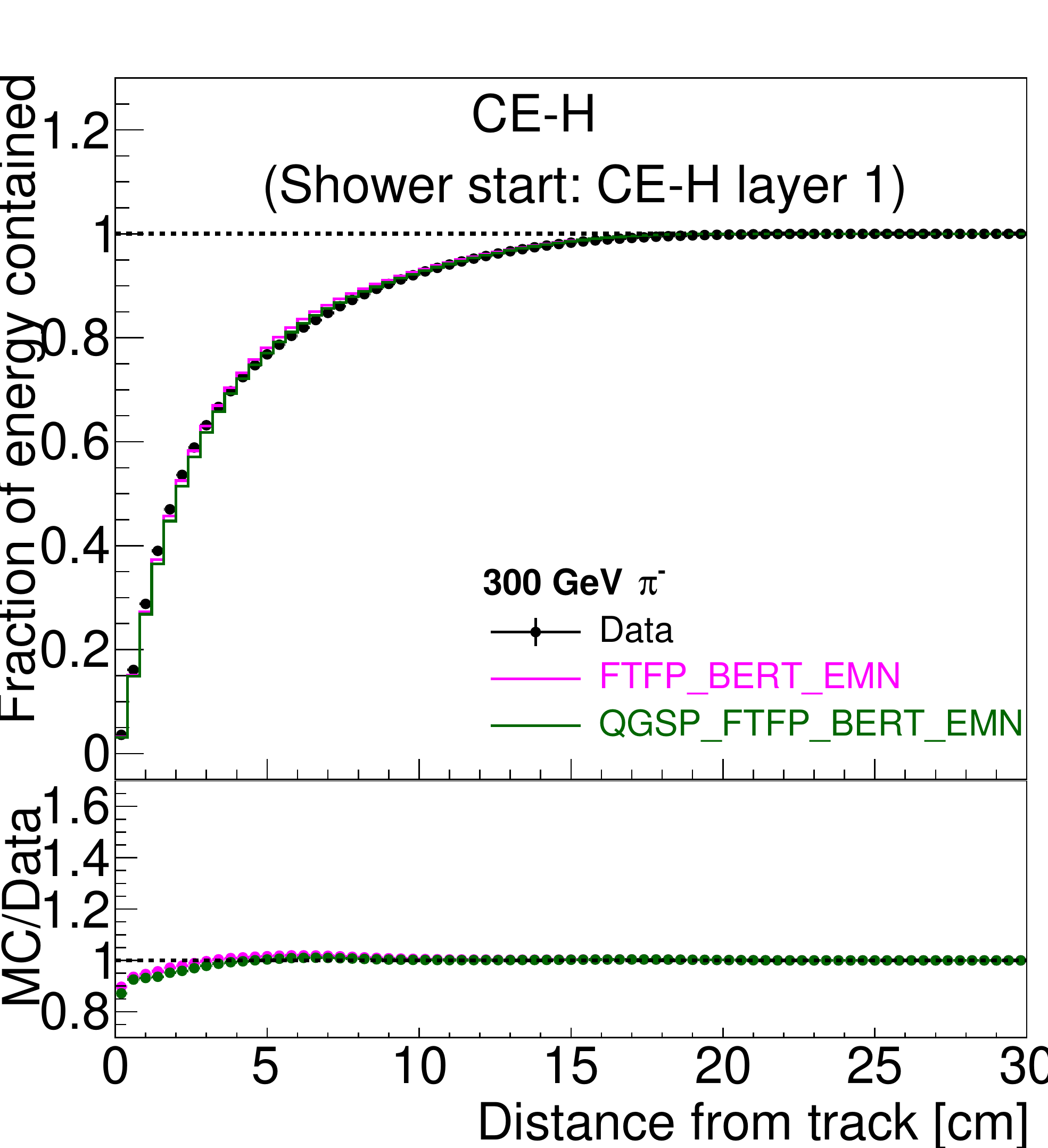}
  \includegraphics[width=0.45\linewidth]{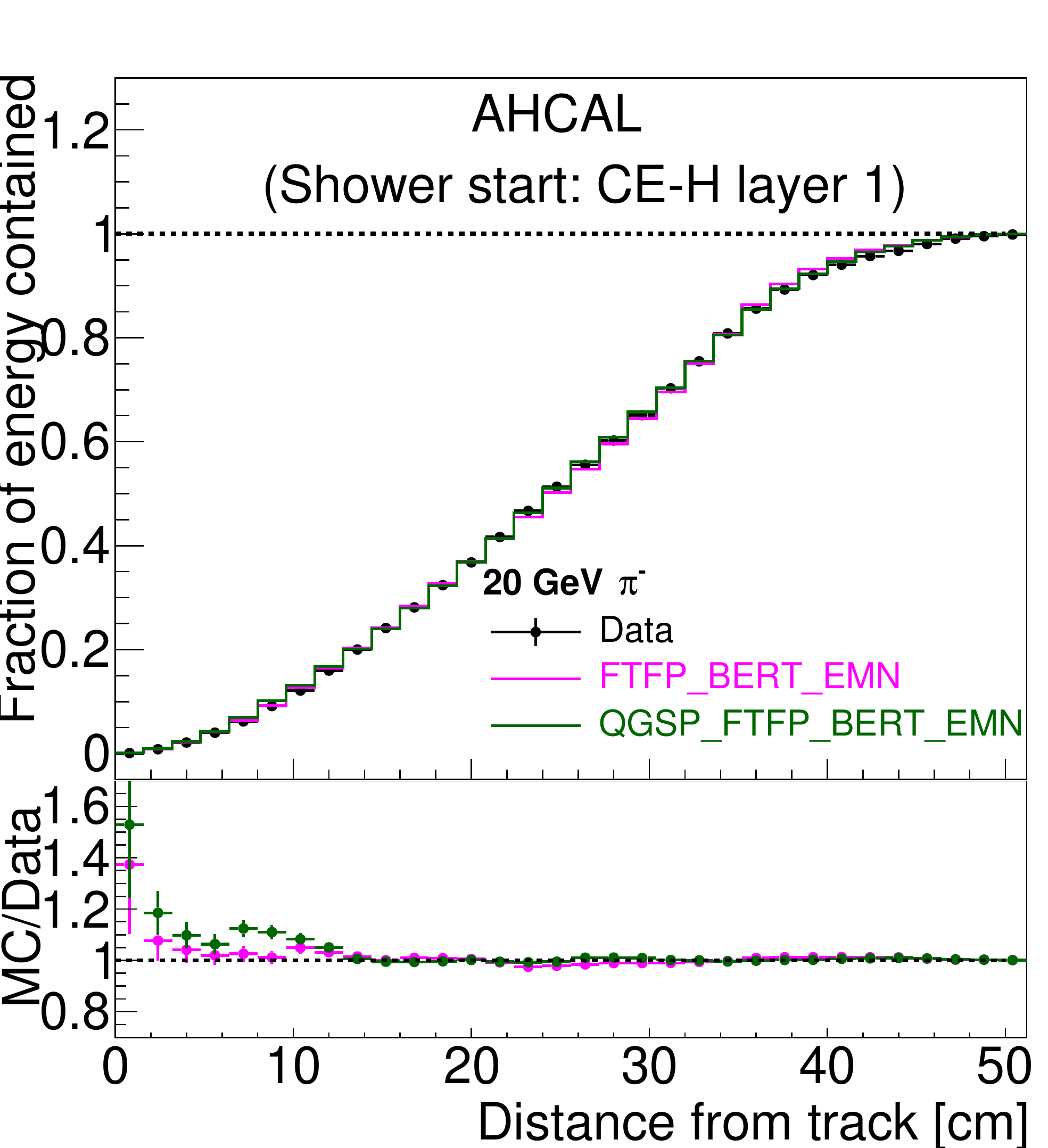}
  \includegraphics[width=0.45\linewidth]{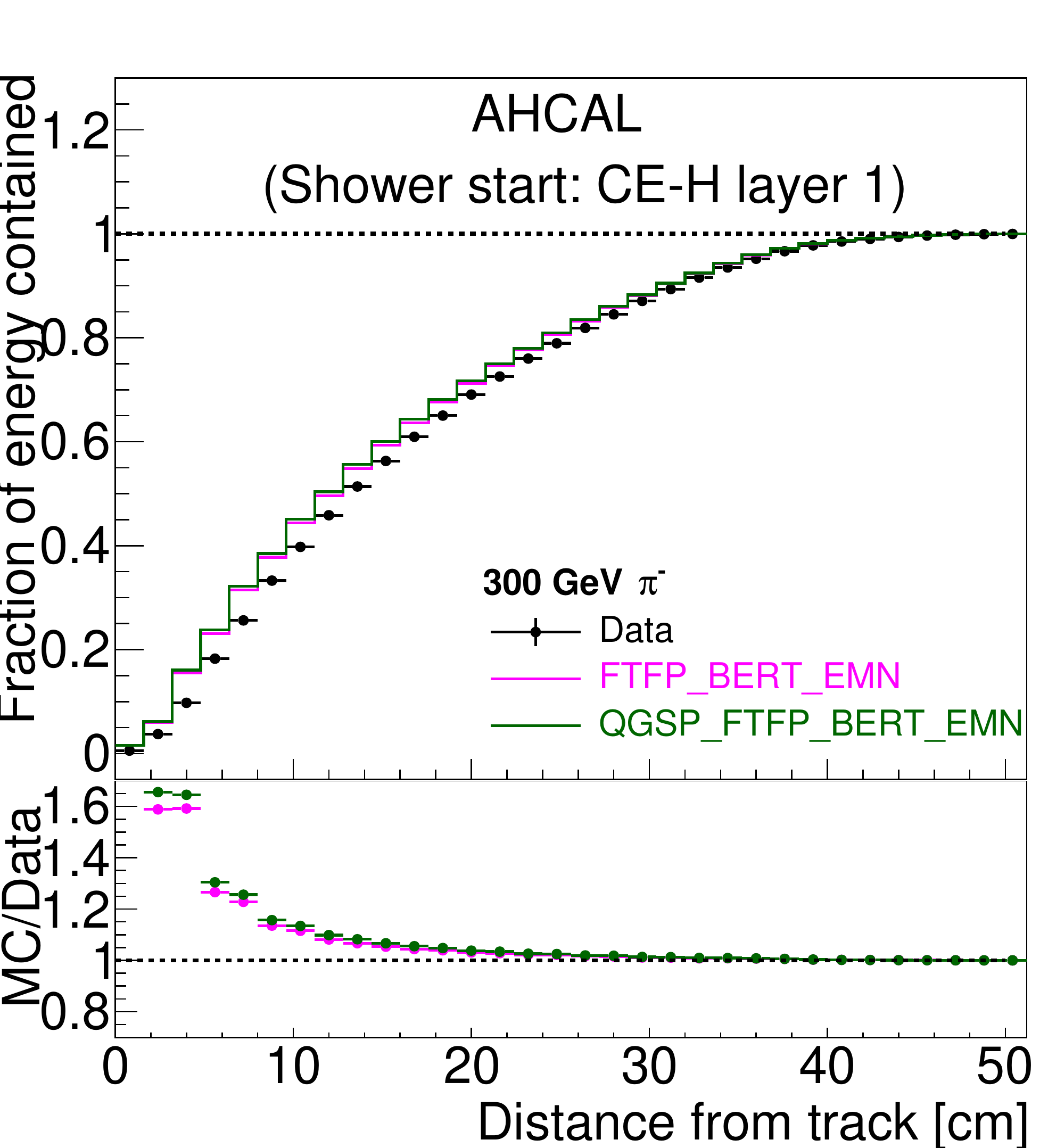}
  \caption{\label{fig:trans-data}{Transverse shower profiles measured in data and simulation for pions of 20 GeV (left) and 300 GeV (right) in CE-E (top), CE-H (middle), and AHCAL (bottom) prototype sections.}}
\end{figure}

\section{Summary and outlook}
\label{sec:summary}

We study the performance of a prototype of the CMS HGCAL detector based on silicon sensors combined with a prototype of CALICE AHCAL detector based on scintillator tiles read out by SiPMs. We report the first measurements of energy response and resolution, and longitudinal and transverse profiles of hadronic showers produced by negatively charged pions in this combined detector. The results are compared with a detailed simulation of hadronic showers using two physics lists, \qgsp{} and \ftfp{}, of the GEANT4. We also present an algorithm to identify the starting point of the hadronic shower in the detector. With the energy scale of the electromagnetic section fixed by 50 GeV positrons and that of the hadronic section fixed by 50 GeV pions, we observe that the response in simulation is higher by 9.5\% for the pions that do not interact in CE-E and 5\% for the pions which start showering in CE-E.

The non-linearity of the response is very well reproduced in the simulation after correction of the above-mentioned difference of energy scale. To compensate for the non-linearity of the calorimeter, energies measured in different compartments are combined using energy dependent weights which are obtained using a $\chi^2$-minimization in data. Here these calibration weights are applied using the beam energy as a reference. We obtained a stochastic term of 131.7$\pm$1.0\% for pions showering in CE-E and 122.1$\pm$1.4\% for pions that behave like MIPs in CE-E for the resolution measured in data. The corresponding constant terms are 8.5$\pm$0.1 and 9.0$\pm$0.2 respectively. Both the physics lists predict the stochastic term and constant terms within 8$-$10\% of those measured in the data. In an alternate approach, we use the energy measured using fixed weights as reference to apply the same energy-dependent calibrations obtained using the data. This results in similar or slightly better resolution of pions as compared to that obtained using beam energy as the reference for the single energy points used in this analysis. We recognize the application of neural networks can add significant performance enhancement in a high granularity calorimeter, and these ideas are currently under investigation.

We then present comparisons of the measured and simulated longitudinal and transverse shower profiles. Overall characteristics of shower development observed in the CE-E, CE-H and AHCAL sections are also closely predicted by the simulation except an over prediction of average energy deposited in CE-E for low beam energies and under prediction in the AHCAL for higher beam energies. The energy deposited close to the particle track extrapolated to the three detector sections also shows large differences compared to simulation. Inputs from a detailed treatment of digitization and electronics effects in simulation are required before concluding the need for further tuning of the physics lists to model hadron showers in this prototype detector setup.

\clearpage
\acknowledgments
We thank the technical and administrative staffs at CERN and at other CMS institutes for their contributions to the success of the CMS upgrade program. We acknowledge the enduring support provided by the following funding agencies and laboratories: BMBWF and FWF (Austria); CERN; CAS, MoST, and NSFC (China); MSES and CSF (Croatia); CEA, CNRS/IN2P3 and P2IO LabEx (ANR-10-LABX-0038) (France); SRNSF (Georgia); BMBF, DFG, and HGF (Germany); GSRT (Greece); DAE and DST (India);  MES (Latvia); MOE and UM (Malaysia); MOS (Montenegro); PAEC (Pakistan); FCT (Portugal); JINR (Dubna); MON, RosAtom, RAS, RFBR, and NRC KI (Russia); MoST (Taipei); ThEP Center, IPST, STAR, and NSTDA (Thailand); TUBITAK and TENMAK (Turkey); STFC (United Kingdom); and DOE (USA).




\end{document}